\documentclass[a4paper,oneside,final,notitlepage,onecolumn,12pt]{article}

\usepackage[T1]{fontenc}
\usepackage[utf8]{inputenc}

\usepackage{amsmath}
\usepackage{amssymb}
\usepackage{authblk}
\usepackage{bbm}
\usepackage{accents}

\usepackage{graphicx}
\usepackage{subcaption}

\usepackage[abs]{overpic}

\DeclareFontFamily{OT1}{rsfs}{} \DeclareFontShape{OT1}{rsfs}{m}{n}{
	<-7> rsfs5 <7-10> rsfs7 <10-> rsfs10}{}
\DeclareMathAlphabet{\mathscr}{OT1}{rsfs}{m}{n}

\usepackage[normalem]{ulem}
\usepackage{color}

\newcounter{mnotecount}%[section]

\newcommand{\mnotex}[1]%{}
{\protect{\stepcounter{mnotecount}}$^{\mbox{\footnotesize $\bullet$\themnotecount}}$
	\marginpar{\color{red}
		\raggedright\tiny\em
		$\!\!\!\!\!\!\,\bullet$\themnotecount: #1} }
\newcommand{\mmnotex}[1]%{}
{
    $^{\mbox{\footnotesize $\bullet$}}$
    \marginpar{
        \color{forest}
        \raggedright\tiny\em
        $\!\!\!\!\!\!$: #1
    }
}

\newcommand{\instar}[1]{\accentset{\smash{\raisebox{-0.12ex}{$\scriptstyle\star$}}}{#1}\rule{0pt}{2.3ex}}

\newcommand{\interior}[1]{\accentset{\smash{\raisebox{-0.12ex}{$\scriptstyle\circ$}}}{#1}\rule{0pt}{2.3ex}}

\newcommand{\NN}{\mathbf{{N}}}

\newcommand{\NNh}{\widehat{N}}
\newcommand{\bgNNh}{{}^{\scriptscriptstyle(\!0\!)}\hspace{-1.2pt}\NNh}
\newcommand{\dNNh}{{}^{\scriptscriptstyle(\!\Delta\!)}\hspace{-1.2pt}\NNh}
\newcommand{\NNt}{\mathbf{\NN}}
\newcommand{\NNtb}{\overline{\NNt}}
\newcommand{\kk}{\mathbf{k}}
\newcommand{\kkb}{\overline{\kk}}
\newcommand{\dkk}{{}^{\scriptscriptstyle(\!\Delta\!)}\hspace{-1.2pt}\kk}
\newcommand{\dkkb}{\overline{\dkk}}
\newcommand{\bgkk}{{}^{\scriptscriptstyle(\!0\!)}\hspace{-1.2pt}\kk}
\newcommand{\bgkkb}{\overline{\bgkk}}
\newcommand{\KK}{\mathbf{K}}
\newcommand{\dKK}{{}^{\scriptscriptstyle(\!\Delta\!)}\hspace{-1.2pt}\KK}
\newcommand{\bgKK}{{}^{\scriptscriptstyle(\!0\!)}\hspace{-1.2pt}\KK}

\newcommand{\Ks}{\overset{\star}{{K}}}
\newcommand{\Kc}{\interior{\mathbf{K}}}
\newcommand{\KKc}{\interior{\KK}}
\newcommand{\Kcqq}{\KKc_{qq}}

\newcommand{\KKh}{\widehat{K}}
\newcommand{\kkappa}{\boldsymbol\kappa}
\newcommand{\dkappa}{{}^{\scriptscriptstyle(\!\Delta\!)}\hspace{-1.2pt}\kkappa}
\newcommand{\bgkappa}{{}^{\scriptscriptstyle(\!0\!)}\hspace{-1.2pt}\kkappa}
\newcommand{\dd}{\mathbf{d}}
\newcommand{\aaa}{\mathbf{a}}
\newcommand{\BB}{\mathbf{B}}
\newcommand{\BBb}{\overline{\BB}}
\newcommand{\bb}{\mathbf{b}}
\newcommand{\bbb}{\overline{\bb}}
\newcommand{\AAA}{\mathbf{A}}
\newcommand{\AAAb}{\overline{\AAA}}
\newcommand{\CC}{\mathbf{C}}
\newcommand{\CCb}{\overline{\CC}}
\newcommand{\ff}{\mathbf{f}}
\newcommand{\dff}{{}^{\scriptscriptstyle(\!\Delta\!)}\hspace{-1.2pt}\ff}
\newcommand{\FF}{\mathbf{F}}
\newcommand{\dFF}{{}^{\scriptscriptstyle(\!\Delta\!)}\hspace{-1.2pt}\FF}
\newcommand{\RR}{{R}}
\newcommand{\RRh}{\widehat{\RR}}
\newcommand{\ethb}{\overline{\eth}}

\definecolor{forest}{RGB}{34,139,34}

\renewcommand{\theequation}{\thesection.\arabic{equation}}
\title{Numerical investigations of the asymptotics of solutions to the evolutionary form of the constraints}

\author[,1]{K\'aroly Csuk\'as \footnote{E-mail address:{\tt csukas.karoly@wigner.mta.hu}}}
\author[,1,2]{Istv\'an R\'acz \footnote{E-mail address:{\tt racz.istvan@wigner.mta.hu}}}

\affil[1]{Wigner RCP, H-1121 Budapest, Konkoly Thege Mikl\'{o}s \'{u}t  29-33, Hungary}
\affil[2]{Faculty of Physics, University of Warsaw, Ludwika Pasteura 5, 02-093 Warsaw, Poland}

\begin{document}
\maketitle
\begin{abstract}
	Systematic numerical investigations of the asymptotics of near Schwarzschild vacuum initial data sets is carried out by  inspecting solutions to the parabolic-hyperbolic and to the algebraic-hyperbolic forms of the constraints, respectively. One of our most important findings is that the concept of near Schwarzschild configurations, applied previously in \cite{Beyer:2017njj,Beyer:2019kty}, is far too restrictive. It is demonstrated that by relaxing the conditions on the freely specifiable part of the data a more appropriate notion of near Schwarzschild initial data configurations can be defined which allows us to generate asymptotically flat initial data configurations.
\end{abstract}

\section{Introduction}
\setcounter{equation}{0}

Near Schwarzschild vacuum initial data configurations are automatically regarded as being asymptotically flat. This expectation may stem from the fact that when the constraints are solved by using elliptic method suitable fall off (or boundary) conditions at infinity can be imposed to guarantee asymptotic flatness \cite{York:1971}. Indeed, under suitable conditions,  requiring e.g., the trace of the extrinsic curvature to be (almost) constant, or, alternatively by demanding smallness of the `TT' (transverse-traceless) part of the rescaled extrinsic curvature the existence of solutions to the corresponding elliptic boundary value problem can be shown \cite{anderson:2018,maxwell:2017}. As alternatives to the elliptic approach, recently, two evolutionary formulations of the constraint equations---a parabolic-hyperbolic and an algebraic-hyperbolic formulation---were introduced in \cite{Racz2016,Racz:2014jra,Racz:2014gea}. Since there is a one to one correspondence between space of solutions to the constraints produced by the elliptic method and the space of solutions yielded by either of the evolutionary methods, one would expect that the asymptotically flat solutions could, in principle, be generated by making use of the evolutionary methods, as well. The first numerical studies of the related issues were carried out in \cite{Beyer:2017njj,Beyer:2019kty}, where axially symmetric near Schwarzschild initial data configurations were considered. The investigations in \cite{Beyer:2017njj,Beyer:2019kty} made it clear that it is not at all obvious how one could generate, by making use of the evolutionary method, asymptotically flat initial data configurations. Clearly, the main technical difficulty originates from the fact that while applying either of the evolutionary methods, in generating near Schwarzschild configurations, data for the constrained fields can only be chosen as free at the initial $2$-sphere. This, in principle, may not allow one to acquire control on the asymptotic behavior of the yielded solutions since, while applying either of the evolutionary methods, we cannot impose the usual fall off conditions on the to be solution that can be done in applying the elliptic method. It is worth, however, to be emphasized here the evolutionary methods do also have some complementary advantages which are not available in the elliptic method. Namely, in advance of solving the elliptic equations one has to fix all of the freely specifiable fields once and for all globally. As opposed to this when either of the evolutionary methods is applied we have an enormous freedom in adjusting the freely specifiable variables simply in the interim of the ``time integration''. Indeed, the freely specifiable fields can be fixed while proceeding leaves by leaves according to the needs of the desired behavior of constrained fields.
All these make it of obvious interest to know if asymptotic flatness can be guaranteed by choosing suitably the freely specifiable fields and the initial data for the evolutionary form of the constraints.  In this paper, by utilizing the aforementioned freedom, the asymptotics of near Schwarzschild initial data configurations are investigated by integrating numerically the parabolic-hyperbolic and algebraic-hyperbolic form of the constraints, respectively. One of our most important findings is that the notion of near Schwarzschild initial data configurations, used in all the previous investigations \cite{Beyer:2017njj,Beyer:2019kty}, is far too restrictive. In particular, it is shown that once we can get a control on the  monopole part of the trace  ${\rm\bf K}={\rm\bf K}^l{}_l=\widehat \gamma{}^{kl}K_{kl}$ of the three-dimensional extrinsic curvature with respect to the metric induced on the foliating two-spheres---see subsection \ref{subsec: basics} for definitions---even the strong asymptotic flatness of the solutions to the evolutionary form of the constraints can be guaranteed.

Notably, the authors in \cite{Beyer:2020kty}  have found a very effective relaxation of the notion of the strictly near Schwarzschild initial data configurations relevant for the parabolic-hyperbolic system. They set the otherwise freely specifiable scalar field $\boldsymbol{\kappa}$ to be proportional to the constraint variable ${\bf K}$, i.e.~for some suitably chosen function $\mathcal{R}$ the relation  $\boldsymbol{\kappa}=\mathcal{R}\cdot{\rm\bf K}$ holds. It was shown then in \cite{Beyer:2020kty} by applying this anzatz strongly asymptotically flat near Schwarzschild initial data configurations can be generated.
One of our aims in this paper is to indicate that an analogous process may apply to the algebraic-hyperbolic form of the constraints if the trace free part $\Kc_{ij}$ of the projection of the extrinsic curvature's tensor can be specified in a suitable way in the interim of the ``time integration'' process.

\medskip

This paper is organized as follows. Section \ref{sec: preliminary}  is to introduce the analytic framework. In subsection \ref{subsec: basics},  we start by specifying the basic variables and then, in subsections \ref{subsec: par-hyp-eqs} and \ref{subsec: alg-hyp-eqs}, the evolutionary form of the constraints are recalled. A short review of asymptotic flatness in terms of the applied new variables is given in subsection \ref{subsec: asymp-flatness}. The use of concept of near Schwarzschild configurations is explained in subsection \ref{subsec: near-Schwarzscild}.
In subsection \ref{subsec: spher-symm-sol} purely spherically symmetric initial data configurations are investigated.
Section \ref{sec: numerical-results} is to motivate the use and derive the non-linear perturbative approach.  First the applied multipole expansion and the numerical setup is outlined in subsection \ref{numerical-setup}. Then, in subsection \ref{subsec: results-full-equat}, results based on the use of the full set of evolutionary forms are presented. These results, along with the ones reported in the succeeding section, are to convince the readers that a clear separation of the excitation modes from the Schwarzschild background is necessary in order to identify those modes which may get in the way of asymptotic flatness. For this reason a non-linear perturbative approach is introduced in subsection \ref{subsec: non-linear-pert}.
A systematic investigation based on this non-linear perturbative approach is carried out in section \ref{sec: results-on-non-lin-pert}. In particular, all the relevant modes of the basic variables using either the parabolic-hyperbolic or the algebraic-hyperbolic systems are identified in subsections \ref{subsec: par-hyp-non-lin} and \ref{subsec: alg-hyp-non-lin}. Attempts to alter the asymptotic behavior of the critical variable, ${\rm\bf K}=\widehat \gamma{}^{kl}K_{kl}$ are investigated  in subsection \ref{subsec: revisit} by relaxing slightly the notion of near Schwarzschild initial data configurations. In subsection \ref{subsub: Florian-ansatz} first the efficiency of the ansatz introduced in \cite{Beyer:2020kty} is verified. It is also demonstrated that one can simply generate the desired strongly asymptotically flat initial data configurations by updating, leaves by eaves, $\boldsymbol{\kappa}$, according to the ansatz suggested in \ref{subsub: Florian-ansatz}, in the original parabolic-hyperbolic system introduced in \cite{Racz2016}. In subsection \ref{subsub: alg-hyp-ansatz} it is indicated that by applying a somewhat analogous but more involved process the algebraic-hyperbolic system can also be used to produce asymptotically flat initial data configurations.
The discussions are completed by our final remarks in section \ref{sec: final-remarks}. The paper is closed by an appendix providing the detailed form of the evolutionary forms of the constraints relevant for non-linear perturbations.

\section{Preliminaries}\label{sec: preliminary}
\setcounter{equation}{0}

This section is to introduce the applied analytic setup.

\subsection{Evolutionary form of the equations}\label{subsec: basics}

As the Schwarzschild spacetime is a vacuum solution to Einstein's equations throughout this paper our considerations will be restricted to the vacuum constraints.
Vacuum initial data configurations are represented by two symmetric tensor fields $h_{ij}$ and $K_{ij}$ defined on a $3$-dimensional manifold $\Sigma$ such that  $h_{ij}$ is a Riemannian metric there. These fields are not free as they are subject to the vacuum constraints
\begin{align}
    \label{eq:Ham}
    {}^{\scriptscriptstyle(\!3\!)\!\!}R+\big(K^j{}_j\big)^2-K_{ij}K^{ij}&=0\\
    \label{eq:mom}
    D_jK^j{}_i-D_iK^j{}_j&=0\,,
\end{align}
where ${}^{\scriptscriptstyle(\!3\!)\!\!}R$ and $D_i$ are the scalar curvature and the covariant derivative operator associated with $h_{ij}$.

As the Schwarzschild spacetime is foliated by a $2$-parameter family of metric spheres there exists a high variety of initial data surfaces that can be foliated by a one-parameter family of such surfaces, determined by the level surfaces of the area radius function $r:\mathbb{R}^+\rightarrow \Sigma$. As we are looking for near Schwarzschild initial data sets we shall assume that $\Sigma$ is\footnote{Throughout this paper $\Sigma$ will always be a $t_{KS}=const$ Kerr-Schild time slice.} foliated by the $r=const$ surfaces (denoted also by $\mathscr{S}_r$) that are level surfaces  of a function $r:\mathbb{R}^+\rightarrow \Sigma$ the gradient of which is nowhere vanishing. We shall also assume that a flow $r^i$, intersecting each of the $r=const$ level surfaces precisely once and which is scaled such that $r^i\partial_i r=1$ throughout $\Sigma$, had been fixed. Then all the tensor fields on $\Sigma$ can be decomposed using variables intrinsic to the $r=const$ level surfaces and normal to them. In particular, the flow can be characterized by its lapse and shift, $\widehat{N}$ and $\widehat{N}{}^i$,
and it can be decomposed as
\begin{equation}\label{time-evol}
	r^i=\widehat{N}\widehat{n}{}^i+\widehat{N}{}^i\,,
\end{equation}
where $\widehat{n}{}^i$ denotes the unit normal to the leaves $\mathscr{S}_r$. The lapse and shift, along with the $2$-metric $\widehat{\gamma}_{ij}$, induced on the $r=const$ level surfaces, give an algebraically equivalent representation of the metric $h_{ij}$ \cite{Racz2016,Racz:2014jra,Racz:2014gea}. Analogously, the geometric content of $K_{ij}$  can be represented by its scalar, vector and tensorial projections $\kkappa$, $\mathbf{k}_i$, $\mathbf{K}_{ij}$ via the decomposition (for details see, e.g.  \cite{Racz2016,Racz:2014jra,Racz:2014gea})
\begin{equation}\label{dec_1}
K_{ij}= \boldsymbol\kappa \,\widehat  n_i \widehat  n_j  + \left[\,\widehat  n_i \,{\rm\bf k}{}_j  + \widehat  n_j\,{\rm\bf k}{}_i\,\right]  + {\rm\bf K}_{ij}\,.
\end{equation}
As pointed out in \cite{Racz2016,Racz:2014jra,Racz:2014gea} in getting the evolutionary form of the constraints it is also essential to split $\mathbf{K}_{ij}$ into its trace $\mathbf{K}=\mathbf{K}^i{}_i=\widehat{\gamma}^{ij}\mathbf{K}_{ij}$ and trace-free part $\interior{\mathbf{K}}_{ij}=\mathbf{K}_{ij}-\tfrac12\,\widehat{\gamma}_{ij}\mathbf{K}$.

\medskip

By construction the above set of variables $\widehat{N}$, $\widehat{N}{}^i$, $\widehat{\gamma}_{ij}$, $\kkappa$, $\mathbf{k}_i$, $\mathbf{K}$ and $\interior{\mathbf{K}}_{ij}$ give an algebraically equivalent representation of the content of twelve components of the original pair $h_{ij}$ and $K_{ij}$.
In particular, in coordinates $(r,x^A)$, with $A=1,2$,  adopted to the foliation  $\mathscr{S}_r$ and the flow $r^i$, the $3$-metric takes the form
\begin{equation}\label{hijind}
h_{ij}= \widehat  N^2({\rm d}r)_i({\rm d}r)_j+\widehat \gamma_{AB}\,\big[\,\widehat  N^A\,({\rm d}r)_{i} +({\rm d} x^A)_{i}\,\big]\,\big[\,\widehat  N^B\,({\rm d}r)_{j} +({\rm d} x^B)_{j}\,\big] \,,
\end{equation}
i.e.~the components of $h_{ij}$ read as
\begin{equation}\label{hijind-comp}
h_{ij}=\left(\hskip-.09cm
\begin{array}{cc}
\hskip-.09cm  \hskip-.09cm \widehat  N^2+ {\widehat  \gamma}_{EF}\widehat  N^E\widehat  N^F & \hskip-.09cm {\widehat  \gamma}_{AE} \widehat  N^E \\
\hskip-.09cm {\widehat  \gamma}_{BF}\widehat  N^F  & {\widehat  \gamma}_{AB}
\end{array}
\right)\,.
\end{equation}
Analogously, in these coordinates, the components of $K_{ij}$ read as
\begin{equation}\label{Kijind}
K_{ij}= K_{rr}\,({\rm d}r)_i({\rm d}r)_j+2\,K_{r A}\,({\rm d}r)_{(i}({\rm d} x^A)_{j)} +  K_{AB}\,({\rm d} x^A)_i\,({\rm d} x^B)_j\,,
\end{equation}
where, in virtue of \eqref{dec_1}, the relation
\begin{align}
K_{rr} = {}  & \boldsymbol\kappa\,{\widehat  N}{}^{2}  + 2\,{\widehat  N}\,{\rm\bf k}{}_{A}{\widehat  N}{}^A +  {\rm\bf K}_{AB}{\widehat  N}{}^A{\widehat  N}{}^B \label{rec_expr41} \\
K_{r A} = {}  & {\widehat  N}\,{\rm\bf k}{}_{A}  + {\rm\bf K}_{AB}{\widehat  N}{}^B
\label{rec_expr42}\\
K_{AB} =  {} & {\rm\bf K}_{AB}\,,
\label{rec_expr43}
\end{align}
hold, i.e.~in the adopted coordinates $(r,x^A)$ the components of $K_{ij}$ reads as
\begin{equation}
K_{ij}=\left(\hskip-.09cm
\begin{array}{cc}
\boldsymbol\kappa\,{\widehat  N}{}^{2}  + 2\,{\widehat  N}\,{\rm\bf k}{}_{E} {\widehat  N}{}^E +  {\rm\bf K}_{EF}{\widehat  N}{}^E{\widehat  N}{}^F & \hskip-.09cm {\widehat  N}\,{\rm\bf k}{}_{A}  + {\rm\bf K}_{AE}{\widehat  N}{}^E  \\
\hskip-.09cm {\widehat  N}\,{\rm\bf k}{}_{B}  + {\rm\bf K}_{BE}{\widehat  N}{}^E  &\hskip-.09cm \hskip-.09cm {\rm\bf K}_{AB}
\end{array}
\right)\,.
\end{equation}
Regardless whether $(h_{ij}, K_{ij})$ or $(\widehat{N}$, $\widehat{N}{}^A$, $\widehat{\gamma}_{AB}$, $\kkappa$, $\mathbf{k}_A$, $\mathbf{K}, \interior{\mathbf{K}}_{AB})$ are used as our basic variables we always need to select four functions of the pertinent twelve components such that these four are subject to the constraints whereas the remaining eight are freely specifiable throughout $\Sigma$.

\medskip

In giving the constraint in their evolutionary forms we shall use the extrinsic curvature of the $\mathscr{S}_r$ leaves of the foliation in $\Sigma$ given as
\begin{equation}\label{eq: Khat}
	{\widehat K}{}_{ij} = \tfrac12\,\mathscr{L}_{\widehat n} {\widehat \gamma}{}_{ij} = {\widehat N}^{-1} \big[\,\tfrac12\,\mathscr{L}_{r} {\widehat \gamma}{}_{ij} - D_{(i} {\widehat N}_{j)} \,\big]\,,
\end{equation}
where in the last step \eqref{time-evol} was applied. It turned out that, in addition to ${\widehat K}{}_{ij}$ and its trace ${\widehat K}={\widehat K}{}_{ij}{\widehat \gamma}{}^{ij}$, the term in the square bracket on the right hand side of \eqref{eq: Khat}
\begin{equation}\label{eq: Kstar}
\instar{K}{}_{ij} = \tfrac12\,\mathscr{L}_{r} {\widehat \gamma}{}_{ij} - D_{[i} {\widehat N}_{j]} = {\widehat N}{\widehat K}{}_{ij} \,,
\end{equation}
along with its trace
\begin{equation}\label{eq: Kstar2}
\instar{K} = \instar{K}{}_{ij}{\widehat \gamma}{}^{ij} =  {\widehat N}{\widehat K}\,,
\end{equation}
plays important role \cite{Racz2016,Racz:2014jra,Racz:2014gea}.

\medskip

Although the initial data surface $\Sigma$ is foliated by $r=const$ surfaces in general they are not metric but merely topological $2$-spheres. Nevertheless, a complex dyad field $\{q_i,\overline{q}_i\}$ can be introduced  throughout $\Sigma$ by adapting the construction outlined in \cite{Racz:2015ena,Racz:2016wcs,Racz:2017krc}. This starts by fixing a complex dyad $\{q_i,\overline{q}_i\}$  on the unit sphere $\mathbb{S}^2$ which can be dragged first onto one of the leaves (say onto $\mathscr{S}_{0}$) and in the second step Lie dragged onto all the other $\mathscr{S}_{r}$ leaves along the flow $r^i$. Having the complex dyad $\{q_i,\overline{q}_i\}$ defined throughout $\Sigma$ all the tangential derivatives can be given in terms of the operators $\eth$ and $\ethb$. It is plausible to use then instead of the variables $(\widehat{N}$, $\widehat{N}{}^A$, $\widehat{\gamma}_{AB}$, $\kkappa$, $\mathbf{k}_A$, $\mathbf{K}, \interior{\mathbf{K}}_{AB})$ the ones yielded by contracting the involved tensorial expressions with the dyad vector $q_i$ and/or its complex conjugate $\overline{q}_i$ in their free indices. These new variables possess definite spin-weights whence they are also analogous to the basic variables applied in the Newman-Penrose formalism. For the definitions of the most frequently used spin-weighted variables see Table\,\ref{table:data0}. The derivation of these, along with some other, more involved expressions (which are, however, not applied explicitly in this paper), can be found in \cite{Racz:2015ena,Racz:2016wcs,Racz:2017krc}.
\begin{table}[ht!]
	\centering  \hskip-.15cm
	\begin{tabular}{|c|c|c|}
		\hline notation &  definition  &
		spin-weight \\ \hline \hline

		$\mathbf{a}$ &  $\tfrac12\,q^i\,\overline q^j\,\widehat\gamma_{ij}$  &
		 $0$ \\  \hline

		$\mathbf{b}$ &  $\tfrac12\,q^i q^j\,\widehat\gamma_{ij}$
		&  $2$ \\  \hline

		$\mathbf{d}$ &  $\mathbf{a}^2-\mathbf{b}\,\overline{\mathbf{b}}$
		& $0$ \\  \hline

		$\mathbf{A}$ &  $q^a q^b {C^e}{}_{ab}\,\overline q_e
		= \mathbf{d}^{-1}\left\{ \mathbf{a}\left[2\,\eth\,\mathbf{a}
		-\,\overline{\eth}\,\mathbf{b}\right] -  \,\overline{\mathbf{b}}\,\eth\,\mathbf{b} \right\} $
		&  $1$ \\  \hline

		$\mathbf{B}$ &  $\,\overline q^a q^b {C^e}{}_{ab}\,q_e
		= \mathbf{d}^{-1}\left\{ \mathbf{a}\,\overline{\eth}\,\mathbf{b}
		- \mathbf{b}  \,\eth\,\overline{\mathbf{b}}\right\}$
		&  $1$ \\  \hline

		$\mathbf{C}$ &  $q^a q^b {C^e}{}_{ab}\,q_e
		= \mathbf{d}^{-1}\left\{ \mathbf{a}\,\eth\,\mathbf{b}
		-  \mathbf{b}\left[2\,\eth\,\mathbf{a}
		-\,\overline{\eth}\,\mathbf{b}\right] \right\}$
		&  $3$ \\  \hline

		$\,\widehat{{R}}$ &  $\tfrac12\, {\mathbf{a}}^{-1}\left( 2\,{R}
		 - \left\{ \,  \eth\,\overline{\mathbf{B}} - \overline{\eth}\,\mathbf{A}
		 - \tfrac12\,\left[\, \mathbf{C}\,\overline{\mathbf{C}}
		 - \mathbf{B}\,\overline{\mathbf{B}} \,\right]\, \right\}\,\right)$
		 &  $0$ \\  \hline
		 
		$\mathbf{N}$ &  $q_i\widehat N^i $
		&  $1$ \\  \hline

		$\mathbf{k}$ &  $q^i {\rm\bf k}{}_{i}$  &  $1$
		\\  \hline

	\end{tabular}
	\caption{\small Some of the basic variables we shall use in recasting the constraints are listed. The tensor field ${C^e}{}_{ab}=  \tfrac12\,\widehat\gamma^{ef}\left\{ {\mathbb D}_{a}\widehat\gamma_{fb}
			+ {\mathbb D}_{b}\widehat\gamma_{af}-{\mathbb D}_{f}\widehat\gamma_{ab}\right\}$ above relates the covariant derivative ${\widehat D}_i$, determined by $\widehat{\gamma}{}_{ij}$, to the covariant  derivative, ${\mathbb D}_i$, associated with the unit sphere metric on $\mathbb{S}^2$. Note also that $\widehat{R}$ and $R$ denote the scalar curvatures of $\widehat{\gamma}{}_{ij}$ and that of the unit sphere metric, respectively.
		(For detailed derivations of these, and some other, more involved, expressions
		see references \cite{Racz:2015ena,Racz:2016wcs,Racz:2017krc}.)  }
	\label{table:data0}
\end{table}

\subsubsection{Parabolic--hyperbolic equations}\label{subsec: par-hyp-eqs}

The constraint equations can be seen to form a parabolic-hyperbolic system, for the dependent variables $(\NNh,\kk,\KK)$,  given as \cite{Racz:2017krc}
\begin{multline}
    \label{eq:phN}
    \instar{K}\left[\partial_r\NNh-\tfrac12\,\NNt\,\ethb\NNh-\tfrac12\,\NNtb\,\eth\NNh\right]\\
    -\tfrac12\,\dd^{-1}\NNh^2\left[\,\aaa\left\{\eth\ethb\NNh-\BB\,\ethb\NNh\right\}-
    \bb\left\{\ethb^2\!\NNh-\tfrac12\,\AAAb\,\ethb\NNh-\tfrac12\,\CCb\,\eth\NNh\right\}+cc.\right]\\
    -\mathcal{A}\,\NNh-\mathcal{B}\,\NNh^{\,3}=0\,,
\end{multline}
\begin{equation}
    \label{eq:phk}
    \partial_r\kk-\tfrac12\,\NNt\,\ethb\kk-\tfrac12\,\NNtb\,\eth\kk-\tfrac12\,\NNh\,\eth\KK+\ff=0\,,
\end{equation}
\begin{equation}
    \label{eq:phK}
    \partial_r\KK-\tfrac12\,\NNt\,\ethb\KK-\tfrac12\,\NNtb\,\eth\KK-
    \tfrac{1}{2}\,\NNh\,\dd^{-1}\Big\{\,\aaa(\eth\kkb+\ethb\kk)-\bb\,\ethb\kkb-\bbb\,\eth\kk\,\Big\}+\FF=0\,,
\end{equation}
where the coefficients $\mathcal{A}$, $\mathcal{B}$, in \eqref{eq:phN}, and the source terms $\ff$, $\FF$, in \eqref{eq:phk} and \eqref{eq:phK}, read as
	\begin{equation}\label{eq:phA}
	\mathcal{A}=\partial_r\instar{{K}}-\tfrac12\,\NNt\,\ethb\instar{{K}}-\tfrac12\NNtb\,\eth\instar{{K}}+
	\tfrac12\,\Big[\instar{{K}}{}^2+\instar{{K}}{}_{kl}\instar{{K}}{}^{kl}\Big]\,,
	\end{equation}
	\begin{equation}\label{eq:phB}
	\mathcal{B}=-\tfrac12\,\Big[\RRh+2\,\kkappa\,\KK+\tfrac12\,\KK^2-\dd^{-1}[2\,\aaa\,\kk\,\kkb-\bb\,\kkb^2-\bbb\,\kk^2]-\Kc{}_{kl}\Kc{}^{kl}\Big]\,,
	\end{equation}
\begin{multline}\label{eq:phf}
    \ff=-\tfrac12\,\Big[\kk\,\eth\NNtb+\kkb\,\eth\NNt\Big]-\left[\kkappa-\tfrac12\,\KK\right]\eth\NNh
    +\instar{{K}}\,\kk-\NNh\Big[\eth\kkappa+q^i\dot{\widehat{n}}^l\interior{\mathbf{K}}{}_{li}-q^i\widehat{D}^l\interior{\mathbf{K}}{}_{li}\Big]\,,
\end{multline}
\begin{align}\label{eq:phF}
    \FF=\tfrac{1}{4}\NNh\,\dd^{-1}\Big\{2\,\aaa\,\BB\,\kkb-\bb(\,\CCb\,\kk+\AAAb\,\kkb)+cc.\Big\}
    & -\dd^{-1}\Big[(\aaa\,\kkb-\bbb\,\kk)\,\eth\NNh+cc.\Big]\nonumber \\ & +\Big[\interior{\mathbf{K}}_{ij}{\Ks}{}^{ij}-\left(\kkappa-\tfrac12\,\KK\right)\instar{{K}}\Big]\,,
\end{align}
where ‘cc.’ stands everywhere for the complex conjugate of the preceding terms within the same parentheses, $\dot{\widehat{n}}^l=\widehat{D}^l\ln \widehat{N}$, and the explicit form of the terms $q^i\dot{\widehat{n}}^l\interior{\mathbf{K}}{}_{li}$, $q^i\widehat{D}^l\interior{\mathbf{K}}{}_{li}$, $\interior{\mathbf{K}}_{ij}{\Ks}{}^{ij}$, $\Kc{}_{kl}\Kc{}^{kl}$, $\instar{{K}}{}_{kl}\instar{{K}}{}^{kl}$, in terms of elementary spin-weighted variables, can be found, e.g.~in \cite{Racz:2017krc}.

\subsubsection{Algebraic-hyperbolic equations}\label{subsec: alg-hyp-eqs}

Analogously, the constraint equations can be seen to form an algebraic-hyperbolic system, for the variables $(\KK, \kk, \kkappa)$, given as \cite{Racz:2017krc}
\begin{equation}
    \label{eq:ah1}
    \partial_r\KK-\tfrac12\,\NNt\,\ethb\KK-\tfrac12\,\NNtb\,\eth\KK-\tfrac{1}{2}\,\NNh\,\dd^{-1}\Big\{\aaa(\eth\kkb+\ethb\kk)-\bb\,\ethb\kkb-\bbb\,\eth\kk\Big\}+\FF=0\,,
\end{equation}
\begin{multline}
    \label{eq:ah2}
    \partial_r\kk-\tfrac12\,\NNt\,\ethb\kk-\tfrac12\,\NNtb\,\eth\kk+\NNh\,\KK^{-1}\Big\{\kkappa\,\eth\KK-\dd^{-1}\big[(\aaa\kk-\bb\kkb)\,\eth\kkb
    +(\aaa\kkb-\bbb\kk)\,\eth\kk\big]\Big\}+\ff=0\,,
\end{multline}
\begin{equation}
    \label{eq:ah3}
    \kkappa=\tfrac{1}{2}\KK^{-1}\Big[\dd^{-1}\big(2\aaa\,\kk\kkb-\bb\,\kkb^2-\bbb\,\kk^2\big)-\tfrac12\KK^2-\kkappa_0\Big]\,,
\end{equation}
where
\begin{equation}\label{eq:ah4}
    \kkappa_0={}^{\scriptscriptstyle(\!3\!)}\!R-\Kc_{kl}\Kc{}^{kl}\,,
\end{equation}
and the pertinent source terms $\FF$, $\ff$ read as
\begin{align}\label{eq:ah5}
    \FF=  \tfrac{1}{4}\,\NNh\,\dd^{-1}\Big\{2\,\aaa\,\BB\,\kkb-\bb(\,\CCb\,\kk+\AAAb\,\kkb)+cc.\Big\}
    & -\dd^{-1}\Big[(\aaa\kkb-\bbb\kk)\,\eth\NNh+cc.\Big] \nonumber \\ & +\Big[\Kc_{ij}{\Ks}{}^{ij}-\left(\kkappa-\tfrac12\,\KK\right)\instar{{K}}\Big]\,,
\end{align}
\begin{align}\label{eq:ah6}
    \ff = &-\tfrac12\,\Big[\kk\,\eth \NNtb +\kkb\,\eth \NNt\Big]\nonumber \\
      &+\tfrac{1}{2}\,\NNh\,(\dd\cdot\KK)^{-1}\Big[(\aaa\,\kk-\bb\,\kkb)(\BBb\,\kk+\BB\,\kkb)+(\aaa\,\kkb-\bbb\,\kk)(\CC\,\kkb+\AAA\,\kk)\Big] \nonumber \\ &
    -\left[\kkappa-\tfrac12\,\KK\right]\eth\NNh+\NNh\Big[\tfrac{1}{2}\KK^{-1}\eth\kkappa_0+\KKh\,\kk-q^i\dot{\widehat{n}}{}^l\Kc_{li}+q^i\widehat{D}^l\Kc_{li}\Big]\,.
\end{align}

The system comprised by \eqref{eq:ah1}--\eqref{eq:ah2} is known to be a symmetrizable hyperbolic system when the inequality $\boldsymbol{\kappa}\cdot \bf K<0$ holds \cite{Racz2016}. Notably, this condition was shown to be satisfied by near Schwarzschild initial data configurations \cite{Racz:2015ena}.

\subsection{The strong and weak forms of asymptotic flatness}\label{subsec: asymp-flatness}

Since our aim is to study the asymptotics of near Schwarzschild initial data configurations it suffices to consider initial data surfaces with a single asymptotically flat end that is diffeomorphic to a region complementing a ball $\mathscr{B}$ in $\mathbb{R}^3$.
Hereafter an initial data set $(\Sigma, h_{ij}, K_{ij})$ is called   asymptotically flat in the ``strong sense'', if the complement of a compact set in $\Sigma$ can
be mapped by an admissible Cartesian coordinate system $\{x_i\}$ diffeomorphically onto the complement of $\mathscr{B}$ in $\mathbb{R}^3$, and also there exist a positive constant $C$, such that in these coordinates
\begin{equation} \label{af-h}
h_{ij}=\left(1+\tfrac{C}{r}\right)\delta_{ij}+\mathscr{O}(r^{-2})\,,
%h_{ij}-\delta_{ij}=\mathscr{O}(r^{-1})
\end{equation}
\begin{equation} \label{af-K}
K_{ij}=\mathscr{O}(r^{-2})
\end{equation}
hold while  $r=\sqrt{x_1^2+x_2^2+x_3^2}\longrightarrow \infty$, where 
the indices $i, j, k...$ are also assumed to denote coordinate indices, taking the values $1, 2, 3$, whereas $\delta_{ij}$ denotes the components of the flat metric in the admissible Cartesian coordinates $\{x_i\}$.
Notably the above conditions are known to guarantee that the mass, the momentum, and the angular momentum of the initial data set are well  defined \cite{Sergio-Helmut}.

\medskip

Note also that there exist various weaker notions of asymptotic flatness \cite{ch-ch,ch}.  In what follows an initial data set $(\Sigma, h_{ij}, K_{ij})$ is called asymptotically flat in the ``weak sense'' if for some positive constant $C$ and for some arbitrarily small positive $\varepsilon$
\begin{equation} \label{waf-h}
h_{ij}=\left(1+\tfrac{C}{r}\right)\delta_{ij}+\mathscr{O}(r^{-3/2-\varepsilon})\,,
%h_{ij}-\delta_{ij}=\mathscr{O}(r^{-\alpha})
\end{equation}
\begin{equation} \label{waf-K}
K_{ij}=\mathscr{O}(r^{-3/2-\varepsilon})
\end{equation}
hold.
It was shown in \cite{ch} that if \eqref{waf-h} and \eqref{waf-K} holds, and also $h_{ij}$ and $K_{ij}$ satisfy the constraints, then the ADM mass and the linear momentum are still well-defined.

\subsubsection{The fall off properties of the new variables}

The arguments in this subsection will only be outlined for strongly asymptotically flat initial data sets as their straightforward modifications can also be applied to deduce the fall off conditions relevant for weakly asymptotically flat initial data configurations.

\medskip

To start off recall first that the coordinate basis fields $\{(\partial_{x_i})^a\}$ of admissible Cartesian coordinates $\{x_i\}$ can always be expressed as linear combinations of the coordinate basis fields  $(\partial_{r})^a$ and $(\partial_{\theta})^a,(\partial_{\phi})^a$ of spherical coordinates $(r, \theta,\phi)$, where the coefficients in these linear combinations are of order $\mathscr{O}(r^{0})$, $\mathscr{O}(r^{-1})$ in $r$, respectively.
\footnote{To see this recall that in $\mathbb{R}^3$ the Cartesian coordinates $(x_1,x_2,x_3)$ and the spherical coordinates $(r, \theta,\phi)$ are related as
\begin{align}
 x_1= r \sin\theta\cos\phi\,, \quad  x_2= r \sin\theta \sin \phi\,, \quad x_3= r \cos\theta\,,
\end{align}
whereas the corresponding coordinate basis fields as
\begin{align}
& (\partial_{x_1})^a = \sin\theta\cos\phi \,(\partial_{r})^a +{r}^{-1} \Big[ \cos\theta \cos\phi \,(\partial_{\theta})^a - (\sin\theta)^{-1}\sin\phi \,(\partial_{\phi})^a \Big] \\
& (\partial_{x_2})^a = \sin\theta\sin\phi \,(\partial_{r})^a +{r}^{-1} \Big[ \cos\theta\sin\phi \,(\partial_{\theta})^a + (\sin\theta)^{-1} \cos\phi \,(\partial_{\phi})^a \Big] \\
& (\partial_{x_3})^a = \cos\theta \,(\partial_{r})^a - {r}^{-1} \sin\theta \,(\partial_{\theta})^a \,.
\end{align}}

Accordingly, from $h_{ij}=h_{ab}(\partial_{x_i})^a(\partial_{x_j})^b$ we get
\begin{equation}
h_{ij}= h_{rr} + 2\,{r}^{-1} \,h_{r A} + {r}^{-2} \,h_{A B} \,,
\end{equation}
which, along with \eqref{hijind-comp} and \eqref{af-h}, implies that
\begin{align}
(\widehat  N^2+ \widehat  N_E\widehat  N^E)-1 & \sim \mathscr{O}(r^{-1}) \label{h1} \\
 \widehat  N_A = \widehat \gamma_{AB} \widehat  N^B & \sim \mathscr{O}(r^{-1}) \label{h2} \\
 \widehat \gamma_{AB}-r^2\,\interior{\gamma}_{AB} &  \sim  \mathscr{O}(r^{0})\,, \label{h3}
\end{align}
where $\interior{\gamma}_{AB}$ denotes the standard unit sphere metric.

In virtue of \eqref{h3} it follows that
\begin{equation}\label{hgfall}
\widehat \gamma_{AB} \sim  \mathscr{O}(r^{2}) \quad {\rm and} \quad \widehat \gamma^{AB} \sim  \mathscr{O}(r^{-2})\,,
\end{equation}
which, along with \eqref{h2}, gives then $\widehat  N^A$ should fall off as
\begin{equation}\label{hNvfall}
\widehat  N^A \sim  \mathscr{O}(r^{-3})\,.
\end{equation}
Finally, the last two relations, along with  \eqref{h1}, gives that $\widehat  N^2-1 \sim \mathscr{O}(r^{-1})$, which, by $\widehat  N^2-1=(\widehat  N+1)(\widehat  N-1)$ and since $\widehat  N$ cannot change sign, thereby, it may, without loss of generality, be assumed to be positive, implies that
\begin{equation}\label{hNfall}
\widehat  N -1 \sim  \mathscr{O}(r^{-1})\,.
\end{equation}

Analogously, from
\begin{equation}
K_{ij} = K_{rr} + 2\,{r}^{-1} \,K_{r A} + {r}^{-2} \,K_{A B}
\end{equation}
we get,  along with \eqref{rec_expr41}, \eqref{rec_expr42}, \eqref{rec_expr43}  and  \eqref{af-K} that
\begin{align}
\boldsymbol\kappa\,{\widehat  N}{}^{2}  + 2\,{\widehat  N}\,{\rm\bf k}{}_{A}{\widehat  N}{}^A +  {\rm\bf K}_{AB}{\widehat  N}{}^A{\widehat  N}{}^B \sim  \mathscr{O}(r^{-2}) \label{K1} \\
{\widehat  N}\,{\rm\bf k}{}_{A}  + {\rm\bf K}_{AB}{\widehat  N}{}^B \sim  \mathscr{O}(r^{-1})
\label{K2}\\
{\rm\bf K}_{AB} \sim  \mathscr{O}(r^{0})\,.
\label{K3}
\end{align}
Correspondingly, \eqref{K3}, along with the relation ${\rm\bf K}_{AB} = \interior{\rm\bf K}_{AB} + \tfrac12\, \widehat\gamma_{AB} \,{\rm\bf K}$ and \eqref{hgfall}, gives then that ${\rm\bf K}_{AB}, \interior{\rm\bf K}_{AB}$ and ${\rm\bf K} = \widehat \gamma^{EF}{\rm\bf K}_{EF}$ should fall off as
\begin{equation}\label{bfKfall}
{\rm\bf K}_{AB}, \interior{\rm\bf K}_{AB} \sim  \mathscr{O}(r^{0}) \quad {\rm and} \quad {\rm\bf K}  \sim  \mathscr{O}(r^{-2})\,.
\end{equation}
Then, \eqref{K2}, along with \eqref{hNfall}, \eqref{hNvfall} and \eqref{bfKfall}, implies \
\begin{equation}\label{bfkfall}
{\rm\bf k}{}_{A}  \sim  \mathscr{O}(r^{-1})\,,
\end{equation}
and, finally, in virtue of \eqref{K1}, that
\begin{equation}
\boldsymbol\kappa \sim  \mathscr{O}(r^{-2})\,.
\end{equation}

The fall off properties relevant for the constraint variables---these appear in either of the evolutionary forms of the constraint equations---are collected in Table \ref{table:falloff}.
\begin{table}[ht!]
\centering  \hskip-.15cm
	\begin{tabular}{|c|c|c|}
		\hline
		variable & strong form & weak form \\ \hline
		$\NNh-1$   & $r^{-1}$ & $r^{-1}$  \\ \hline
		$\boldsymbol{\kappa}$   & $r^{-2}$ & $r^{-3/2-\varepsilon}$  \\ \hline
		$\kk$   & $r^{-1}$ & $r^{-1/2-\varepsilon}$  \\ \hline
		$\KK$   & $r^{-2}$ & $r^{-3/2-\varepsilon}$  \\ \hline
	\end{tabular}
\caption{\small 
	The fall off conditions compatible with the strong and weak forms of asymptotic flatness, respectively, are collected for the four distinguished dependent variables, the fall off rates of which will be monitored in sections \ref{sec: numerical-results} and \ref{sec: results-on-non-lin-pert}.}
\label{table:falloff}
\end{table}

\subsection{``Near Schwarzschild configurations''}\label{subsec: near-Schwarzscild}

In advance of the determination of near Schwarzschild initial data configurations recall first that among the twelve scalar variables stored either in $(h_{ij}, K_{ij})$ or in $(\widehat{N}, \widehat{N}{}^i, \widehat{\gamma}_{ij}, \kkappa, \mathbf{k}_i, \mathbf{K}, \interior{\mathbf{K}}_{ij})$, there are eight which can always be specified freely throughout $\Sigma$, whereas whenever the evolutionary forms of the constraints are used, the initial data for the four constrained variables can also be freely specified but only on one of the leaves, say on $\mathscr{S}_{r_0}$, in $\Sigma$. Notice that $\mathscr{S}_{r_0}$ plays the role of an ``initial data surface'' in solving either of the evolutionary form of the constraints. 

In \cite{Beyer:2017njj,Beyer:2019kty} initial data specifications were considered to be near Schwarzschild if each of the eight otherwise completely freely specifiable variables was fixed to take exactly its Schwarzschild functional form, which means that an initial data specification became near Schwarzschild simply by allowing at least one of the four constrained variables to differ, in a non-spherical way, from the corresponding Schwarzschild values, at the initial leaf $\mathscr{S}_{r_0}$.
Clearly, this definition is considerably restrictive and  there is a high variety of possible determinations of near Schwarzschild initial data configurations.  Nevertheless, hereafter, apart from subsection \ref{subsec: revisit}, we shall explore the implications of the use of strictly near Schwarzschild initial data specification, applied in  \cite{Beyer:2017njj,Beyer:2019kty}.

Before determining these solutions to the evolutionary form of the constraints it is rewarding to recall the functional form of the basic variables deduced from the four-dimensional Schwarzschild solution on a $t_{KS}=const$ Kerr-Schild time slice \cite{Racz:2015ena}. The non-identically vanishing elements (notably these are all of zero spin-weight) are listed in Table
\ref{near-Schwarzschild},
whereas all the other variables, including $\mathbf{b}$, $\mathbf{A}$, $\mathbf{B}$, $\mathbf{C}$, $\mathbf{N}$, $\mathbf{k}$, $\interior{\mathbf{K}}{}_{ij}$,  $\interior{\widehat{K}}{}_{ij}={\widehat{K}}{}_{ij}-\tfrac12\,{\widehat{K}}\,{\widehat{\gamma}}_{ij}$,  vanish identically \cite{Racz:2015ena}.
\begin{table}[hbt!]
	\centering
	\begin{tabular}{|c|c||c|c|}
		\hline
		variable & functional form & variable & functional form \\ \hline

		$\aaa$  %& $\tfrac12\,\widehat{\gamma}_{ij}q^i\overline{q}^j$
		& $r^2$ & $\instar{K}$
		& $\frac{2}{r}$\\ \hline

		$\dd$  %  & $\aaa^2-\bb\bbb$
		& $r^4$ & $\KKh$ % & $\frac{\instar{\mathbb{K}}}{\NNh}$ $\widehat{K}{}^l{}_l$
		& $\frac{2}{r\sqrt{1+2M/r}}$ \\ \hline

		$\NNh$ % & $\widehat{N}$
		& $\sqrt{1+2M/r}$ & $\RRh$  % & $\frac{1}{2\aaa}\big(2\RR-\{\eth\BBb-\ethb\AAA-\tfrac12\,[\CC\CCb-\BB\BBb]\}\big)$
		& $\frac{2}{r^2}$ \\ \hline

		$\KK$   % & $\mathbb{K}=\widehat{\gamma}^{ij}\mathbf{K}_{ij}$
		& $-\frac{4M}{r^2\sqrt{1+2M/r}}$ & $\RR$ & $2$ \\ \hline

		$\kkappa$ % & $\kkappa$
		& $\frac{2M(1+M/r)}{r^2(1+2M/r)^{3/2}}$ & ${}^{(3)\!\!}R$ & $ \frac{8 M^2}{r^4 \left(1+{2	M}/{r}\right)^2}$  \\ \hline
	\end{tabular}
	\caption{\small The non-identically vanishing elements of the basic variables relevant for the  initial data deduced from the four-dimensional Schwarzschild solution on a $t_{KS}=const$ time slice.}
	\label{near-Schwarzschild}
\end{table}

\subsection{The spherically symmetric solutions}\label{subsec: spher-symm-sol}

In proceeding it is rewarding to have a glance at the spherically symmetric solutions to the evolutionary form of the constraints. In the succeeding subsections spherically symmetric near Schwarzschild solutions will be considered. In both---the parabolic-hyperbolic and the algebraic-hyperbolic---cases we start by inspecting the strictly near Schwarzschild configurations. Then, it is shown that by relaxing the selection rules for some of the freely specifiable fields strongly asymptotically flat spherically symmetric solutions can be deduced from the evolutionary form of the constraints.

\subsection{Spherical solutions to the parabolic-hyperbolic system}
\label{sec:strict_spher_ph}

Since any smooth vector field must vanish somewhere on a topological two-sphere it is straightforward to see that spherical symmetry requires the vanishing of the vectorial projection of the extrinsic curvature $\mathbf{k}_A$, or equivalently that of $\kk$ throughout $\Sigma$. Then \eqref{eq:phN} and  \eqref{eq:phK} can be seen to reduce to the system
\begin{align}
\frac{{\rm d}\widehat{N}}{{\rm d}r} & = \frac{1}{2\,
	r}\,\widehat{N}-\frac{ 4 \,\boldsymbol{\kappa}\,
	\mathbf{K}\, r^2+\mathbf{K}^2
	r^2+4}{8 \,r}\,\widehat{N}^3 \label{eqr-Nh} \\
	\frac{{\rm d}\mathbf{K}}{{\rm d}r} & = \Big(\boldsymbol{\kappa}-\tfrac12 \mathbf{K}\Big)\,\frac{2}{r}\,. \label{Kbf-par-hyp}
\end{align}

\subsubsection{The strict case}

Note first that equations \eqref{eqr-Nh} and \eqref{Kbf-par-hyp} decouple and---provided that the strict notion of near Schwarzschild initial data is adapted such that $\boldsymbol{\kappa}=\boldsymbol{\kappa}_{\text{Schw}}$---the generic solutions to \eqref{Kbf-par-hyp} can be seen to take the form
\begin{equation}\label{eq: sper-K-par-hyp}
\mathbf{K} = \mathbf{K}_{\text{Schw}} +\frac{C_\mathbf{K}}{r}\,,
\end{equation}
where $\mathbf{K}_{\text{Schw}}=-\frac{4 M}{r^2 \sqrt{1+\frac{2 M}{r}}}$ is the Schwarzschild form of  $\mathbf{K}$ as given in Table \ref{near-Schwarzschild}, and ${C_\mathbf{K}}$ is a constant of integration. This means that, whenever the strict notion of near Schwarzschild initial data is applied, the above solution to \eqref{eq: sper-K-par-hyp}, in virtue of Table\,\ref{table:falloff}, cannot even fit to the weakly asymptotically flat scenario unless ${C_\mathbf{K}}=0$.

By substituting $\mathbf{K} = \mathbf{K}_{\text{Schw}}$, into \eqref{eqr-Nh} and solving the yielded equation for $\widehat{N}$ we get 
\begin{equation}\label{eq: Nhat-radial-PH}
\widehat{N} = \frac{\sqrt{r (2
		M+r)}}{\sqrt{2 M \left(C_{\widehat{N}}+r\right)+r
		\left(C_{\widehat{N}}+r\right)+4 M^2}}\,,
\end{equation}
where ${C_{\widehat{N}}}$ is another constant of integration. This solution decays as
\begin{equation}
\widehat{N}=1-\frac{C_{\widehat{N}}}{2
	r}+\frac{\frac{3 C_{\widehat{N}}^2}{8}-2
	M^2}{r^2}+\mathscr{O}\left({r}^{-9/4}\right)\,.
\end{equation}
Note that, in virtue of \eqref{eq: Nhat-radial-PH}, $\widehat{N}$, in general, differs from its Schwarzschild form, $\widehat{N}=\sqrt{1+{2 M}/{r}}$, which occurs only for the particular choice $C_{\widehat{N}}=-2M$.

In summing up note that whenever the strict notion of  near Schwarzschild initial data is applied, then even the spherically symmetric solutions to \eqref{eq:phN} and \eqref{eq:phK} fail to be weakly asymptotically flat unless $\mathbf{K}$ is chosen to be of the form $\mathbf{K}_{\text{Schw}}=-\frac{4 M}{r^2 \sqrt{1+\frac{2 M}{r}}}$.

\subsubsection{Modifying the background for the parabolic-hyperbolic system}
\label{subsub: Florian-ansatz-radial}

It was shown in \cite{Beyer:2020kty} that by relaxing the selection rules applied previously in \cite{Beyer:2017njj,Beyer:2019kty} even strongly asymptotically flat near Schwarzschild initial data solutions exist to the parabolic-hyperbolic system \eqref{eqr-Nh} and \eqref{Kbf-par-hyp}. As  $\boldsymbol{\kappa}$ is freely specifiable in this setup the authors in \cite{Beyer:2020kty} chose  $\boldsymbol{\kappa}$ to be proportional to the dependent field ${\bf K}$, i.e.\,that
\begin{equation}\label{eq: ansatz}
	\boldsymbol{\kappa}= \mathcal{R}\cdot {\bf K}%\,,
\end{equation}
holds, where the factor of proportionality $\mathcal{R}$ was assumed to be a smooth function such that it tends to $-\tfrac12$ at infinity. Start, for instance, by choosing
\begin{equation}\label{eq: ansatz2}
	\mathcal{R} = -\frac{M + r}{2\,(2 M +  r)}\,.
\end{equation}
The generic solution to \eqref{Kbf-par-hyp} reads as
\begin{equation}
	{\bf K}= \frac{C_{\bf K}}{r^{3/2}\sqrt{2 M +  r}} \,,
\end{equation}
where $C_{\bf K}$ is an integration constant. It is straightforward to see that this ${\bf K}$ reduces to $\mathbf{K}_{\text{Schw}}$ for the choice $C_{\bf K}=-4\,M$, and also that for any non-vanishing $C_{\bf K}$ this solution falls off as $r^{-2}$.
Note that, in virtue of \eqref{eq: ansatz} and as $\mathcal{R}$ tends to $-\tfrac12$ at infinity,  $\boldsymbol{\kappa}$ falls off with the same rate as ${\bf K}$ does. The corresponding generic solution to \eqref{eqr-Nh} is
\begin{equation}\label{eq: beyer-Nhat}
{\widehat N}= \frac{\sqrt{2 M r + r^2}}{\sqrt{(2 M + r) (2 M +C_{\widehat N} + r)  + \tfrac14\,C_{\bf K}^2  }}  \,,
\end{equation}
where $C_{\widehat N}$ is a constant of integration. This  solution reduces to the Schwarzschild form, ${\widehat N}=\sqrt{1+2\,M/r}$, if $C_{\bf K}=C_{\widehat N}=-4\,M$. The generic solution \eqref{eq: beyer-Nhat} decays as
\begin{equation}
	{\widehat N}= 1 - \frac{2\,M+C_{\widehat N}}{2\,r}  + \frac{ 12\, M^2 - C_{\bf K}^2+ 12\, M\, C_{\widehat N} + 3\, C_{\widehat N}^2 }{8\, r^2} + \mathscr{O}(r^{-3})\,,
\end{equation}
which implies that the ADM mass $M_{ADM}=-M - C_{\widehat N}/2$ of this solution is affected by the choice of $C_{\widehat N}$ and it is positive provided that $C_{\widehat N}<-2\,M$. More importantly, for any choice of $C_{\widehat N}<-2\,M$ the non-trivial part of the initial data represented by the fields ${\widehat N}, \boldsymbol{\kappa}, {\bf K}$ all decays in accordance with the conditions indicated in Table\,\ref{table:falloff} for the strong case, i.e.\,the yielded spherically symmetric solution to the parabolic-hyperbolic system \eqref{eqr-Nh} and \eqref{Kbf-par-hyp} is strongly asymptotically flat.

\subsection{Spherical solutions to the algebraic-hyperbolic system}

Note that as in case of the parabolic-hyperbolic system the algebraic-hyperbolic equations \eqref{eq:ah1}--\eqref{eq:ah3} does not admit spherical symmetric solutions either unless $\mathbf{k}_A$, or equivalently $\kk$, vanishes identically. If this happens \eqref{eq:ah1} and \eqref{eq:ah3} reduce to the form
\begin{align}
\frac{{\rm d}\mathbf{K}}{{\rm d}r} & =  \Big(\boldsymbol{\kappa}-\tfrac12 \mathbf{K}\Big)\,\frac{2}{r}\,, \label{eq: Kbfr-alg-hyp}  \\
\boldsymbol{\kappa} & =-\frac{\mathbf{K}}{4}-\frac{\boldsymbol{\kappa}_0}{2
	\mathbf{K}}\,. \label{eq: kappa-alg-hyp}
\end{align}
By applying again the strict notion of near Schwarzschild initial data and by substituting
\begin{equation}\label{eq: kappa0-sphere}
\boldsymbol{\kappa}_0={}^{(3)\!\!}R = \frac{8 M^2}{r^4 \left(1+\frac{2
		M}{r}\right)^2}
\end{equation}
into \eqref{eq: Kbfr-alg-hyp} one gets the generic solution in the form (see also \cite{Beyer:2017njj})
\begin{equation}
\mathbf{K} = -\frac{\sqrt{C_{\mathbf{K}}\,(2\,M+r)+16
		M^2}}{r^2 \sqrt{1+2M/r}}\,,
\end{equation}
where ${C_\mathbf{K}}$ is a constant of integration.
The asymptotic form of $\mathbf{K}$ and that of the corresponding $\boldsymbol{\kappa}$, determined by \eqref{eq: kappa-alg-hyp}, are
\begin{equation}\label{eq: fall-KK-alg-hyp}
\mathbf{K} =-\sqrt{C_{\mathbf{K}}}\,
r^{-3/2}-\frac{8 M^2}
	{\sqrt{C_{\mathbf{K}}}}\, r^{-5/2}+\mathscr{O}\left(r^{-7/2}\right)\,,
\end{equation}
\begin{equation}
\boldsymbol{\kappa} =\frac{1}{4} \sqrt{C_{\mathbf{K}}}\,
r^{-3/2}+\frac{6 M^2
	}{\sqrt{C_{\mathbf{K}}}}\,r^{-5/2}+\mathscr{O}\left(r^{-7/2}\right)\,.
\end{equation}
Whenever ${C_\mathbf{K}}\not=0$ both $\mathbf{K}$ and $\boldsymbol{\kappa}$ are at the borderline to fit the weakly asymptotically flat requirements.
If, however, ${C_\mathbf{K}}=0$ both $\mathbf{K}$ and $\boldsymbol{\kappa}$ take their Schwarzschild form $\mathbf{K}_{\text{Schw}}=-\frac{4 M}{r^2 \sqrt{1+\frac{2 M}{r}}}$ and $\boldsymbol{\kappa}_{\text{Schw}}=\frac{2M(1+M/r)}{r^2(1+2M/r)^{3/2}}$, respectively.

\medskip

Notably these findings, likewise the ones at the end of subsection\,\ref{sec:strict_spher_ph}, indicate that whenever the strict notion of near Schwarzschild initial data is used then not even the spherically symmetric solutions to the algebraic-hyperbolic equations \eqref{eq:ah1}--\eqref{eq:ah3} will admit asymptotically flat solution other than the Schwarzschild one with $\mathbf{K}_{\text{Schw}}=-\frac{4 M}{r^2 \sqrt{1+\frac{2 M}{r}}}$.

\subsubsection{Modifying the background for the algebraic-hyperbolic system}
\label{subsub: ansatz-alg-hyp-radial}

In subsection \ref{subsub: Florian-ansatz-radial} the ansatz proposed in \cite{Beyer:2020kty} was found to be very powerful. Therefore, it could be natural to look for a possible analogue of this trick. Note, however, that  we have very limited freedom in altering the functional relation of $\boldsymbol{\kappa}$ and ${\bf K}$. This is so as $\boldsymbol{\kappa}$ is the algebraic solution to the spherically symmetric form of the Hamiltonian constraint \eqref{eq: kappa-alg-hyp}, thereby, it depends on the solution ${\bf K}$ to \eqref{eq: Kbfr-alg-hyp}. Nevertheless, there is an opportunity to have some control on the relation of $\boldsymbol{\kappa}$ and ${\bf K}$. For instance, by invoking the freedom we have in choosing the freely specifiable field $\interior{\bf K}{}_{ij}$ which is known to vanish on time slices in Schwarzschild spacetime foliated by spheres which respects spherical symmetry. In proceeding note that, the generic form of $\boldsymbol{\kappa}_0$ in \eqref{eq:ah4} involve the contraction $\interior{\bf K}{}_{ij}\interior{\bf K}{}^{ij}$ and by assuming that the relation
\begin{equation}
	\interior{\bf K}{}_{ij}\interior{\bf K}{}^{ij}  = (\tfrac12+2\,\mathcal{R})\KK^2 + {}^{(3)\!\!}R
\end{equation}
holds, we also get
\begin{equation}\label{eq: ansatz-ah}
\boldsymbol{\kappa}= \mathcal{R}\cdot {\bf K}\,.
\end{equation}
Here $\mathcal{R}$ is assumed to be a smooth function tending to a constant value $-\gamma$, while $r\rightarrow \infty$, where $\gamma$ takes values from the interval $1/4 <\gamma \leq 1/2$. At the moment we shall simply assume that such a freely specifiable field $\interior{\bf K}{}_{ij}$ exists. It will be shown in subsection \ref{subsub: alg-hyp-ansatz} that under suitable conditions this choice can always be made. For instance, by choosing
\begin{equation}\label{eq: ansatz2-ah}
\mathcal{R} = -\gamma+\frac{\gamma\,M}{r+2 M }\,,
\end{equation}
the generic solution to \eqref{eq: Kbfr-alg-hyp} reads as
\begin{equation}\label{eq: sloKr-ah}
	\KK = \frac{C_{\KK} }{r^{\left(1 + 2\,\gamma\right)}\,(1+2\,M/r)^{\gamma}} \,,
\end{equation}
where $C_{\bf K}$ is an integration constant. It is straightforward to see that this ${\bf K}$ reduces to $\mathbf{K}_{\text{Schw}}$ if $C_{\bf K}=-4\,M$ and $\gamma=1/2$, and also that, for any non-vanishing $C_{\bf K}$, this solution falls off as $r^{-(1+2\,\gamma)}$. Since $\mathcal{R}$ is negative everywhere, in virtue of the relation $\boldsymbol{\kappa} = \mathcal{R}\cdot \KK$, the algebraic condition $\boldsymbol{\kappa}\,{\bf K} <0$---guaranteeing that the algebraic-hyperbolic system is indeed a symmetrizable hyperbolic one \cite{Racz2016}---is automatically satisfied. In addition, as $\mathcal{R}$ tends to $-\gamma$, which is non-vanishing, $\boldsymbol{\kappa}$ is guaranteed to fall off exactly as fast as ${\bf K}$ does.

\medskip

Notably, the solution ${\bf K}$ \eqref{eq: sloKr-ah}, along with the corresponding $\boldsymbol{\kappa}$, and along with the freely specifiable fields, including $\interior{\bf K}{}_{ij}$, comprise a weakly asymptotic flat initial data for any value of $\gamma$ from the interval $1/4 <\gamma < 1/2$. Note also that the corresponding spherically symmetric initial data configuration is strongly asymptotically flat for $\gamma=1/2$.

\section{The numerical setup}\label{sec: numerical-results}
\setcounter{equation}{0}

This section is to introduce the applied numerical scheme and to present some results relevant for the use of the full set of the evolutionary form of the constraints. These latter results indicate the need for a clear separation of the excitation modes from the Schwarzschild background. For this reason a non-linear perturbative approach will be applied which is introduced in the last subsection.

\subsection{Multipole expansion}\label{numerical-setup}

In the applied numerical setup all the basic variables are expanded by making use of spin-weighted spherical harmonics. As in equations \eqref{eq:phN}--\eqref{eq:phF}  and \eqref{eq:ah1}--\eqref{eq:ah6} the angular derivatives are given in terms of the $\eth$ and $\overline{\eth}$ operators
this allows us to evaluate all of these derivatives analytically, whereas the evolution of the expansion coefficients, in the radial direction, is determined by applying a fourth order accurate finite differencing numerical integrator.

\medskip

Accordingly, the basic variables will be expanded in terms of spin-weighted spherical harmonics. For instance, if ${}^{(s)}\mathbf{P}$ is a spin-weight $s$ variable it is replaced by the expansion
\begin{equation}\label{eq: mult-exp}
{}^{(s)}\mathbf{P}(r,\vartheta,\varphi) = \sum_{\ell=|s|}^{\ell_{max}}\sum_{m=-\ell}^{\ell}\mathbf{P}_\ell{}^{m}(r)\cdot {}_s{Y_\ell}{}^m(\vartheta,\varphi)\,,
\end{equation}
where  ${}_s{Y_\ell}{}^m$ denote the spin-weight $s$ spherical harmonics.
This way the evolutionary forms of the constraints can be reduced to a set of coupled ordinary differential equations (ODEs) for the expansion coefficients---$\mathbf{P}_\ell{}^{m}(r)$ for the variable ${}^{(s)}\mathbf{P}$---, whereas all the angular derivatives are evaluated analytically by using the action of the operators $\eth$ and $\overline{\eth}$ on the spin-weight $s$ spherical harmonics  ${}_s{Y_\ell}{}^m$ (for a more detailed discussion of the applied analytic background see, e.g.~appendix B of \cite{cskritgzs-2019}). As indicated in \eqref{eq: mult-exp} the summation in $\ell$ goes from $\ell=|s|$ (instead of to infinity only) up to some
$\ell=\ell_{max}$ value. In practice, $\ell_{max}$ was chosen to be $\ell_{max}=5$
which was found to be satisfactory in keeping the truncation error tolerably small.
The coupled set of ODEs were solved numerically by applying a $4^{th}$ order accurate adaptive Runge--Kutta--Fehlberg integrator.

\subsubsection{Convergence tests}

\begin{figure}[ht!]
	\begin{centering}
		{\tiny
			\begin{subfigure}{0.48\textwidth}
				\includegraphics[width=\textwidth]{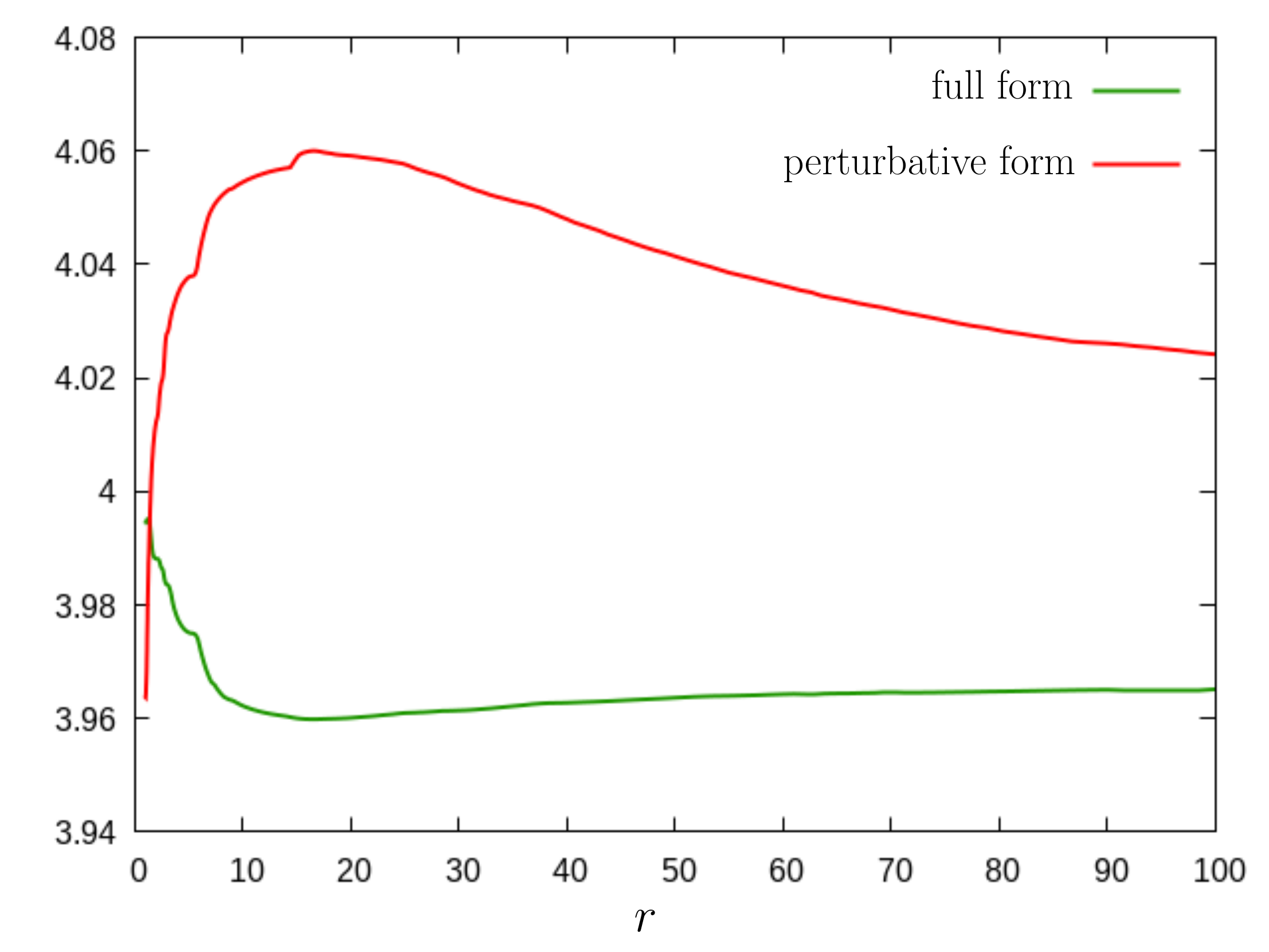}
				\caption{\footnotesize The algebraic-hyperbolic systems. }
				\label{fig:ah_conv}
			\end{subfigure}
			\hskip.02\textwidth
			\begin{subfigure}{0.48\textwidth}
				\includegraphics[width=\textwidth]{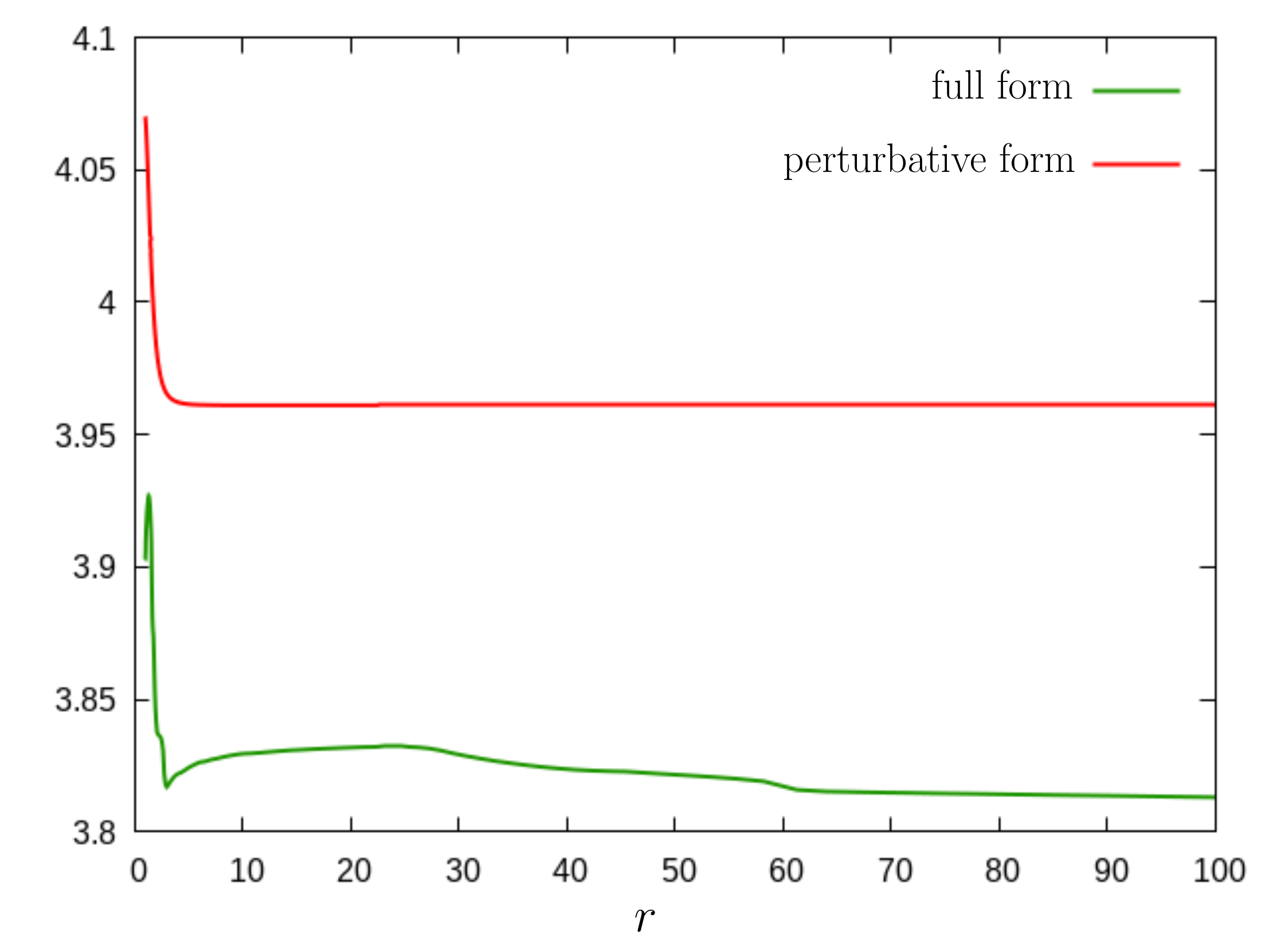}
				\caption{\footnotesize  The parabolic-hyperbolic systems.}
				\label{fig:ph_conv}
			\end{subfigure}
		}
	\end{centering}\vskip0.3cm
	\caption{{\footnotesize The convergence rates relevant for the evolution of the monopole ($\ell=m=0$) parts of $\KK$ and $\dKK=\KK-\KK_{Schw}$ are plotted, for the full and the non-linear perturbative form of the algebraic-hyperbolic system on the left, while for the full and the non-linear perturbative form of the parabolic hyperbolic system on the right. The pertinent solutions are yielded by using strictly near Schwarzschild initial data configurations and by applying the excitation  $\mathbf{K}|_{\mathscr{S}_{r_0}}=\mathbf{K}_{\text{Schw}}|_{\mathscr{S}_{r_0}} - \alpha \cdot {}_0{Y_1}{}^0$, with $\alpha=0.1$ and with $\kk|_{\mathscr{S}_{r_0}}=\kk_{\text{Schw}}=0$ and by $\dKK|_{\mathscr{S}_{r_0}}=- \alpha \cdot {}_0{Y_1}{}^0$, $\dkk|_{\mathscr{S}_{r_0}}=0$, respectively, for the algebraic-hyperbolic system, and this initial data choice was supplemented by $\widehat N|_{\mathscr{S}_{r_0}}=\widehat N_{\text{Schw}}|_{\mathscr{S}_{r_0}}$ and $\dNNh|_{\mathscr{S}_{r_0}}=0$, respectively, for the parabolic-hyperbolic systems.
	As expected, in both cases solutions to the non-linear perturbative equations converge slightly better than they do for the full forms. }}\label{fig: conv}
\end{figure}

This subsection is to demonstrate---regardless whether the full or non-linear perturbative form of the equations are used, and also regardless whether the algebraic hyperbolic or the parabolic-hyperbolic system are applied---that our adaptive numerical integrator always produces the expected fourth-order convergence rate. In particular, the convergence rates relevant for the evolution of the monopole ($\ell=m=0$) part of the variables $\KK$ and $\dKK=\KK-\KK_{Schw}$ are plotted for the full and non-linear perturbative forms of the  algebraic-hyperbolic system on  Fig.\,(\ref{fig:ah_conv}), whereas  for the full and non-linear perturbative forms of the  parabolic-hyperbolic system on Fig.\,(\ref{fig:ph_conv}), respectively. In all the investigated cases the initial data for $\mathbf{K}$ was chosen to be the form $\mathbf{K}|_{\mathscr{S}_{r_0}}=\mathbf{K}_{\text{Schw}}|_{\mathscr{S}_{r_0}} - \alpha \cdot {}_0{Y_1}{}^0$, i.e.~we had $\dKK|_{\mathscr{S}_{r_0}}=- \alpha \cdot {}_0{Y_1}{}^0$, with $\alpha=0.1$ at $r_0=1$, whereas the initial data for the other constraint fields was fixed by choosing their Schwarzschild value at ${\mathscr{S}_{r_0}}$. The corresponding initial values can be evaluated by replacing $r$ by $r_0$ in the functional forms of these variables as given in   Table\,\ref{table:data0}. 

Since our numerical scheme is adaptive in the radial direction the convergence rate was evaluated such that the numerical values, $\mathscr{N}_1$,  for the roughest grid were determined by applying adaptive time steps guided merely by the built in error tolerance. Once this happened, two additional numerical solutions $\mathscr{N}_2$ and $\mathscr{N}_3$ were also determined such that the ``time steps'' for $\mathscr{N}_2$ were exactly half of those for $\mathscr{N}_1$, and analogously the ``time steps'' for $\mathscr{N}_3$ were half of those applied in determining $\mathscr{N}_2$. Accordingly, neither of the grids was of uniform type. The convergence rates plotted on panels of Fig.\,\ref{fig: conv} were then calculated by evaluating the expression
\begin{equation}
-\log_2\frac{|\mathscr{N}_3-\mathscr{N}_2|}{|\mathscr{N}_2-\mathscr{N}_1|}
\end{equation}
at the grid points associated with $\mathscr{N}_1$.

\subsection{Solutions to the full set of equations}\label{subsec: results-full-equat}

This subsection is to present our numerical results relevant for the use of the full set of evolutionary form of the constraint equations. These results are to demonstrate that without a clear separation of the excitation modes from the Schwarzschild background it is really hard to identify those modes which play the key role in determining the asymptotics of the yielded initial data configurations.

\begin{figure}[ht!]
	\begin{centering}
		{\tiny
			\begin{subfigure}{0.48\textwidth}
				\includegraphics[width=\textwidth]{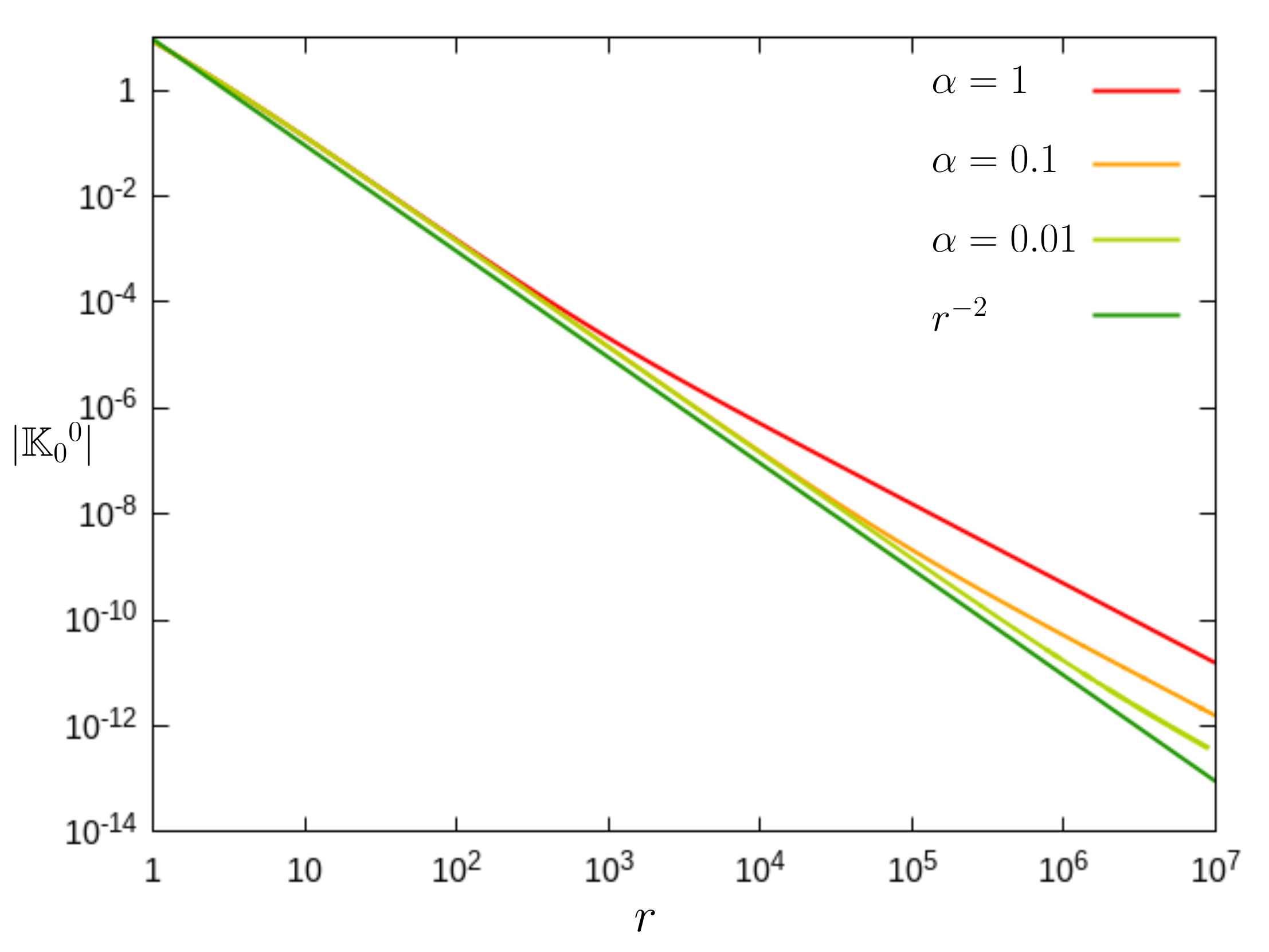}
				\caption{\footnotesize The $r$-dependence of the $\KK_0{}^0$  mode of $\KK$ evolved by the algebraic-hyperbolic system. }
				\label{fig:ah_pert}
			\end{subfigure}
			\hskip.02\textwidth
			\begin{subfigure}{0.48\textwidth}
				\includegraphics[width=\textwidth]{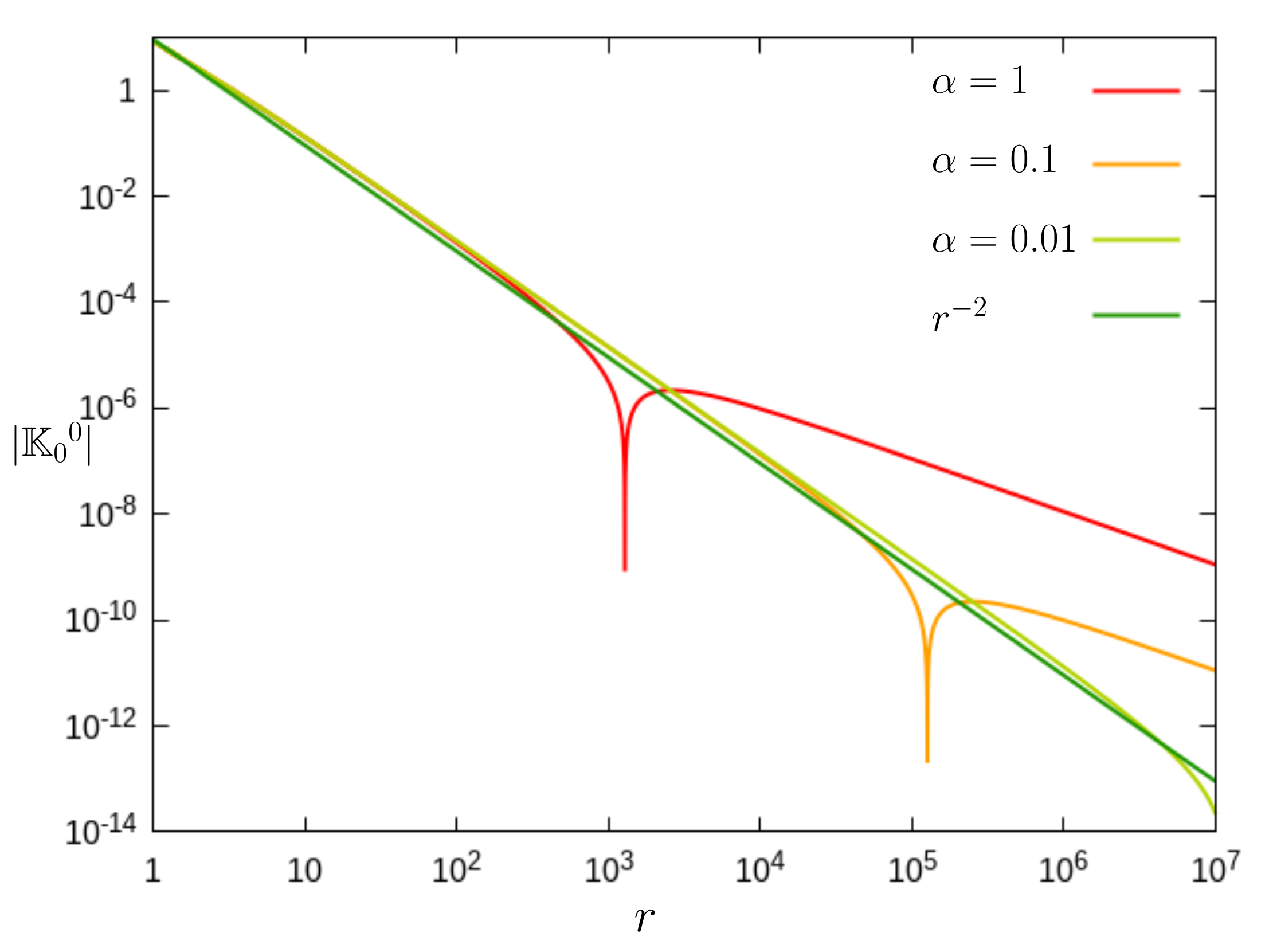}
				\caption{\footnotesize The $r$-dependence of the $\KK_0{}^0$  mode of $\KK$ evolved by the parabolic-hyperbolic system.}
				\label{fig:ph_pert}
			\end{subfigure}
		}
	\end{centering}
\caption{\footnotesize The $r$-dependence of the absolute value of the monopole ($\ell=m=0$) part of $\KK$ is shown for solutions yielded by the excitation $\mathbf{K}|_{\mathscr{S}_{r_0}}=\mathbf{K}_{\text{Schw}}|_{\mathscr{S}_{r_0}} - \alpha \cdot {}_0{Y_1}{}^0$, with $\alpha=1,\,0.1,\,0.01$ and evolved on the left by the algebraic-hyperbolic, whereas on the right by the parabolic-hyperbolic system. The initial data for $\kk$ was $\kk|_{\mathscr{S}_{r_0}}=\kk_{\text{Schw}}=0$ , in both cases, while, in the parabolic-hyperbolic case, for $\widehat N$ it was $\widehat N|_{\mathscr{S}_{r_0}}=\widehat N_{\text{Schw}}|_{\mathscr{S}_{r_0}}$. The smaller the amplitude $\alpha$ of the excitation is the later its effect is showing up in the monopole part of $\KK$. In both the  algebraic-hyperbolic and the parabolic-hyperbolic cases for $\alpha=0.01$ the deviation from the desired fall off rate remain almost completely hidden until relatively high values of $r$ is reached.}
\end{figure}

To see that this is indeed the case we start by inspecting the $r$-dependence of the absolute value of the monopole ($\ell=m=0$) part of $\KK$ is shown for solutions yielded by a non-spherical excitation. Fig.\,\ref{fig:ah_pert} is relevant for integrating the full form of the algebraic-hyperbolic system while Fig.\,\ref{fig:ph_pert} for the full form of the parabolic-hyperbolic system. In both cases, the initial data specified at $\mathscr{S}_{r_0}$, with $r_0=1$, was chosen to be of the form
$\mathbf{K}|_{\mathscr{S}_{r_0}}=\mathbf{K}_{\text{Schw}}|_{\mathscr{S}_{r_0}} - \alpha \cdot {}_0{Y_1}{}^0$, where $\alpha$ is a positive constant, i.e.~the excitation is for $\mathbf{K}$ is of a pure axially symmetric, $m=0$, spherical mode, with  $\kk|_{\mathscr{S}_{r_0}}=\kk_{\text{Schw}}=0$, which was also supplemented by the choice $\widehat N|_{\mathscr{S}_{r_0}}=\widehat N_{\text{Schw}}|_{\mathscr{S}_{r_0}}$ for the parabolic-hyperbolic system. Three different solutions are plotted in each cases corresponding to the excitation amplitudes $\alpha=1,\,0.1,\,0.01$. By inspecting the corresponding graphs it is immediately transparent that the smaller is the amplitude the later the deviation from the desired $\sim{r^{-2}}$ fall off rate is showing up in the  monopole ($\ell=m=0$) part of $\KK$. In particular, for the  algebraic-hyperbolic and the parabolic-hyperbolic systems the effect of the excitation with amplitude $\alpha=0.01$ remains almost completely hidden until the value $r\approx 10^4$ and $r\approx 10^7$ is reached, respectively.

Note also that on Figs.\,\ref{fig:ah_pert} and \ref{fig:ph_pert} the absolute value of various multipole coefficients are plotted against $r$. In addition, always $log-log$ scale is applied which allows us to indicate the polynomial character of the pertinent fall off rates immediately. Since all the remaining plots will be used to indicate the fall off rates of various modes, hereafter in all of the figures in the present paper, the $log-log$ scale will be applied.

There are some remarks in order now. It is important to be mentioned that even though initially only one of the higher $\ell$-modes ($\ell\geq 1$) of $\KK$ is excited---likewise the mode $\KK_1{}^0$ was excited in producing the solutions indicated on  Figs.\,\ref{fig:ah_pert} and \ref{fig:ph_pert}---due to the non-linear couplings of various modes, for strictly near Schwarzschild initial data configurations, the monopole part $\KK_0{}^0$ gets always to be triggered. Note also that exactly the same type of argument applies when various modes of $\kk$ or (in the parabolic-hyperbolic case) that of $\NNh$ are excited. As the difference between the distinct choices of excitations affects only the specific $r$-value where the monopole part $\KK_0{}^0$ of $\KK$ gets excited the use of  $\ell=1,2,3$ excitations of $\KK$ turned out to be optimal.

\medskip

It is worth to be emphasized again that even if the decay rate appears to be optimal on convincingly large $r$-intervals---in virtue of the implications Figs.\,\ref{fig:ah_pert} and \ref{fig:ph_pert}---there is no guarantee that the observed favorable behavior will remain for arbitrarily large values of $r$.
In practice, the above observations have another unfavorable consequence. As the smallest monopole perturbations of $\KK$ decay as $r^{-1}$ in certain situations even numerical errors may affect the integration of the full form of the constraint equations. This means that it may not possible to produce or investigate strictly near Schwarzschild initial data configurations unless such a numerical noise can be suppressed effectively.  Indeed, the desire to overcome the negative consequences of these two points motivated us to develop the non-linear perturbative scheme of the evolutionary forms of the constraint equations outlined in the next subsection.

\subsection{Non-linear perturbative form of the constraint equations}
\label{subsec: non-linear-pert}

As it is clearly indicated by Figs.\,\ref{fig:ah_pert} and \ref{fig:ph_pert} in many circumstances there may be an obvious need to investigate the evolution of separate mode excitations, meanwhile the effect of the background is suppressed. If this can be done one can get a much clearer picture since once the excitations separate from the Schwarzschild background even the small amplitude and slowly decaying modes can get to be transparent as they are not any more covered, for considerably long radial intervals, by the much higher amplitude contribution of the Schwarzschild background.

Note that the basic ideas outlined here had already been applied in using a parabolic-hyperbolic solver in generating Kerr-Schild type black hole initial data in \cite{Nakonieczna:2017eev}. Nevertheless, as the application and the explicit form of the non-linear perturbative approach is definitely new in case of the algebraic-hyperbolic system we decided to give here a systematic review of the basic ideas and in the appendix we provide the nonlinear perturbative form of the constraint equations, using the notation applied in this paper, for both the parabolic-hyperbolic and the algebraic-hyperbolic systems.

To start off note first that the non-linear perturbations can essentially be defined with respect to any fixed background, though, in the present paper it will be the  Schwarzschild initial data induced on $t_{KS}=const$ timeslices. In proceeding note that in both cases the evolutionary form of the constraints can be regarded as a set of equations of the form
\begin{equation}\label{eq: schem-evolv}
\partial_r \mathbf{f}_{(i)}=\mathscr{R}_{(i)}\left(\partial_A\mathbf{f}_{(j)},\mathbf{f}_{(j)}\right)\,,
\end{equation}
where $\mathbf{f}_{(i)}$ denote the dependent variables, $i$ takes the values $1,2,3$ for the parabolic-hyperbolic system and $1,2$ for the algebraic-hyperbolic system, respectively, and the explicit dependence of $\mathscr{R}_{(i)}$ on the background fields and on the coordinates on $\Sigma$ are suppressed.\footnote{It would be more appropriate to indicate second order spatial derivatives in favor of the parabolic equations. It is assumed, however, that the loosely notation applied here will not affect the core of the pertinent argument.} Replacing these dependent variables by the sum
${}^{\scriptscriptstyle(0)\!}\mathbf{f}_{(i)}+{}^{\scriptscriptstyle(\Delta)\!}\mathbf{f}_{(i)}$, where ${}^{\scriptscriptstyle(0)\!}\mathbf{f}_{(i)}$ stand for the unperturbed background variables whereas the variables
${}^{\scriptscriptstyle(\Delta)\!}\mathbf{f}_{(i)}$ denote the deviations from this background, i.e.~${}^{\scriptscriptstyle(\Delta)\!}\mathbf{f}_{(i)}=\mathbf{f}_{(i)}-{}^{\scriptscriptstyle(0)\!}\mathbf{f}_{(i)}$, we get
\begin{align}\label{eq: non-linear-pert}
	\partial_r {}^{\scriptscriptstyle(\Delta)\!} \mathbf{f}_{(i)} & = \mathscr{R}_{(i)}\left(\partial_A({}^{\scriptscriptstyle(0)\!}\mathbf{f}_{(j)}+{}^{\scriptscriptstyle(\Delta)\!}\mathbf{f}_{(j)}),{}^{\scriptscriptstyle(0)\!}\mathbf{f}_{(j)}+{}^{\scriptscriptstyle(\Delta)\!}\mathbf{f}_{(j)}\right)- \partial_r {}^{\scriptscriptstyle(0)\!} \mathbf{f}_{(i)} \nonumber \\
	& = \mathscr{R}_{(i)}\left(\partial_A({}^{\scriptscriptstyle(0)\!}\mathbf{f}_{(j)}+{}^{\scriptscriptstyle(\Delta)\!}\mathbf{f}_{(j)}),{}^{\scriptscriptstyle(0)\!}\mathbf{f}_{(j)}+{}^{\scriptscriptstyle(\Delta)\!}\mathbf{f}_{(j)}\right)-\mathcal{R}_{(i)}\left(\partial_A{}^{\scriptscriptstyle(0)\!}\mathbf{f}_{(j)},{}^{\scriptscriptstyle(0)\!}\mathbf{f}_{(j)}\right)\,,
\end{align}
where in the last step it was assumed that the background fields are subject to some equations, analogous to \eqref{eq: schem-evolv}, of the form
\begin{equation}\label{eq: scem-back}
	\partial_r {}^{\scriptscriptstyle(0)\!} \mathbf{f}_{(i)} = \mathcal{R}_{(i)}\left(\partial_A{}^{\scriptscriptstyle(0)\!}\mathbf{f}_{(j)},{}^{\scriptscriptstyle(0)\!}\mathbf{f}_{(j)}\right)\,.
\end{equation}
If the background fields ${}^{\scriptscriptstyle(0)\!}\mathbf{f}_{(i)}$ are also solutions to the constraint equations---i.e.~to the schematic form \eqref{eq: schem-evolv}---then (on the right hand sides of \eqref{eq: scem-back}) $\mathcal{R}_{(i)}$ possess exactly the same functional form as $\mathscr{R}_{(i)}$ do in \eqref{eq: schem-evolv}.

In the particular case of the parabolic-hyperbolic system the above outlined splitting applies to
\begin{align}\label{eq: splitting-ph}
	\NNh&=\bgNNh+\dNNh\,, \quad
	\KK=\bgKK+\dKK\,, \quad
	\kk=\bgkk+\dkk
\end{align}
whereas in case of the algebraic-hyperbolic system to
\begin{align}\label{eq: splitting-ah}
	\kkappa=\bgkappa+\dkappa\,, \quad
	\KK=\bgKK+\dKK\,, \quad
	\kk=\bgkk+\dkk\,.
\end{align}
The explicit form of the non-linear perturbative equations relevant for the splittings used in \eqref{eq: splitting-ph} and \eqref{eq: splitting-ah} are given in the Appendix.

\section{Numerical results based on non-linear perturbations}\label{sec: results-on-non-lin-pert}
\setcounter{equation}{0}

In this section our numerical results relevant for the use of the non-linear perturbative form of the constraint equations are presented. These results are to demonstrate that by separating the excitation modes from the Schwarzschild background one can get a much clearer picture and an effective framework that could also be used (if possible) to get control on the asymptotic flatness of the yielded initial data configurations.

\subsection{Non-linear perturbation with parabolic-hyperbolic system}\label{subsec: par-hyp-non-lin}

\begin{figure}[ht!]
	\vskip-0.1cm
	\begin{centering}
		{\tiny
			\begin{subfigure}{0.48\textwidth}
				\includegraphics[width=\textwidth]{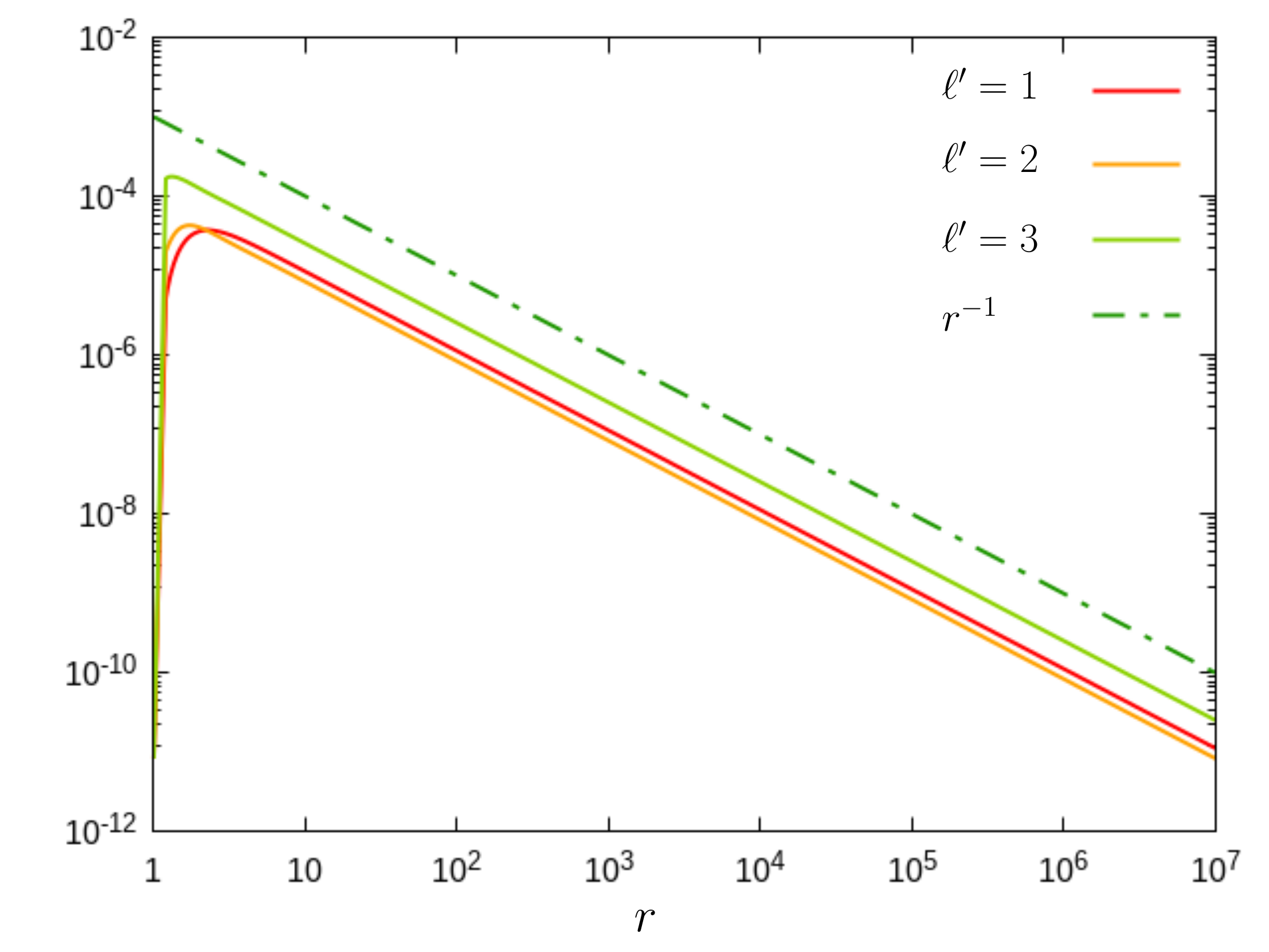}
                \vskip-0.4cm
				\caption{\scriptsize The decay rate of the mode $\dKK{\,}_0{}^0$ of $\dKK$
					is $r^{-1}$. }
				\label{fig:decphK0}
			\end{subfigure}
			\begin{subfigure}{0.48\textwidth}\vskip0.2cm
				\includegraphics[width=\textwidth]{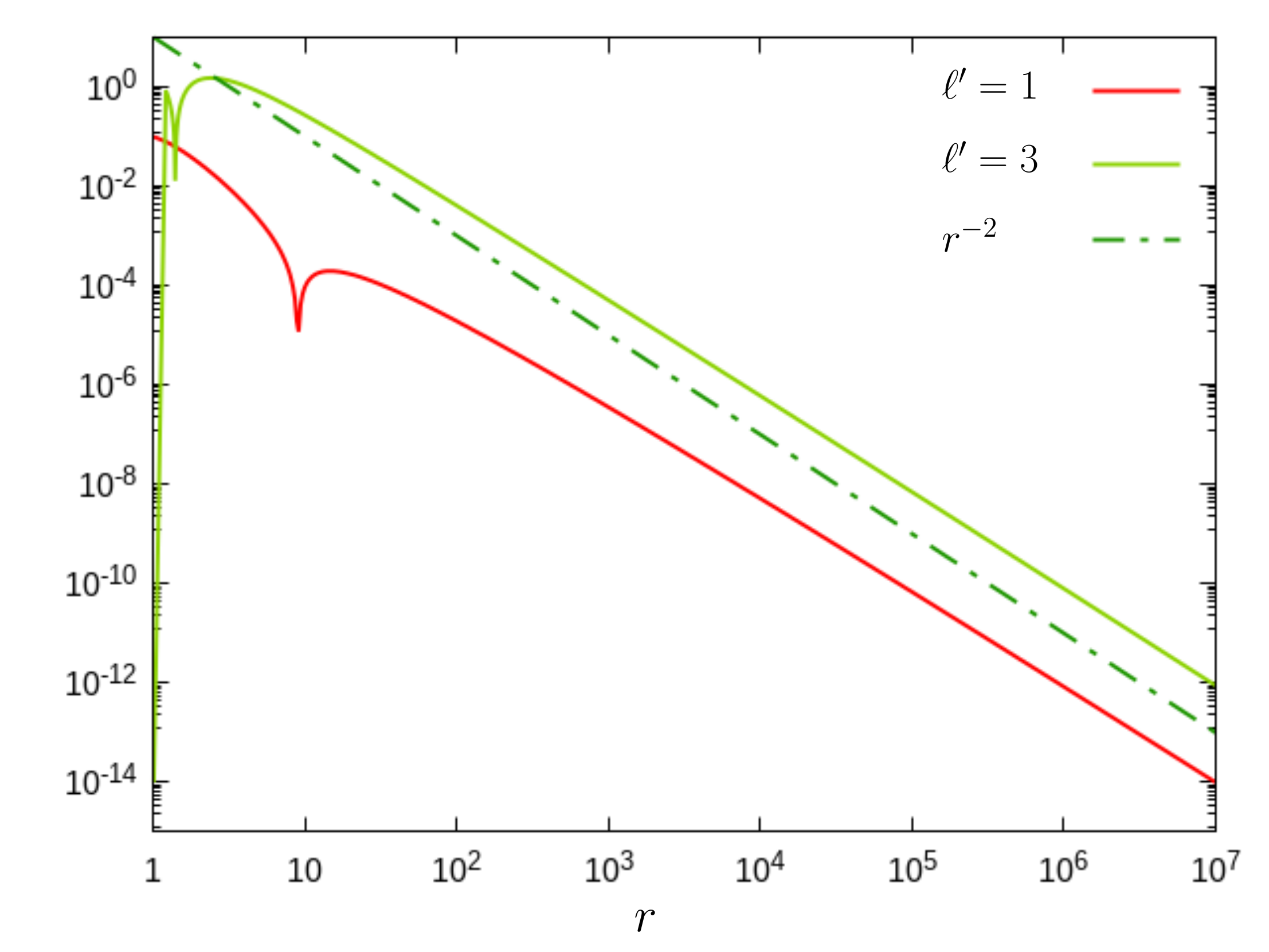}
                \vskip-0.4cm
				\caption{\scriptsize The decay rate of the mode $\dKK{\,}_1{}^0$ of $\dKK$
					is $r^{-2}$.}
				\label{fig:decphK1}
			\end{subfigure}
			\begin{subfigure}{0.48\textwidth}\vskip0.2cm
				\includegraphics[width=\textwidth]{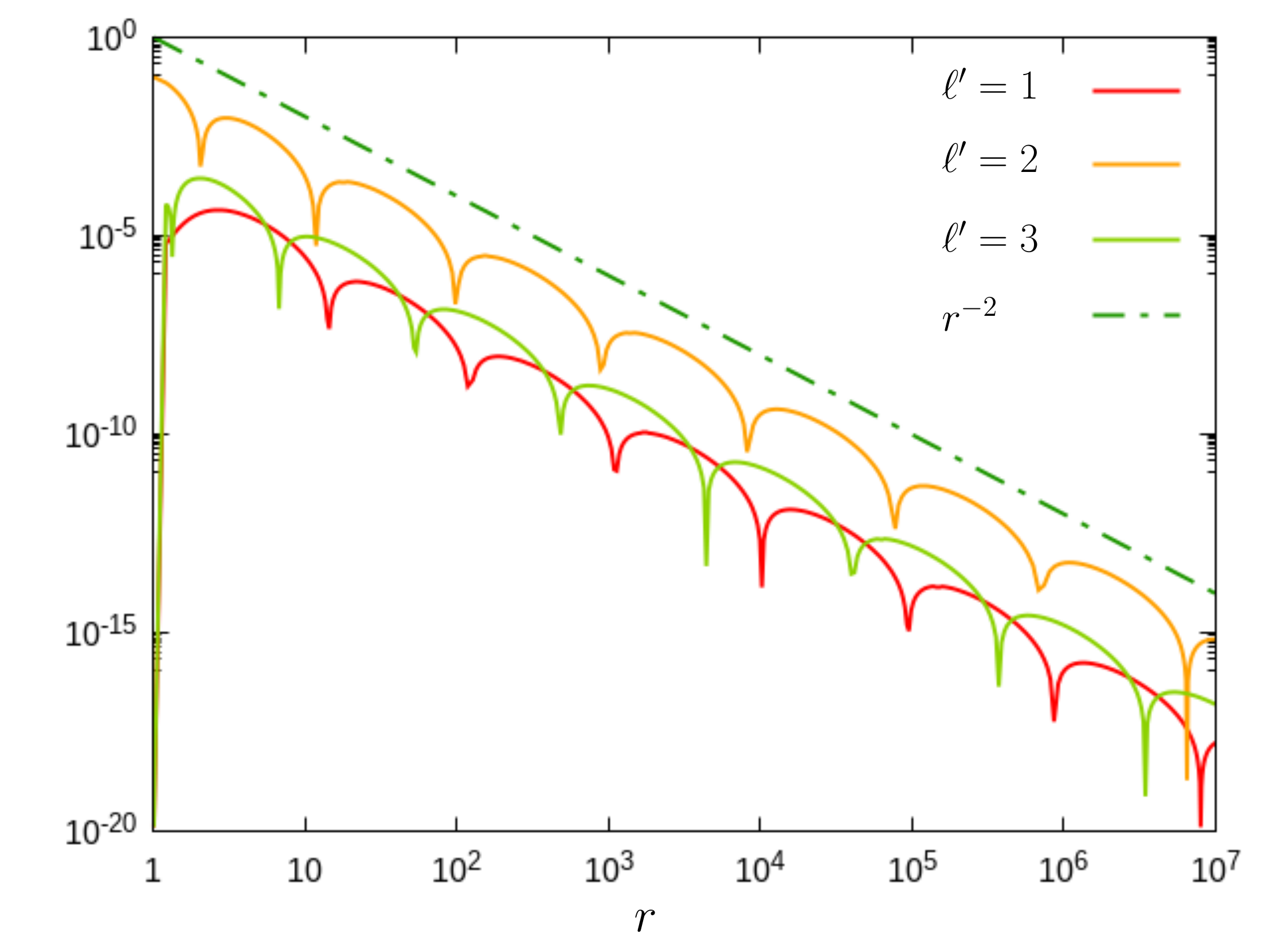}
                \vskip-0.4cm
				\caption{\scriptsize The decay rate of the mode $\dKK{\,}_2{}^0$ of $\dKK$
					is $r^{-2}$.}
				\label{fig:decphK2}
			\end{subfigure}
			\hskip.5cm
			\begin{subfigure}{0.48\textwidth}\vskip0.2cm
				\includegraphics[width=\textwidth]{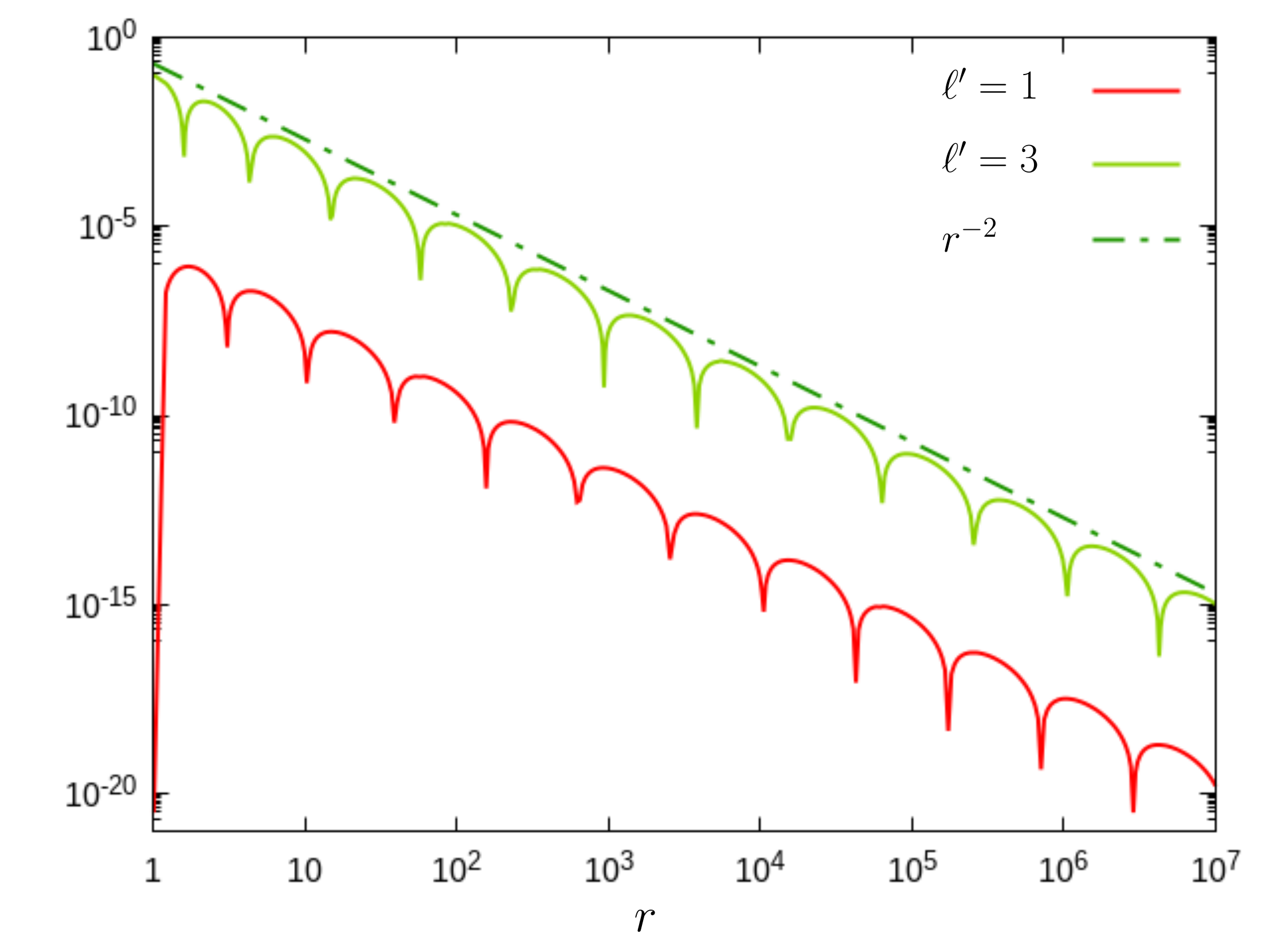}
                \vskip-0.4cm
				\caption{\scriptsize The decay rate of the mode $\dKK{\,}_3{}^0$ of $\dKK$ is $r^{-2}$.}
				\label{fig:decphK3}
			\end{subfigure}
			\begin{subfigure}{0.48\textwidth}\vskip0.2cm
				\includegraphics[width=\textwidth]{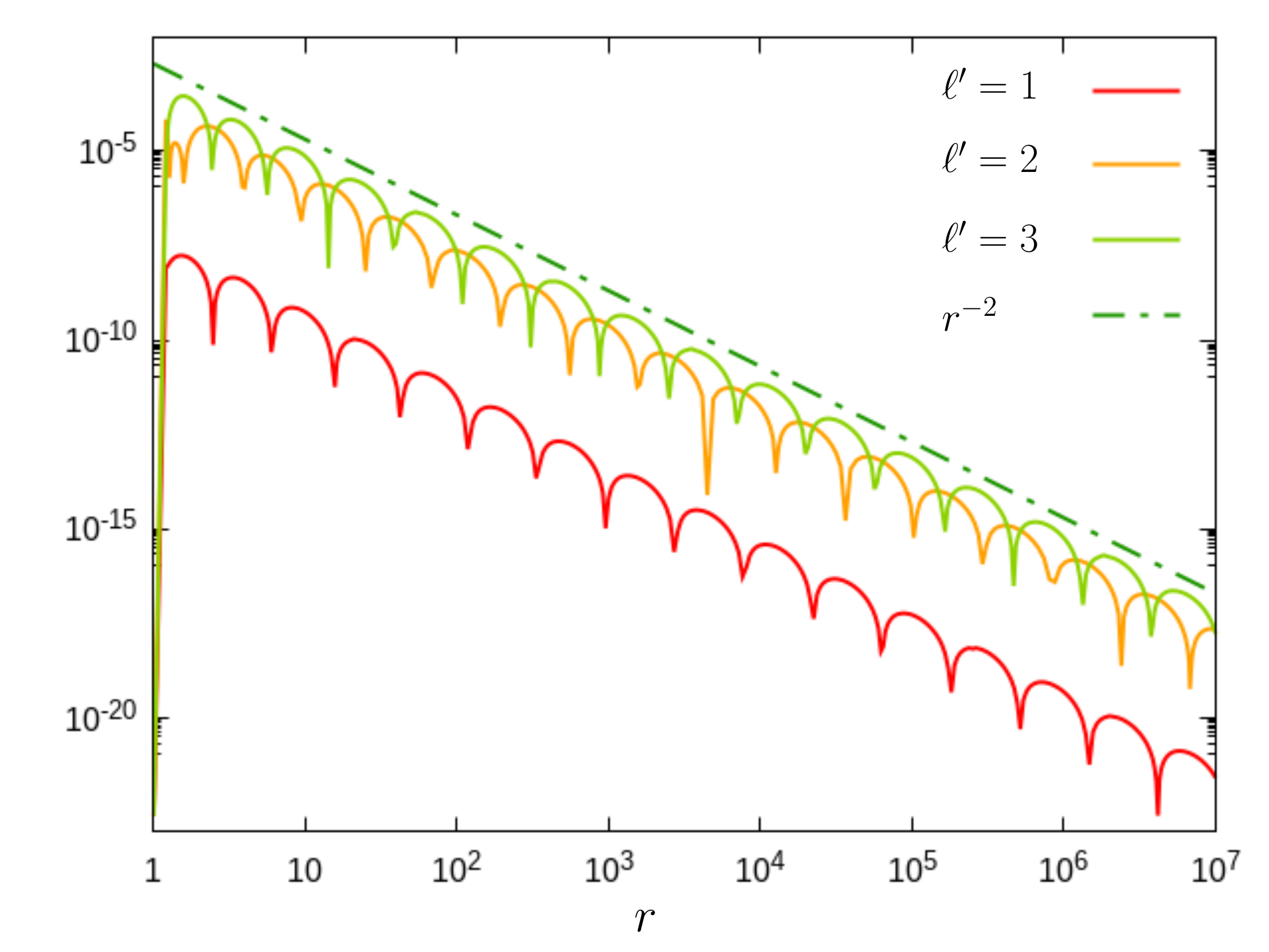}
                \vskip-0.4cm
				\caption{\scriptsize The decay rate of the mode $\dKK{\,}_4{}^0$ of $\dKK$ is $r^{-2}$.}
				\label{fig:decphK4}
			\end{subfigure}
			\hskip.5cm
			\begin{subfigure}{0.48\textwidth}\vskip0.2cm
				\includegraphics[width=\textwidth]{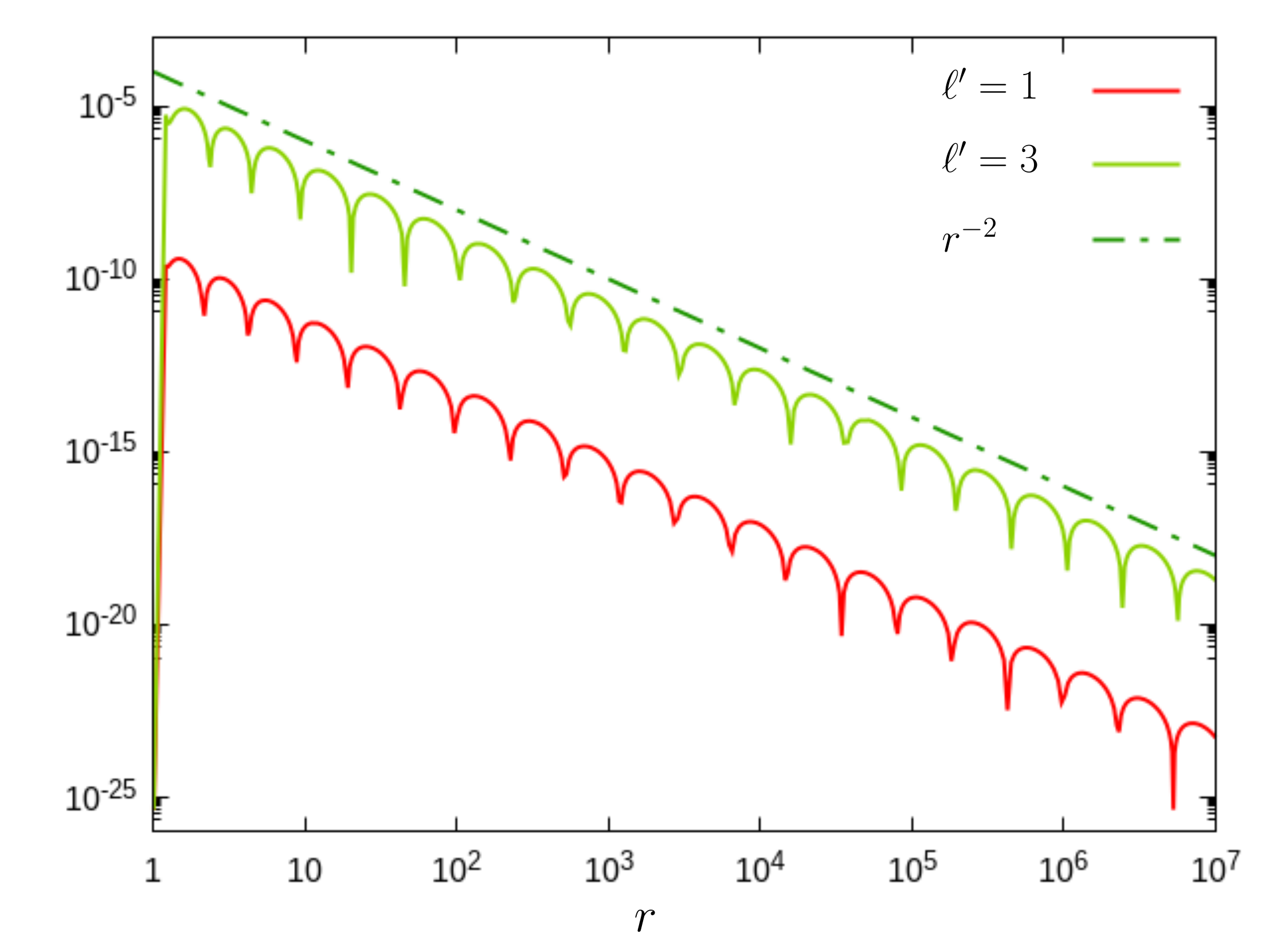}
                \vskip-0.4cm
				\caption{\scriptsize The decay rate of the mode $\dKK{\,}_5{}^0$ of $\dKK$ is $r^{-2}$.}
				\label{fig:decphK5}
			\end{subfigure}
		}
	\end{centering}
\vskip-0.1cm
	\caption{\footnotesize  The non-linear perturbative form of the parabolic-hyperbolic system was integrated numerically by applying the initial perturbation  $\dKK|_{\mathscr{S}_{r_0}}=-10^{-1}\cdot {}_0{Y_{\ell'}}{}^0$ with $\ell'=1,2,3$. The modes $\dKK{\,}_\ell{}^0$, with $\ell=1,2,3,4,5,$ of $\dKK$ decay as $r^{-2}$,  whereas the monopole part $\dKK{\,}_0{}^0$ decays only as $r^{-1}$.}
	\label{fig:decphK}
\end{figure}

\begin{figure}[ht!]
	\vskip-0.1cm
	\begin{centering}
		{\tiny
			\begin{subfigure}{0.48\textwidth}
				\includegraphics[width=\textwidth]{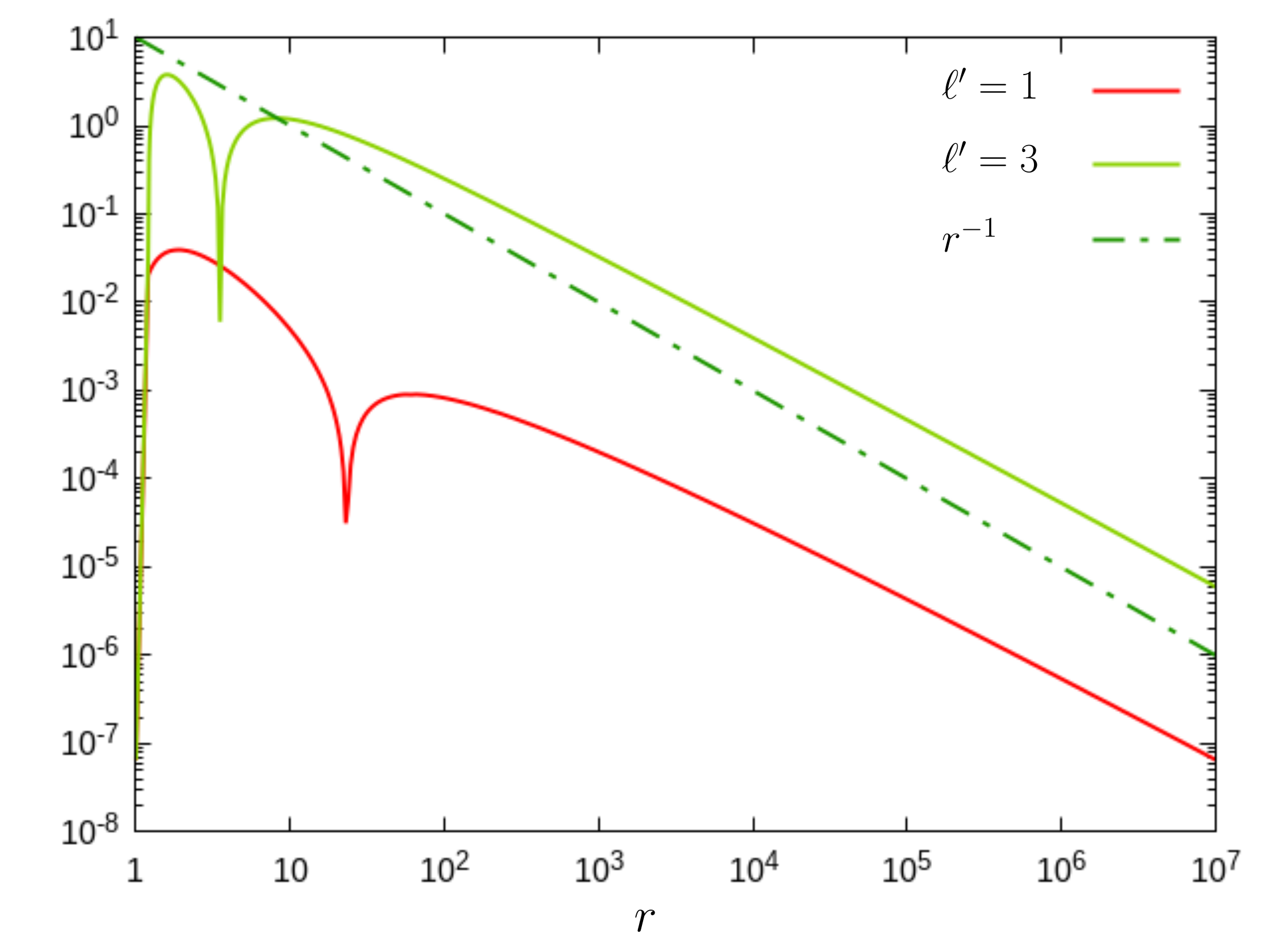}
				\vskip-0.4cm
				\caption{\scriptsize The decay rate of the mode $\dkk{\,}_1{}^0$ of $\dkk$ is $r^{-1}$. }
				\label{fig:decphk1}
			\end{subfigure}
			\begin{subfigure}{0.48\textwidth}
				\includegraphics[width=\textwidth]{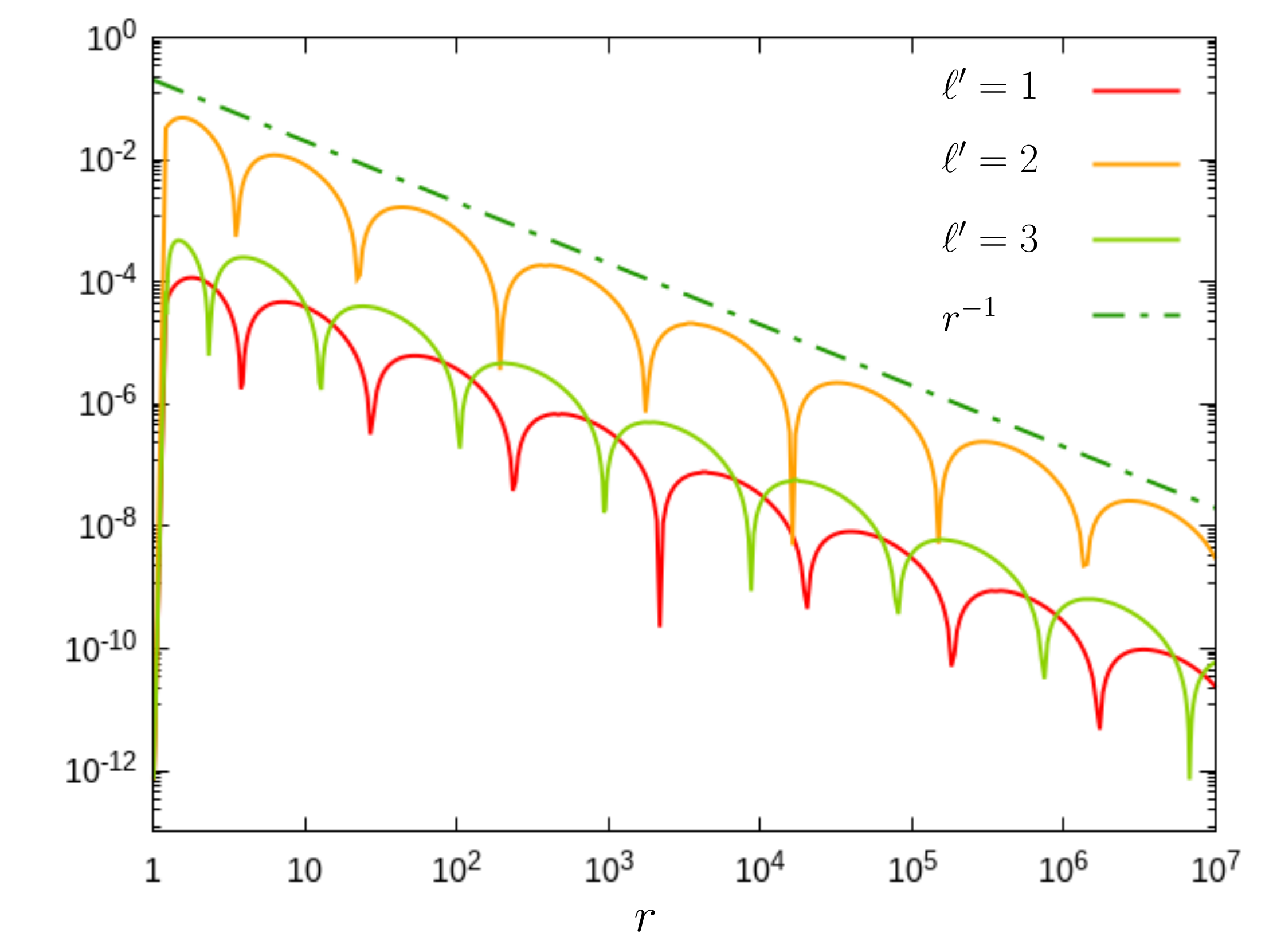}
				\vskip-0.4cm
				\caption{\scriptsize The decay rate of the mode $\dkk{\,}_2{}^0$ of $\dkk$ is $r^{-1}$.   }
				\label{fig:decphk2}
			\end{subfigure}
			\hskip.5cm
			\begin{subfigure}{0.48\textwidth}\vskip0.2cm
				\includegraphics[width=\textwidth]{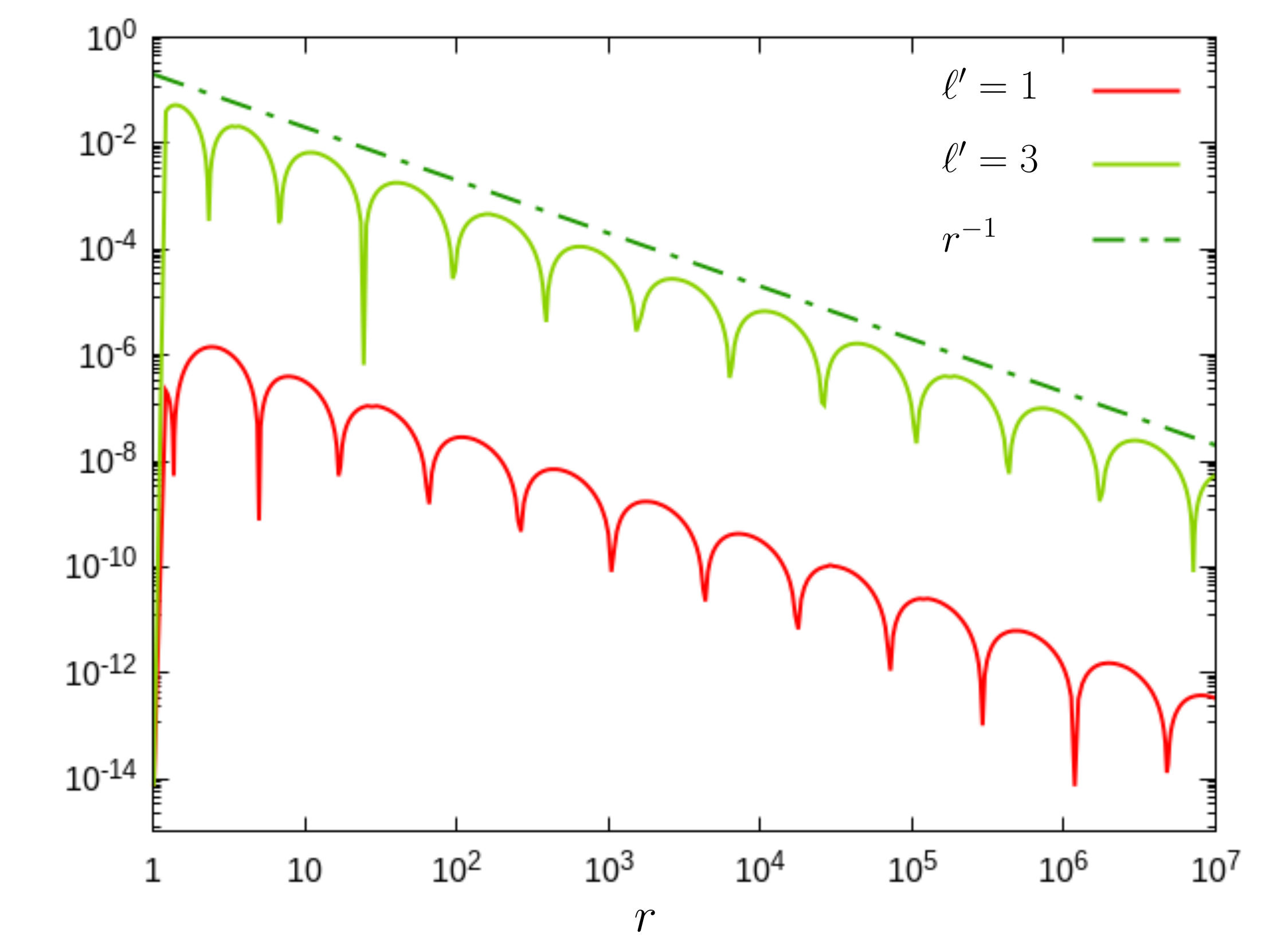}
				\vskip-0.4cm
				\caption{\scriptsize The decay rate of the mode $\dkk{\,}_3{}^0$ of $\dkk$ is $r^{-1}$.}
				\label{fig:decphk3}
			\end{subfigure}
			\begin{subfigure}{0.48\textwidth}\vskip0.2cm
				\includegraphics[width=\textwidth]{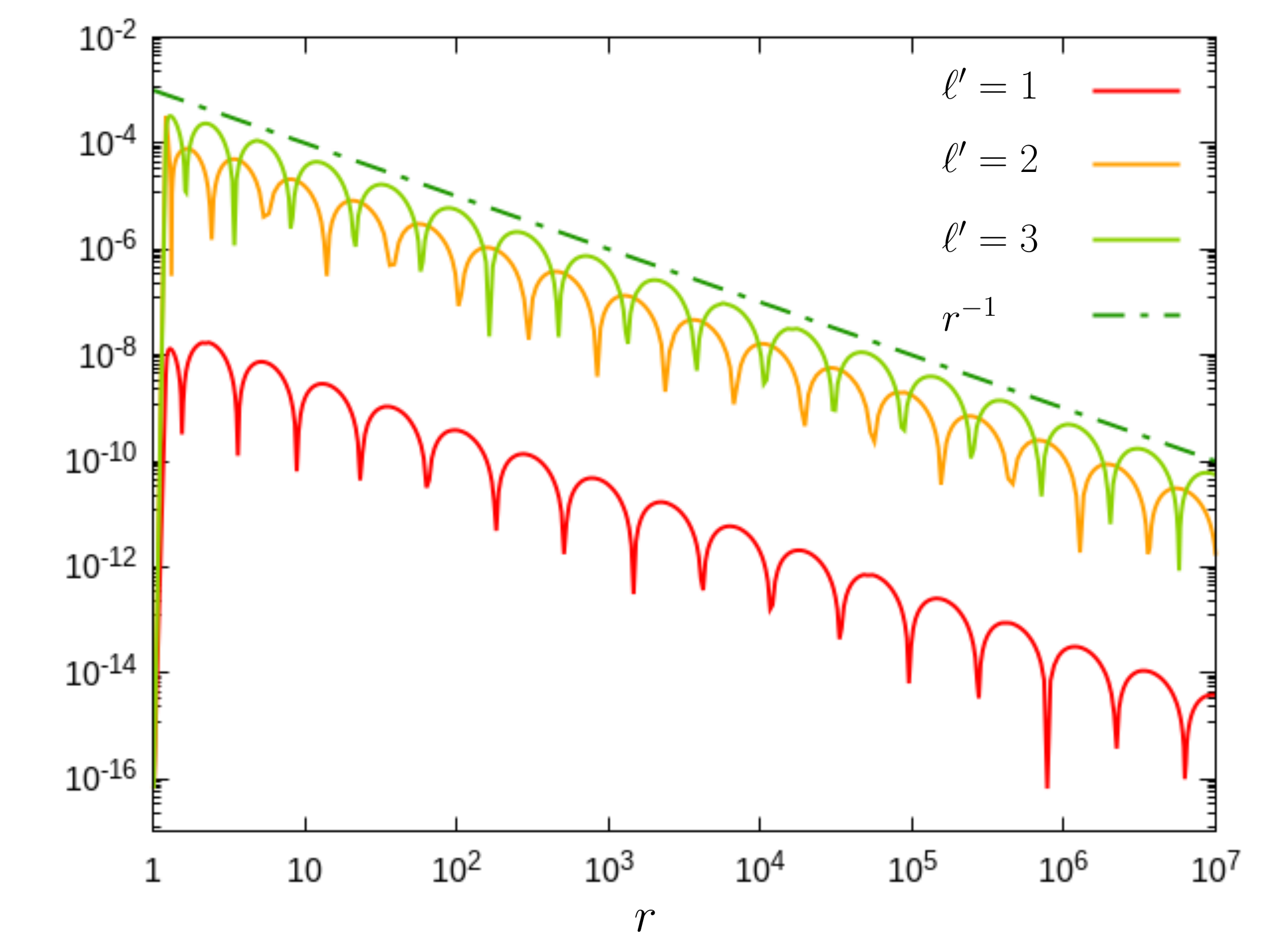}
				\vskip-0.4cm
				\caption{\scriptsize  The decay rate of the mode $\dkk{\,}_4{}^0$ of $\dkk$ is $r^{-1}$.}
				\label{fig:decphk4}
			\end{subfigure}
			\hskip.5cm
			\begin{subfigure}{0.48\textwidth}\vskip0.2cm
				\includegraphics[width=\textwidth]{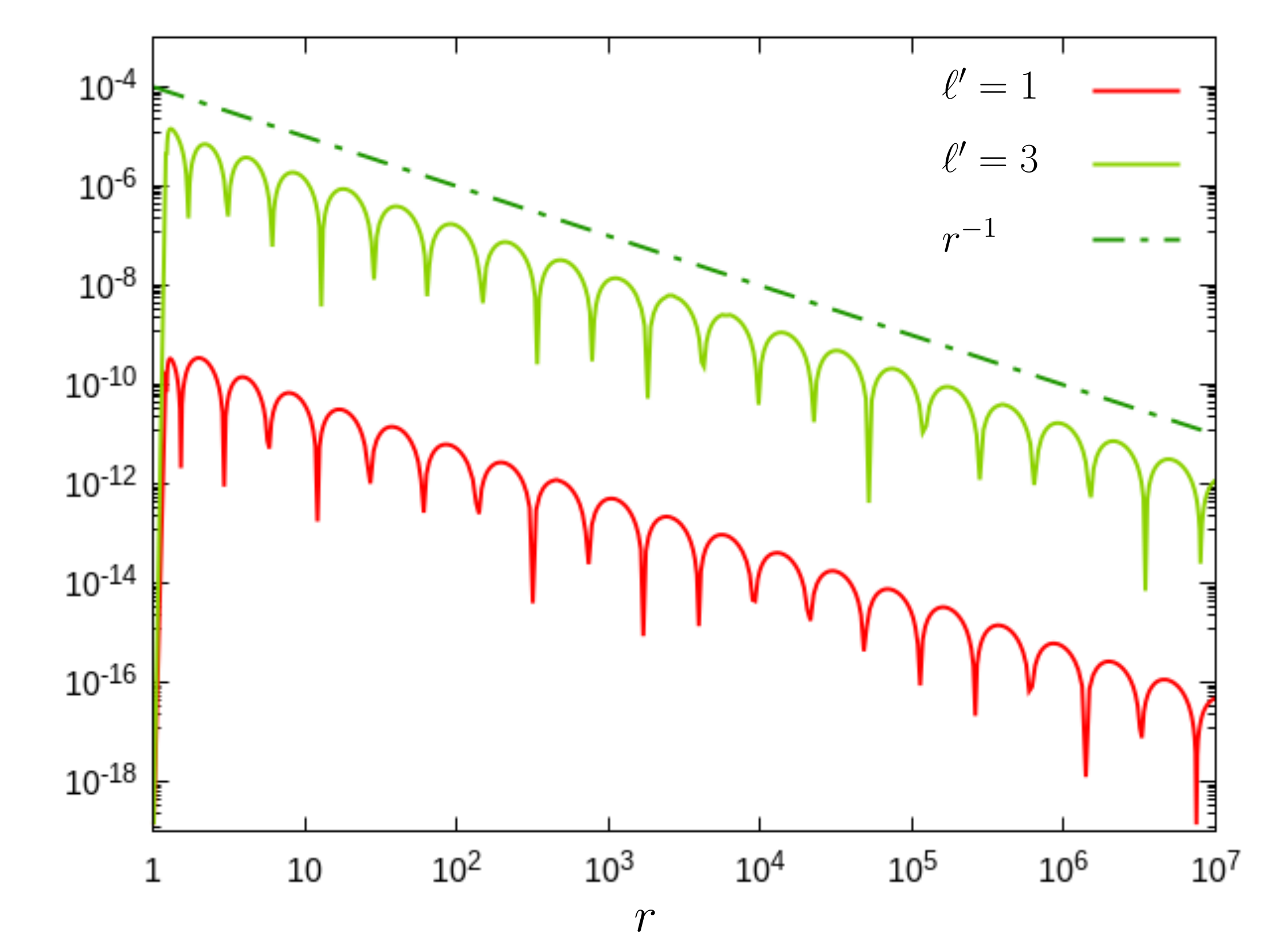}
				\vskip-0.4cm
				\caption{\scriptsize The decay rate of the mode $\dkk{\,}_5{}^0$ of $\dkk$ is $r^{-1}$.}
				\label{fig:decphk5}
			\end{subfigure}
		}
	\end{centering}
\vskip0.1cm
	\caption{\footnotesize  The non-linear perturbative form of the parabolic-hyperbolic system was integrated numerically by applying the initial perturbation  $\dKK|_{\mathscr{S}_{r_0}}=-10^{-1}\cdot {}_0{Y_{\ell'}}{}^0$ with $\ell'=1,2,3$. Neither of the decay rates of the modes $\dkk{\,}_\ell{}^0$ with $\ell=1,2,3,4,5,\, m=0$ of $\dkk$ is slower than $r^{-1}$, though the mode $\dkk{\,}_1{}^0$ decays slightly slower than the others. }
	\label{fig:decphk}
\end{figure}

\begin{figure}[ht!]
	\vskip-0.1cm
	\begin{centering}
		{\tiny
			\begin{subfigure}{0.48\textwidth}
				\includegraphics[width=\textwidth]{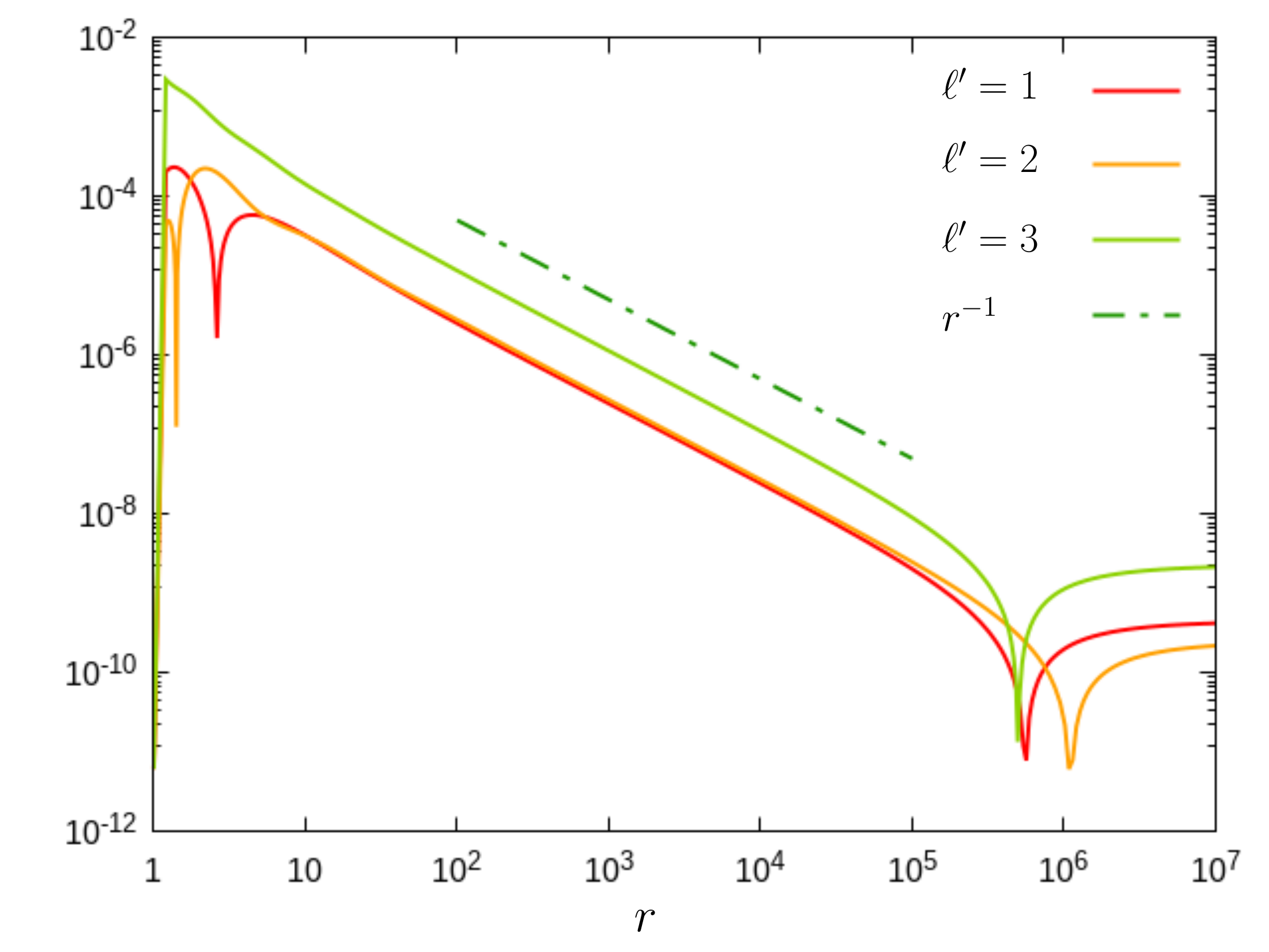}
                \vskip-0.4cm
				\caption{\scriptsize The decay rate of the monopole part $\dNNh_0{}^0$ of $\dNNh$.}
				\label{fig:decphN0}
			\end{subfigure}
			\begin{subfigure}{0.48\textwidth}
				\includegraphics[width=\textwidth]{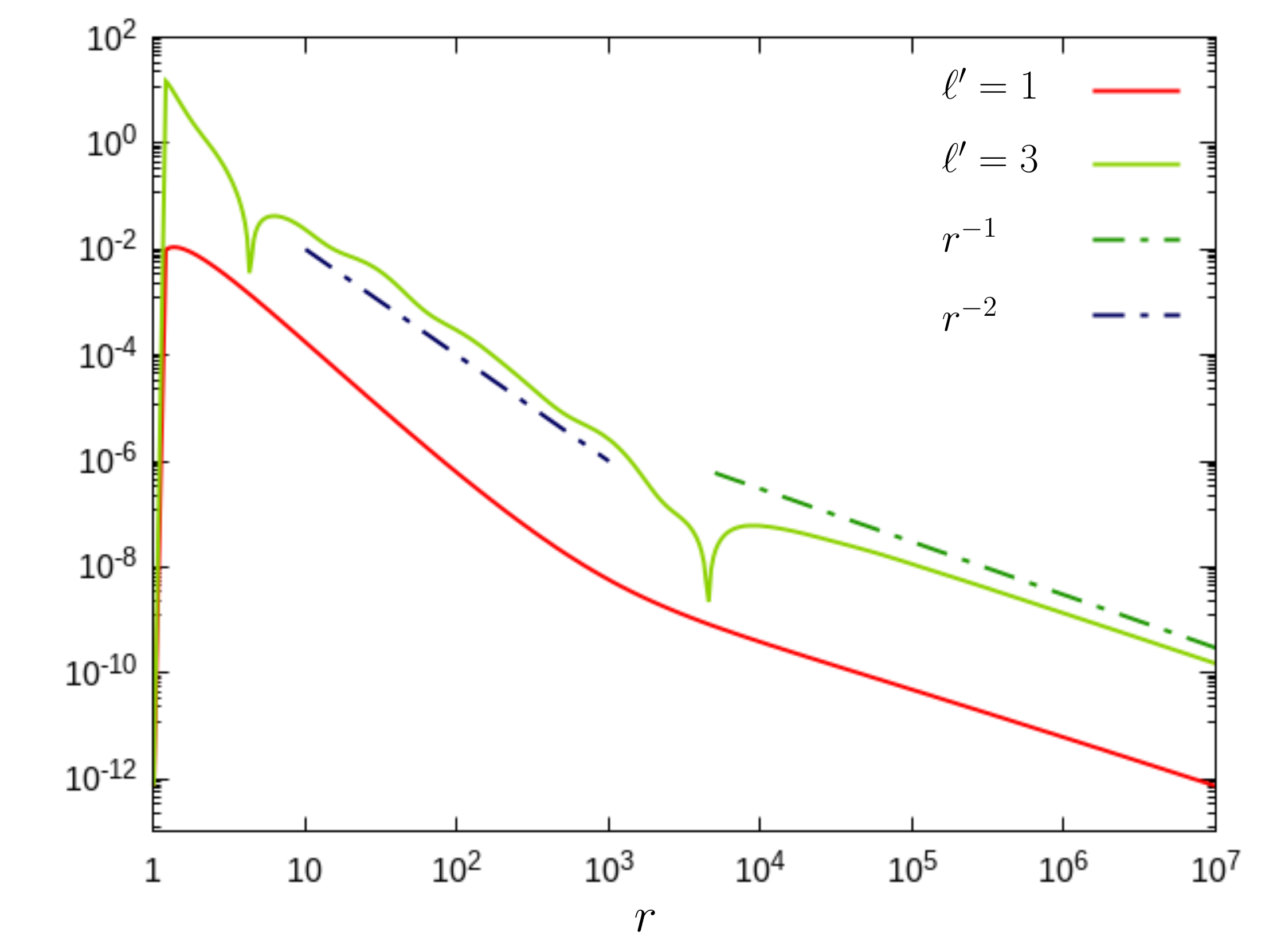}
                \vskip-0.4cm
				\caption{\scriptsize  The decay rate of the mode $\dNNh_1{}^0$ of $\dNNh$ is $r^{-1}$.}
				\label{fig:decphN1}
			\end{subfigure}
			\begin{subfigure}{0.48\textwidth}\vskip0.1cm
				\includegraphics[width=\textwidth]{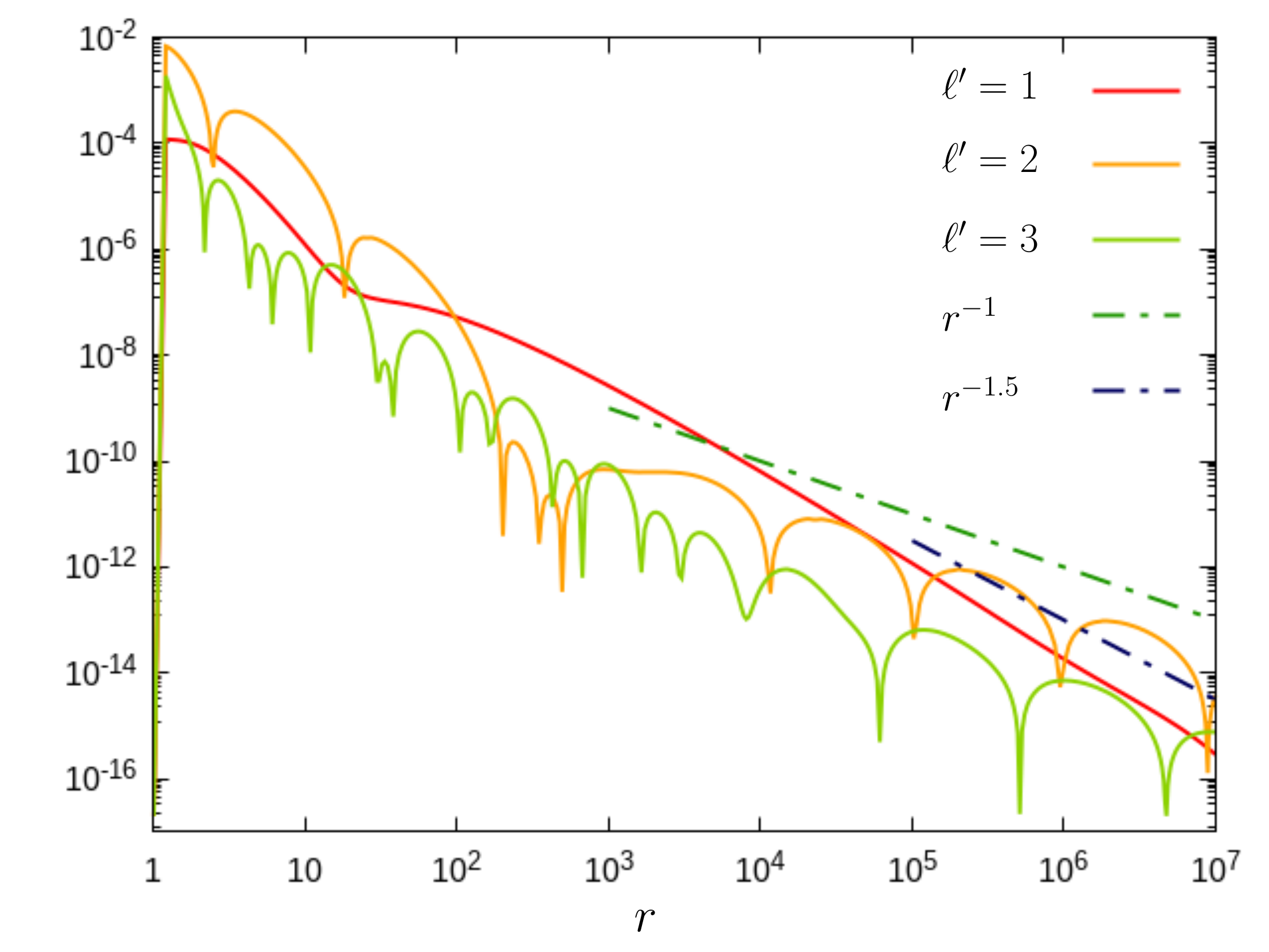}
                \vskip-0.4cm
				\caption{\scriptsize The decay rate of the mode $\dNNh_2{}^0$ of $\dNNh$ is $r^{-1}$.}
				\label{fig:decphN2}
			\end{subfigure}
			\hskip.5cm
			\begin{subfigure}{0.48\textwidth}\vskip0.1cm
				\includegraphics[width=\textwidth]{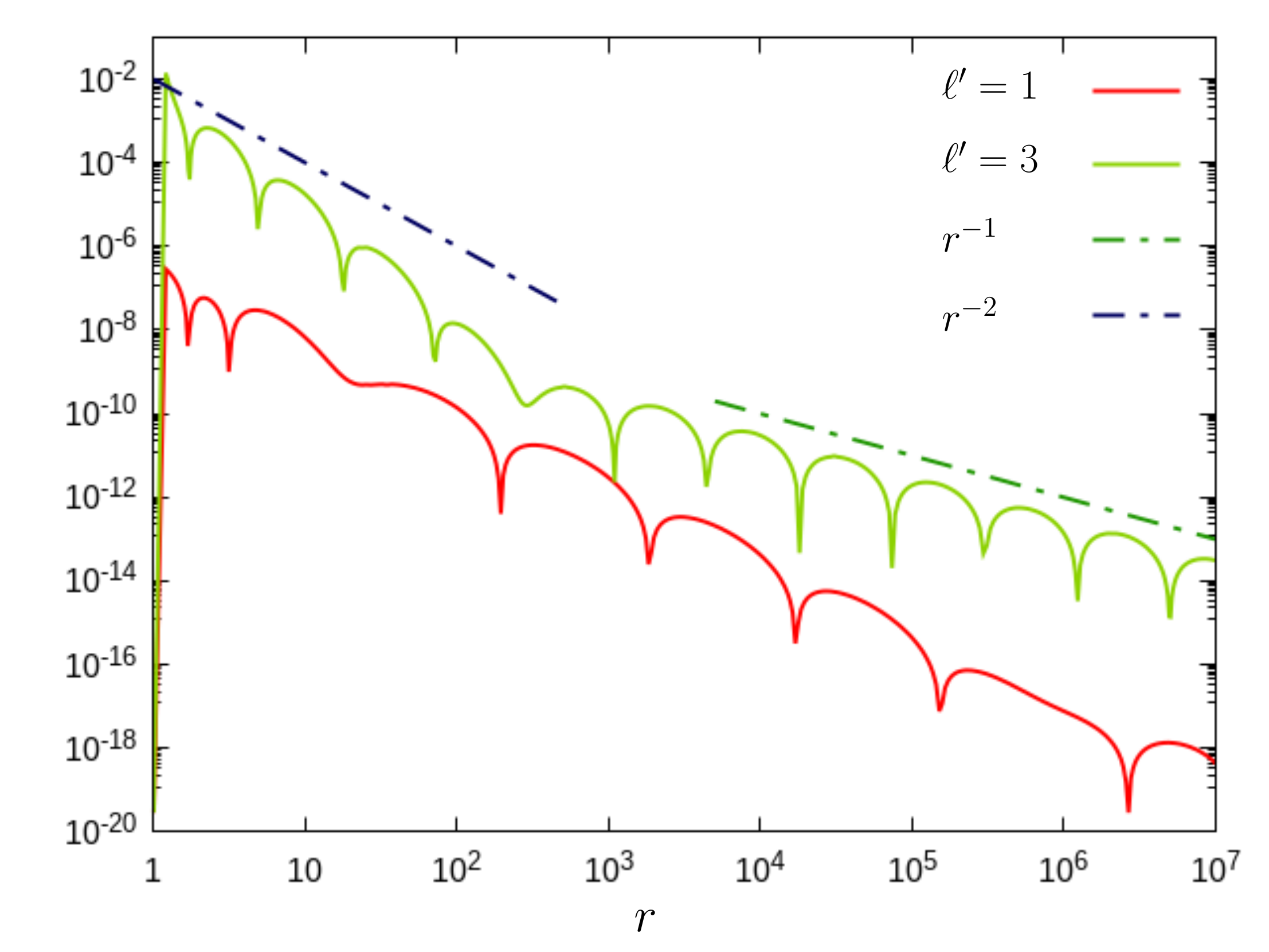}
                \vskip-0.4cm
				\caption{\scriptsize The decay rate of the mode $\dNNh_3{}^0$ of $\dNNh$ is $r^{-1}$.}
				\label{fig:decphN3}
			\end{subfigure}
			\begin{subfigure}{0.48\textwidth}\vskip0.1cm
				\includegraphics[width=\textwidth]{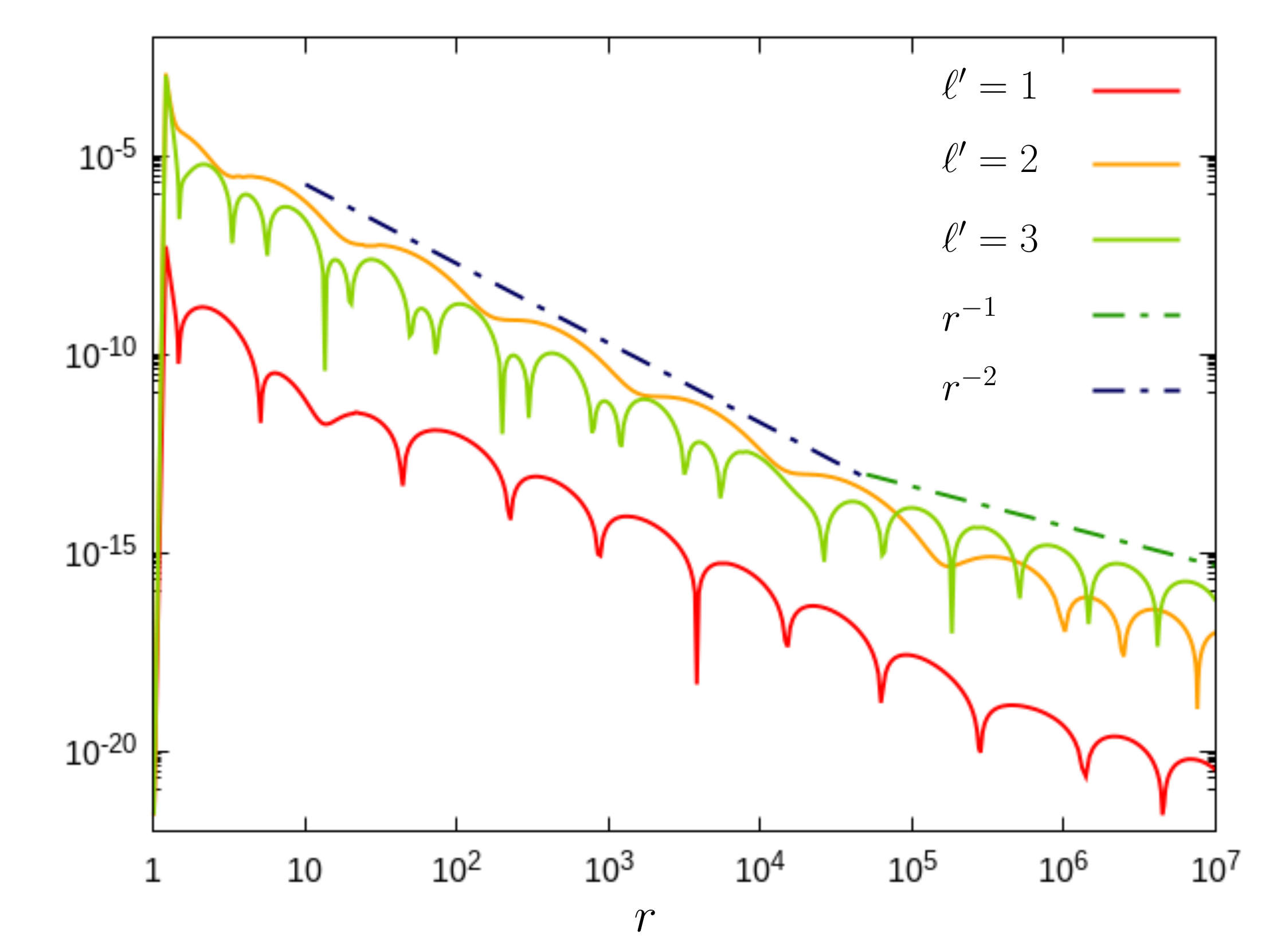}
                \vskip-0.4cm
				\caption{\scriptsize The decay rate of the mode $\dNNh_4{}^0$ of $\dNNh$ is $r^{-1}$.}
				\label{fig:decphN4}
			\end{subfigure}
			\hskip.5cm
			\begin{subfigure}{0.48\textwidth}\vskip0.1cm
				\includegraphics[width=\textwidth]{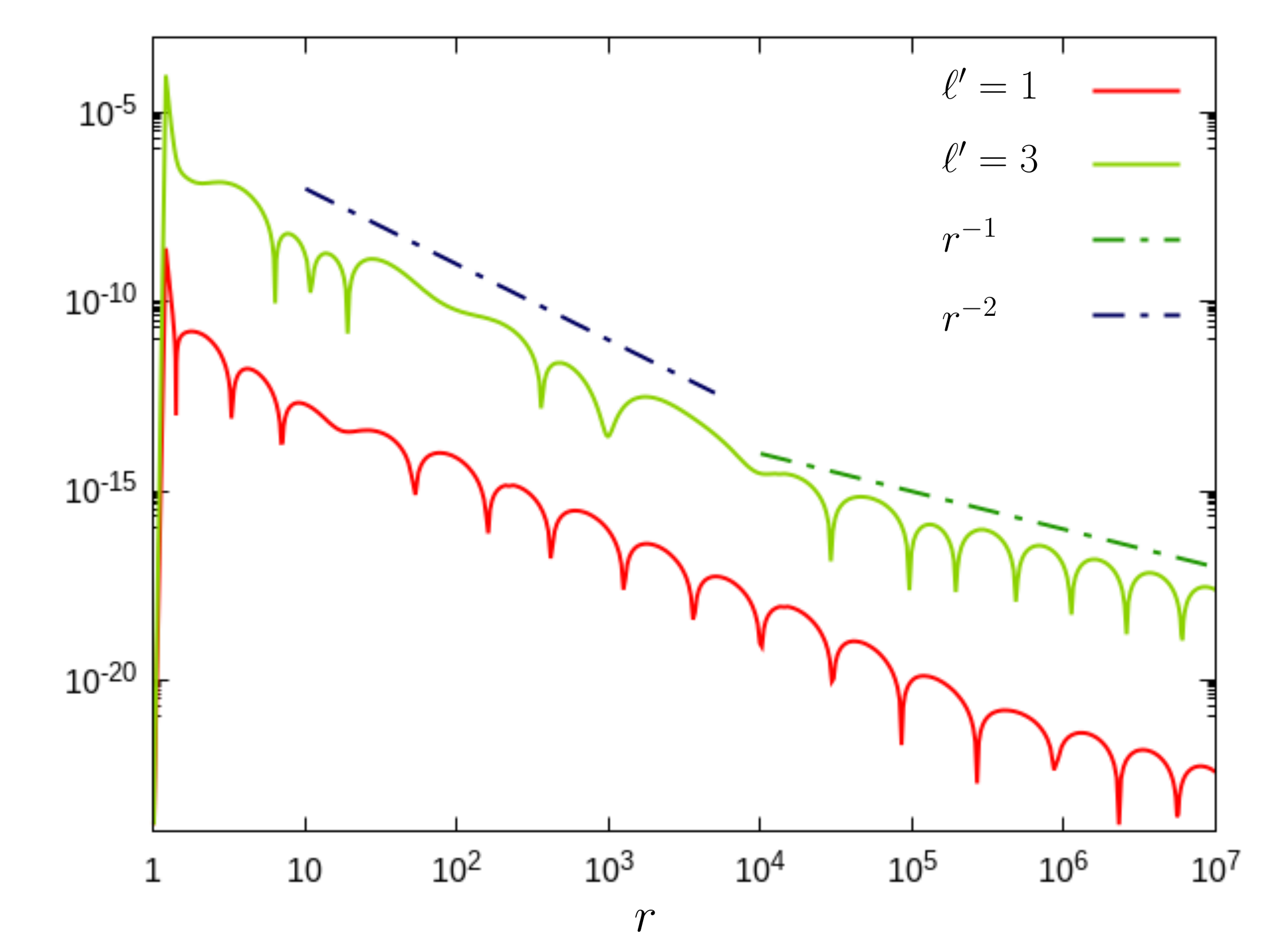}
                \vskip-0.4cm
				\caption{\scriptsize The decay rate of the mode $\dNNh_5{}^0$ of $\dNNh$ is $r^{-1}$. }
				\label{fig:decphN5}
			\end{subfigure}
		}
	\end{centering}
\vskip-0.1cm
	\caption{\footnotesize The non-linear perturbative form of the parabolic-hyperbolic system was integrated numerically by applying the initial perturbation  $\dKK|_{\mathscr{S}_{r_0}}=-10^{-1}\cdot {}_0{Y_{\ell'}}{}^0$ with $\ell'=1,2,3$. The decay rates of all the modes $\dNNh_\ell{}^0$, with $\ell=1,2,3,4,5,\, m=0$, of $\dNNh$ are found to be of order $r^{-1}$. The decay of the mode  $\dNNh_0{}^0$ also starts with  $r^{-1}$ but as explained in the text it needs to tend to a small constant value later.}
	\label{fig:decphN}
\end{figure}

In this subsection we shall inspect the $r$-dependence of the absolute value of various modes of the constrained variables evolved by the non-linear perturbative form of equations deduced from the parabolic-hyperbolic form of the constraints. The initial data specified at $\mathscr{S}_{r_0}$, with $r_0=1$, was chosen to be of the form $\dKK|_{\mathscr{S}_{r_0}}= - \alpha \cdot {}_0{Y_{\ell'}}{}^0$---with $\alpha=0.1$ and $\ell'=1,2,3$---, $\dkk|_{\mathscr{S}_{r_0}}=0$ and $\dNNh|_{\mathscr{S}_{r_0}}=0$.
Three different solutions corresponding to the choices $\ell'=1,2,3$ are plotted in case of each of the monitored modes.\footnote{Hereafter, the $\ell$-indices exciting modes will be indicated by using a prime, i.e.\,by $\ell'$, whereas those of the invoked modes will always come without prime.}
The pertinent set of decay rates for various $\ell$-modes of $\dKK$, $\dkk$ and $\dNNh$ are indicated on Figs.\,\ref{fig:decphK}, \ref{fig:decphk} and \ref{fig:decphN}.

\medskip

Start by a quick inspection of Figs.\,\ref{fig:decphK}, \ref{fig:decphk} and \ref{fig:decphN}.  It gets immediately transparent that the individual modes pick up for relatively small values of the $r$-coordinate their asymptotic decay rates. This, in particular, verifies that the separation of the excitation modes from the Schwarzschild background provide a much clearer picture than the mere inspection of graphs yielded by the set of full evolution equations. It also gets transparent that some of the small amplitude modes decay much slower than the higher amplitude background solution. However small their amplitude may be they inevitably become dominant for sufficiently large values of the $r$-coordinate. There is another very important message conveyed by the panels of Figs.\,\ref{fig:decphK}, \ref{fig:decphk} and \ref{fig:decphN}.
Notably, each of the $\ell=1,2,3,4,5$ modes of $\dKK{\,}_\ell{}^0$ of $\dKK$  decay $r^{-2}$ and, analogously, each of the $\ell=1,2,3,4,5$ modes of $\dkk{\,}_\ell{}^0$ of $\dkk$ and each of the $\ell=1,2,3,4,5$ modes of $\dNNh{\,}_\ell{}^0$
of $\dNNh$ decay as $r^{-1}$, thereby, in virtue of Table\,\ref{table:falloff}, they all would suit even to the requirements of strong asymptotic flatness. Nevertheless, the monopole part $\dKK{\,}_0{}^0$ of $\dKK$ decays only with the rate $r^{-1}$ which, in the case of the yielded strictly near Schwarzschild initial data configurations, is far too slow to allow asymptotic flatness even in the weak sense.

\medskip

Based on the behavior of spherically symmetric near Schwarzschild configurations, studied in subsection \ref{sec:strict_spher_ph} the solution to \eqref{eq: sper-K-par-hyp}, with ${C_\mathbf{K}}\not=0$, is substituted into \eqref{eqr-Nh} the solution for $\widehat{N}$ will not be as simple as the one that was given by \eqref{eq: Nhat-radial-PH}, nevertheless, it can be seen to possess the asymptotic form
\begin{equation}\label{eq: decay-Nh-ph-CK}
			\widehat{N}_{C_\mathbf{K}}=\frac{2}{\sqrt{C_\mathbf{K}^2+4}}+\frac{8\, C_\mathbf{K}\,
				M-4\, {C_{\widehat{N}}}}{\left(C_\mathbf{K}^2+4\right){}^{3/2}\, r}+\mathscr{O}\left(r^{-2}\right)\,.
\end{equation}
Note that the $r^{-1}$ decay of $\dKK_0{}^0$ on panel (a) of Fig.\,\ref{fig:decphK} indicates the non-vanishing of the pertinent ${C_\mathbf{K}}$.
Note also that this $\NNh$ does not tend to one at infinity. As it is also clearly indicated by the graphs on panel (a) of Fig.\,\ref{fig:decphN}, and also by \eqref{eq: decay-Nh-ph-CK}, the monopole part $\dNNh_0{}^0$ of $\dNNh$, around ${r\approx 10^6}$, starts to tend to the constant value $1/\sqrt{1+C_{\KK}/4}-1$, where, in virtue of \eqref{eq: sper-K-par-hyp}, the approximate value of $C_{\KK}$ can be evaluated by extrapolating the value of $\dKK{\,}_0{}^0$ to $r_0=1$. 

\subsection{Non-linear perturbation with algebraic-hyperbolic system}\label{subsec: alg-hyp-non-lin}

\begin{figure}[ht!]
	\vskip-0.1cm
	\begin{centering}
		{\tiny
			\begin{subfigure}{0.48\textwidth}
				\includegraphics[width=\textwidth]{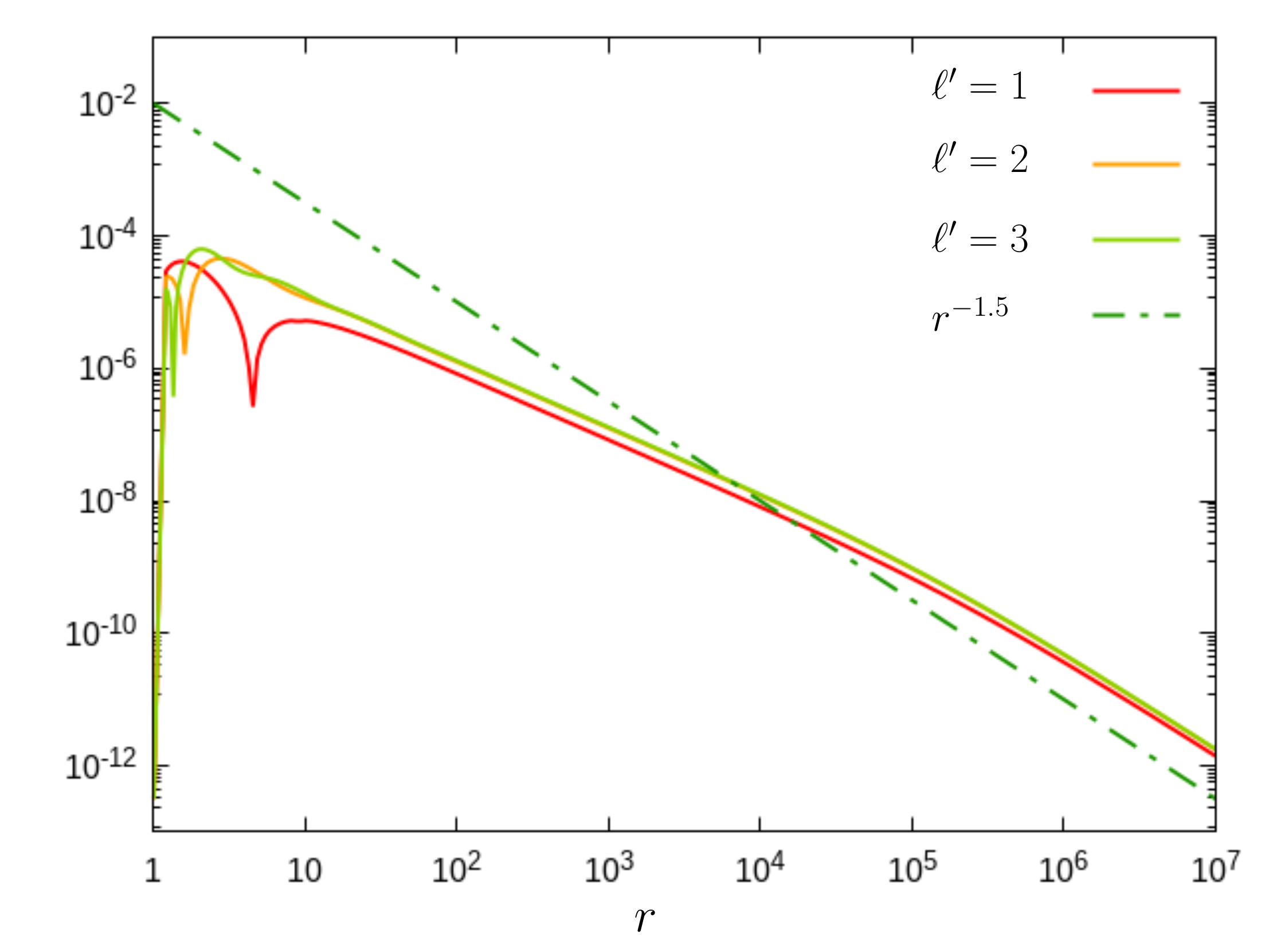}
                \vskip-0.4cm
				\caption{\scriptsize The decay rate of the mode  $\dKK_0{}^0$ of $\dKK$ is $r^{-1.5}$.}
				\label{fig:decahK0}
			\end{subfigure}
			\begin{subfigure}{0.48\textwidth}
				\includegraphics[width=\textwidth]{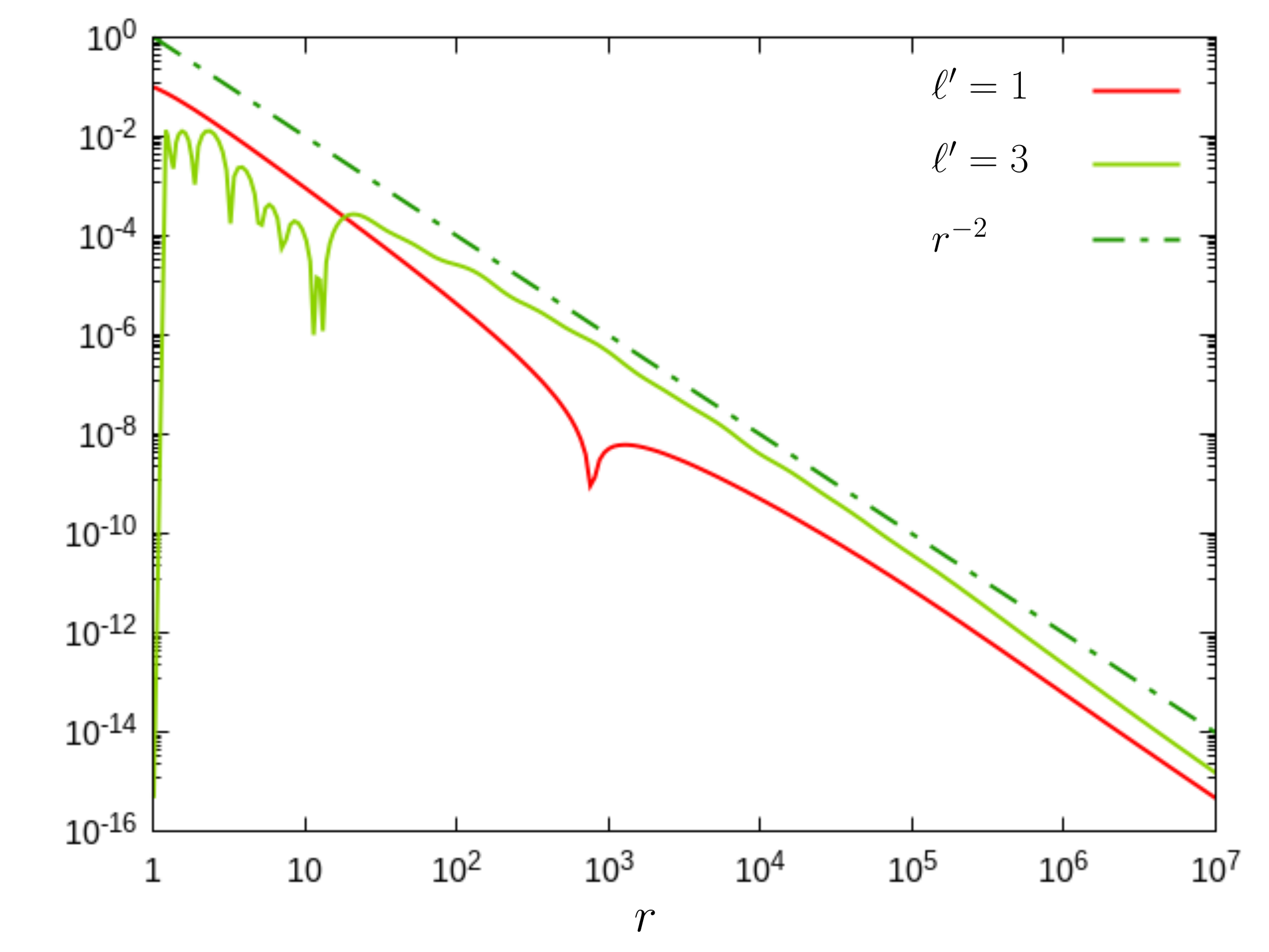}
                \vskip-0.4cm
				\caption{\scriptsize The decay rate of the mode  $\dKK_1{}^0$ of $\dKK$ is $r^{-2}$. }
				\label{fig:decahK1}
			\end{subfigure}
			\begin{subfigure}{0.48\textwidth}\vskip0.2cm
				\includegraphics[width=\textwidth]{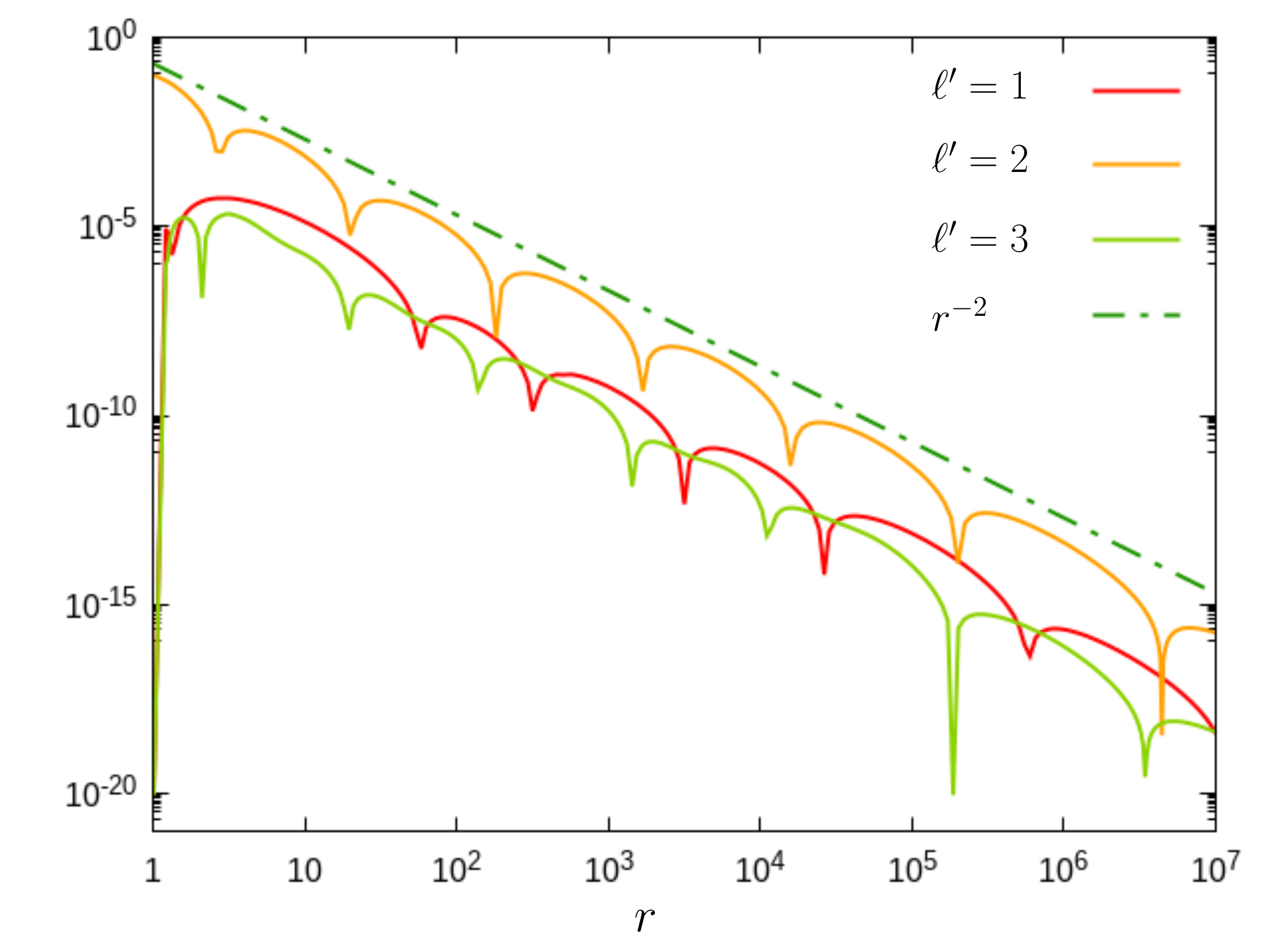}
                \vskip-0.4cm
				\caption{\scriptsize The decay rate of the mode  $\dKK_2{}^0$ of $\dKK$ is $r^{-2}$.}
				\label{fig:decahK2}
			\end{subfigure}
			\hskip.5cm
			\begin{subfigure}{0.48\textwidth}\vskip0.2cm
				\includegraphics[width=\textwidth]{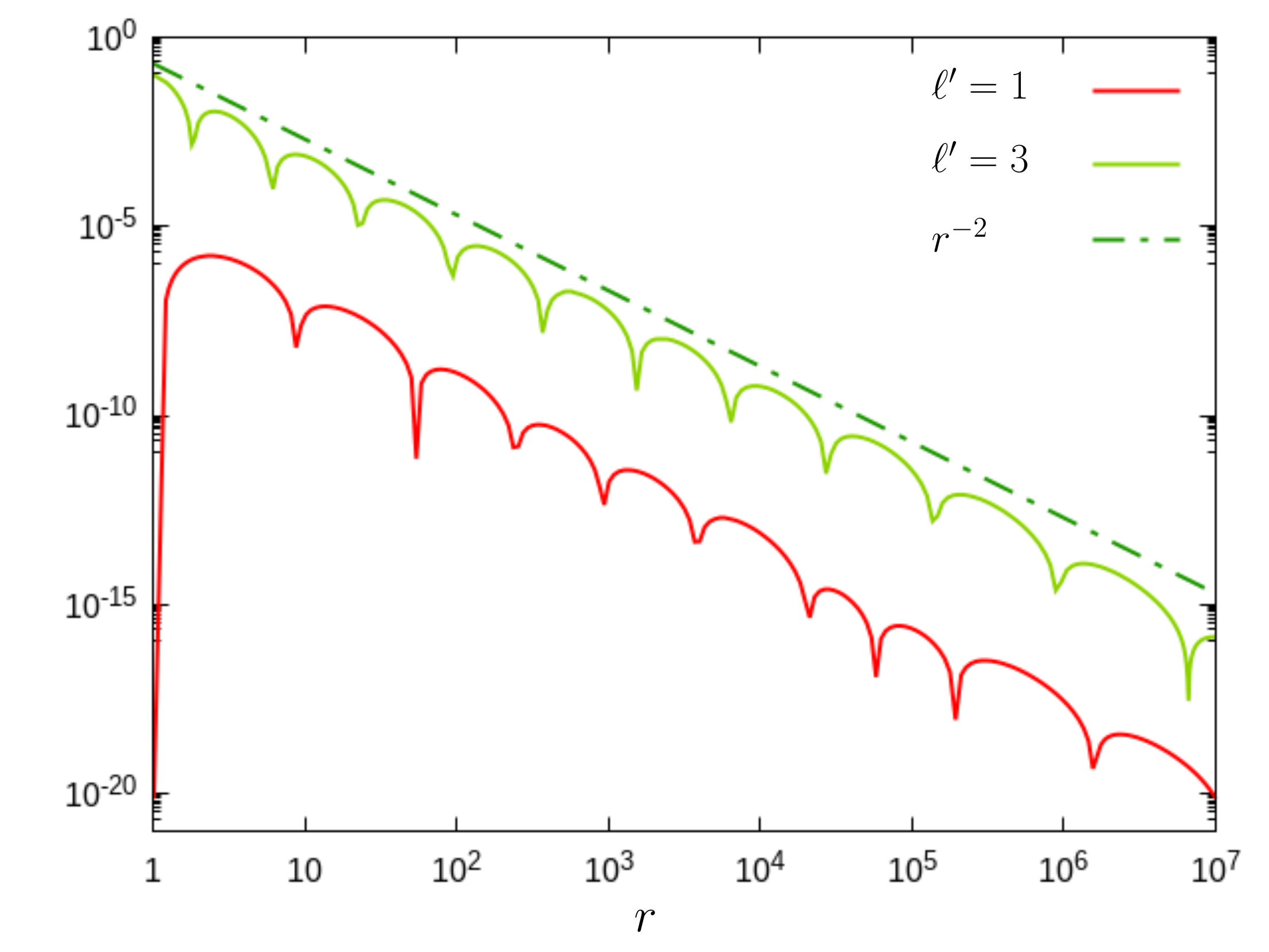}
                \vskip-0.4cm
				\caption{\scriptsize The decay rate of the mode  $\dKK_3{}^0$ of $\dKK$ is $r^{-2}$. }
				\label{fig:decahK3}
			\end{subfigure}
			\begin{subfigure}{0.48\textwidth}\vskip0.2cm
				\includegraphics[width=\textwidth]{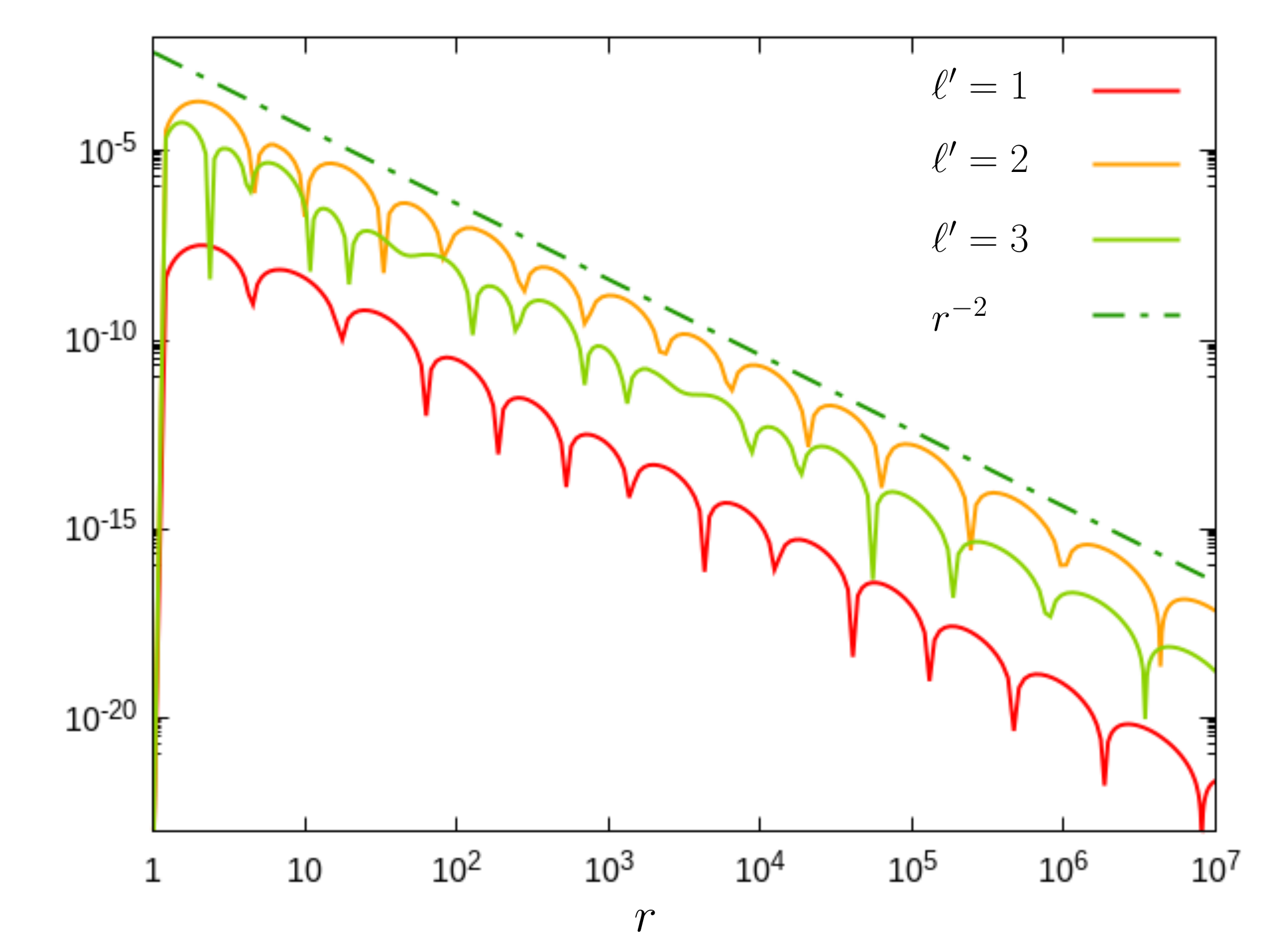}
                \vskip-0.4cm
				\caption{\scriptsize The decay rate of the mode  $\dKK_4{}^0$ of $\dKK$ is $r^{-2}$.}
				\label{fig:decahK4}
			\end{subfigure}
			\hskip.5cm
			\begin{subfigure}{0.48\textwidth}\vskip0.2cm
				\includegraphics[width=\textwidth]{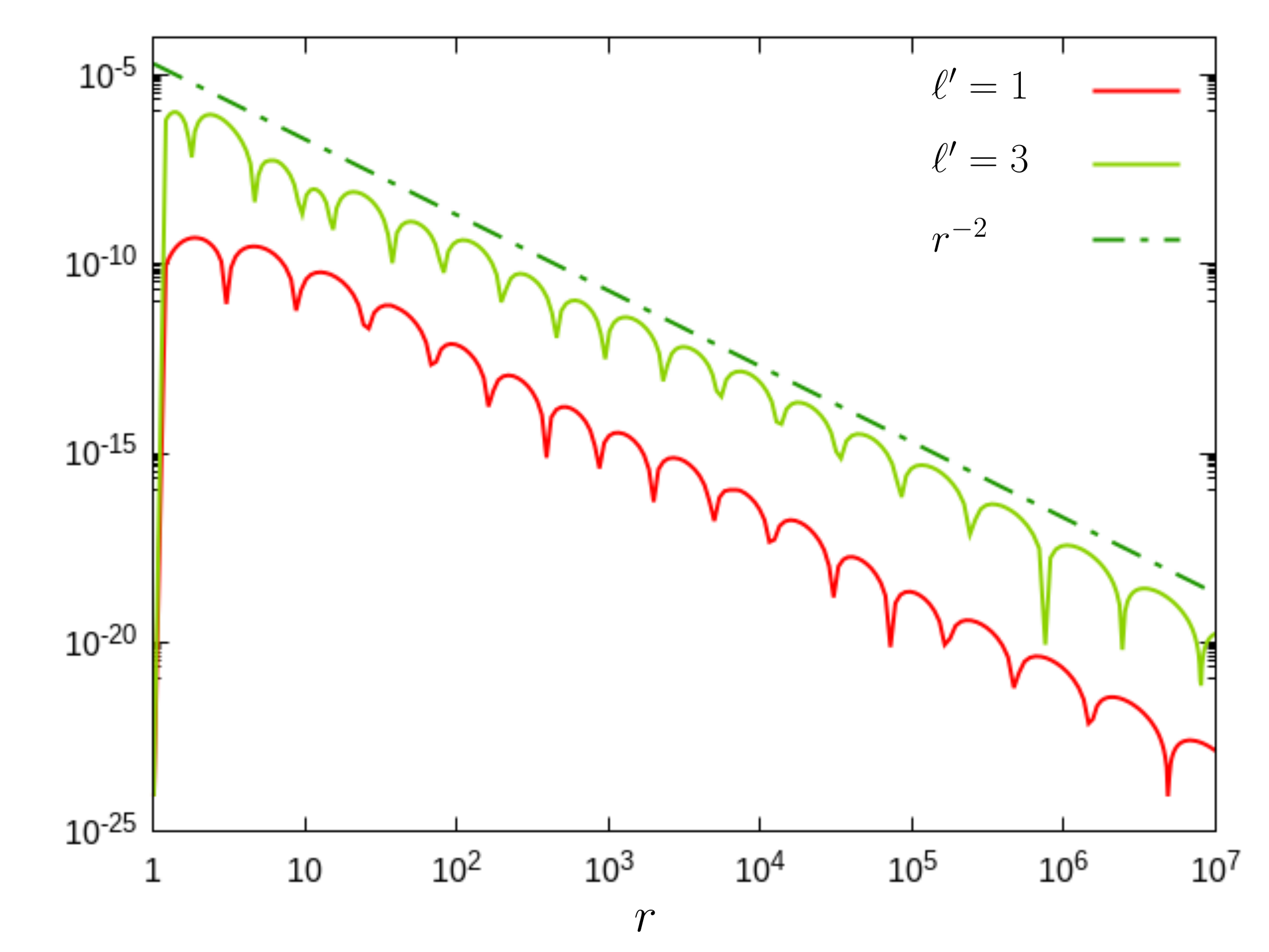}
                \vskip-0.4cm
				\caption{\scriptsize The decay rate of the mode  $\dKK_5{}^0$ of $\dKK$ is $r^{-2}$.}
				\label{fig:decahK5}
			\end{subfigure}
		}
	\end{centering}
\vskip0.1cm
	\caption{\footnotesize The non-linear perturbative form of the algebraic-hyperbolic system was integrated numerically by applying the initial data $\dKK|_{\mathscr{S}_{r_0}}=-10^{-1}\cdot {}_0{Y_{\ell'}}{}^0$ with $\ell'=1,2,3$. The decay rates of all the modes $\dKK_\ell{}^0$, with $\ell=1,2,3,4,5,\, m=0$, of $\dKK$ are found to be of order $r^{-2}$, whereas the mode $\dKK_0{}^0$ decays slightly slower than $r^{-3/2}$.}
	\label{fig:decahK}
\end{figure}

\begin{figure}[ht!]
	\begin{centering}
		{\tiny
			\begin{subfigure}{0.48\textwidth}
				\includegraphics[width=\textwidth]{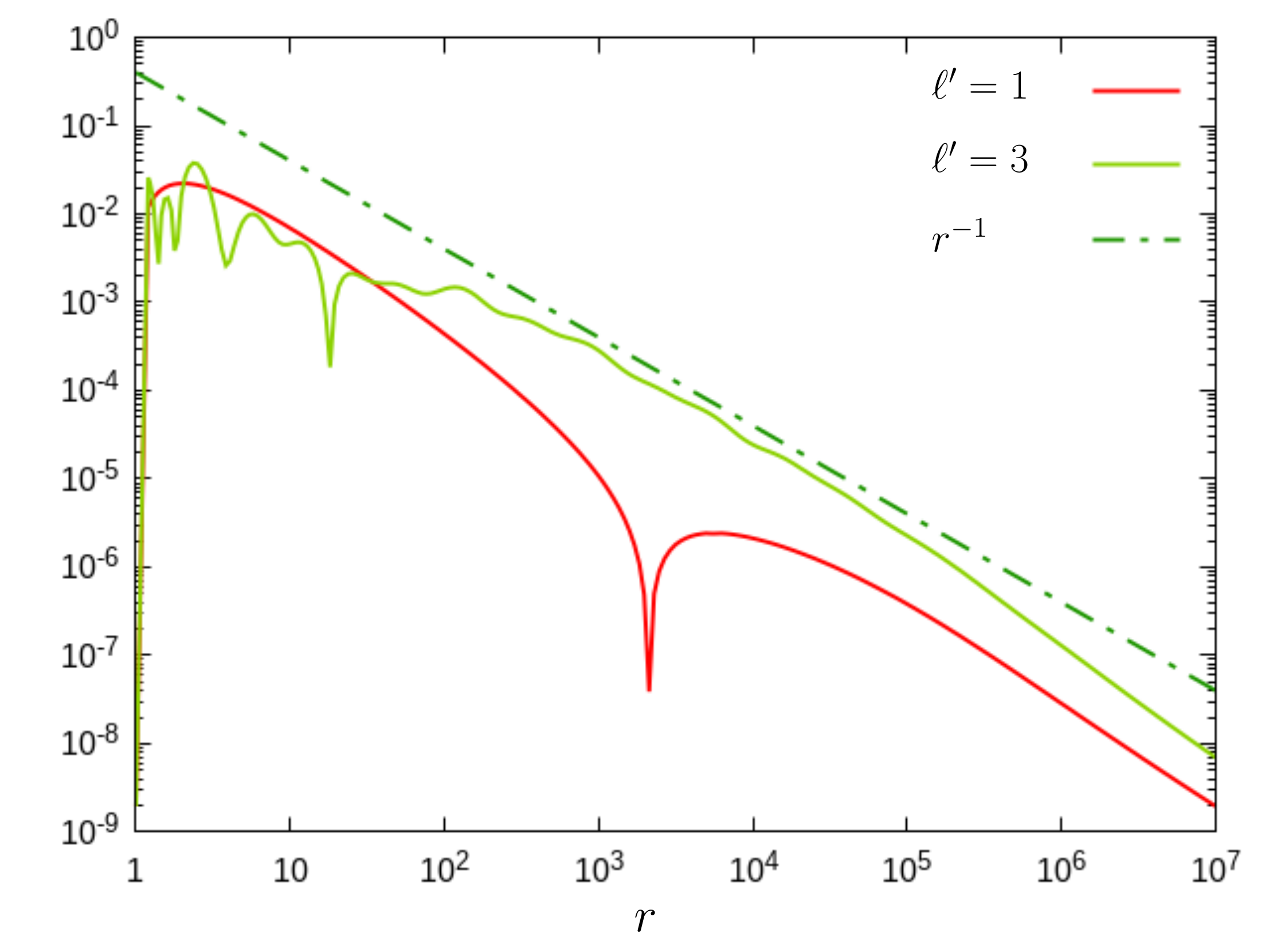}
				\vskip-0.4cm
				\caption{\scriptsize The decay rate of the mode  $\dkk_1{}^0$ of $\dkk$ is $r^{-1}$.}
				\label{fig:decahk1}
			\end{subfigure}
			\begin{subfigure}{0.48\textwidth}
				\includegraphics[width=\textwidth]{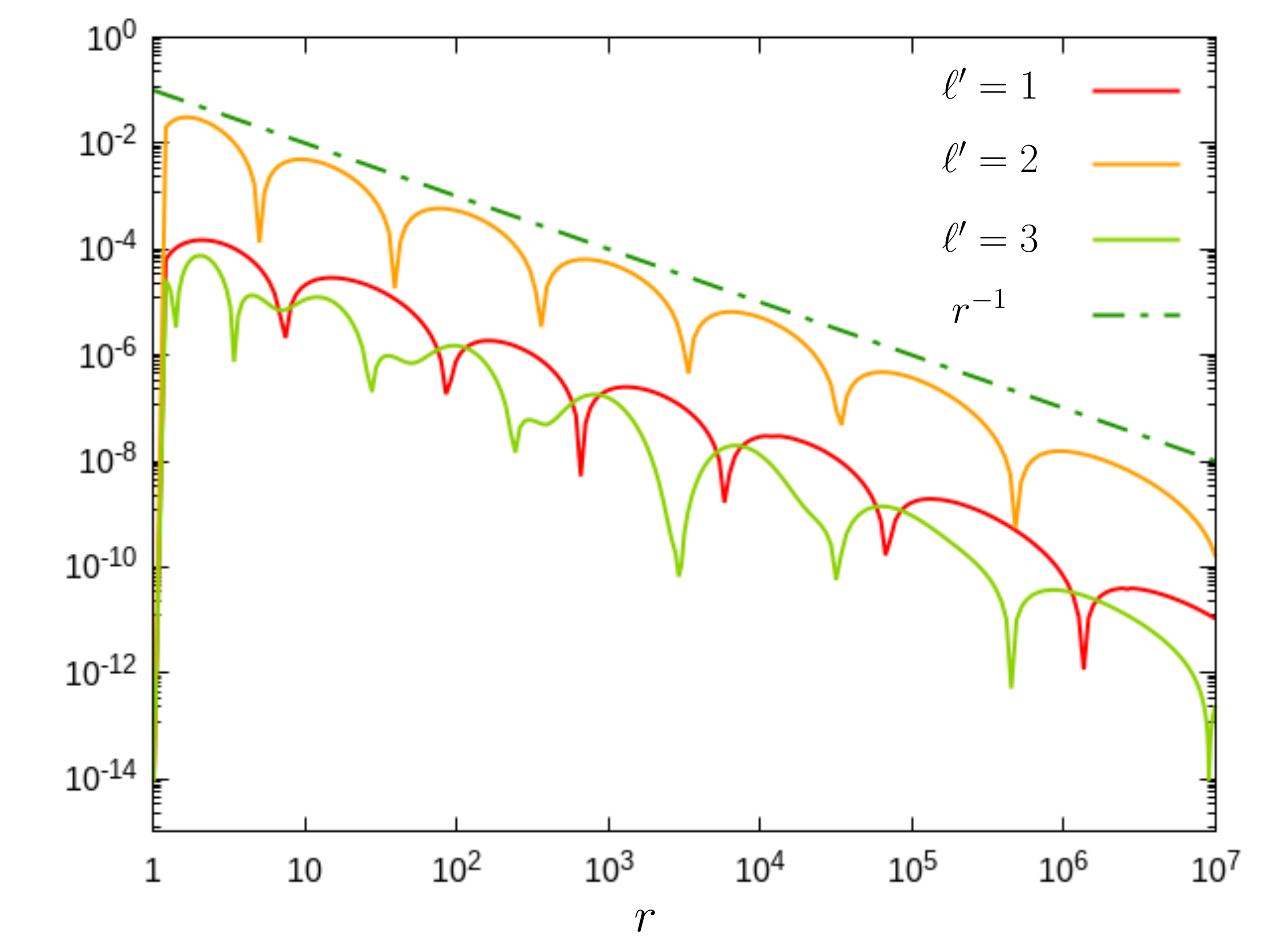}
				\vskip-0.4cm
				\caption{\scriptsize The decay rate of the mode  $\dkk_2{}^0$ of $\dkk$ is $r^{-1}$.}
				\label{fig:decahk2}
			\end{subfigure}
			\hskip.5cm
			\begin{subfigure}{0.48\textwidth}
				\includegraphics[width=\textwidth]{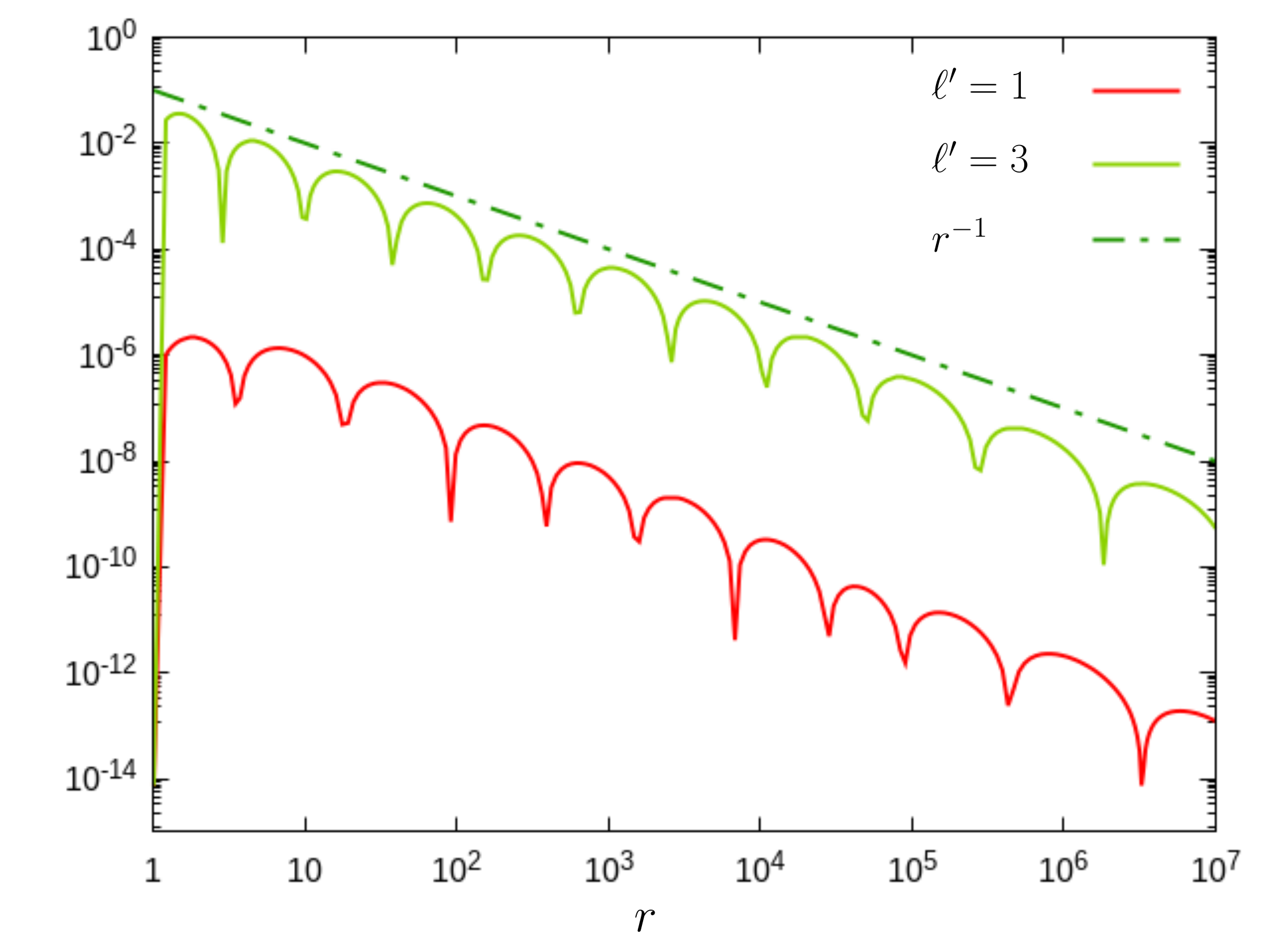}
				\vskip-0.4cm
				\caption{\scriptsize The decay rate of the mode  $\dkk_3{}^0$ of $\dkk$ is $r^{-1}$.}
				\label{fig:decahk3}
			\end{subfigure}
			\begin{subfigure}{0.48\textwidth}
				\includegraphics[width=\textwidth]{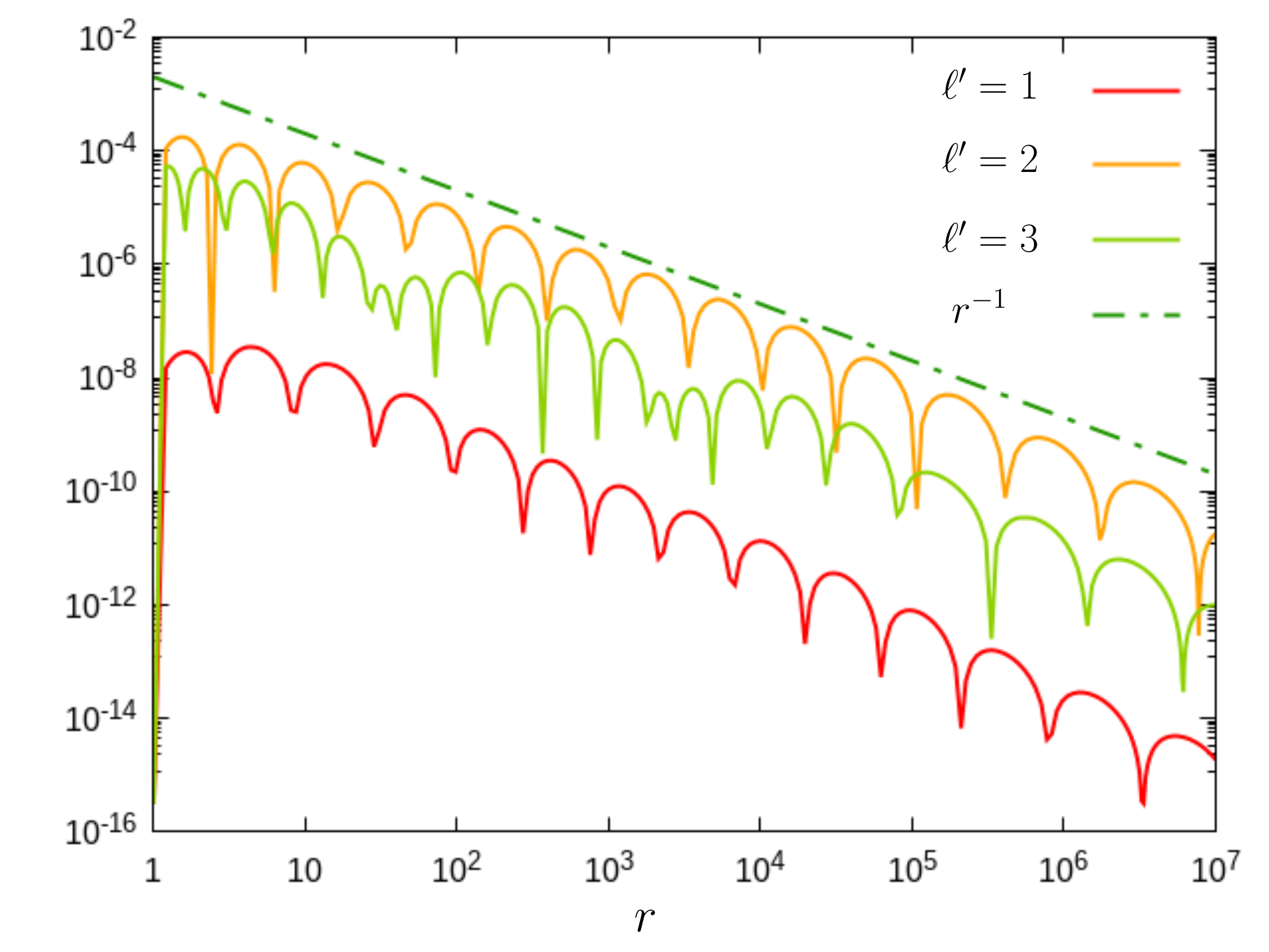}
				\vskip-0.4cm
				\caption{\scriptsize The decay rate of the mode  $\dkk_4{}^0$ of $\dkk$ is $r^{-1}$.}
				\label{fig:decahk4}
			\end{subfigure}
			\hskip.5cm
			\begin{subfigure}{0.48\textwidth}
				\includegraphics[width=\textwidth]{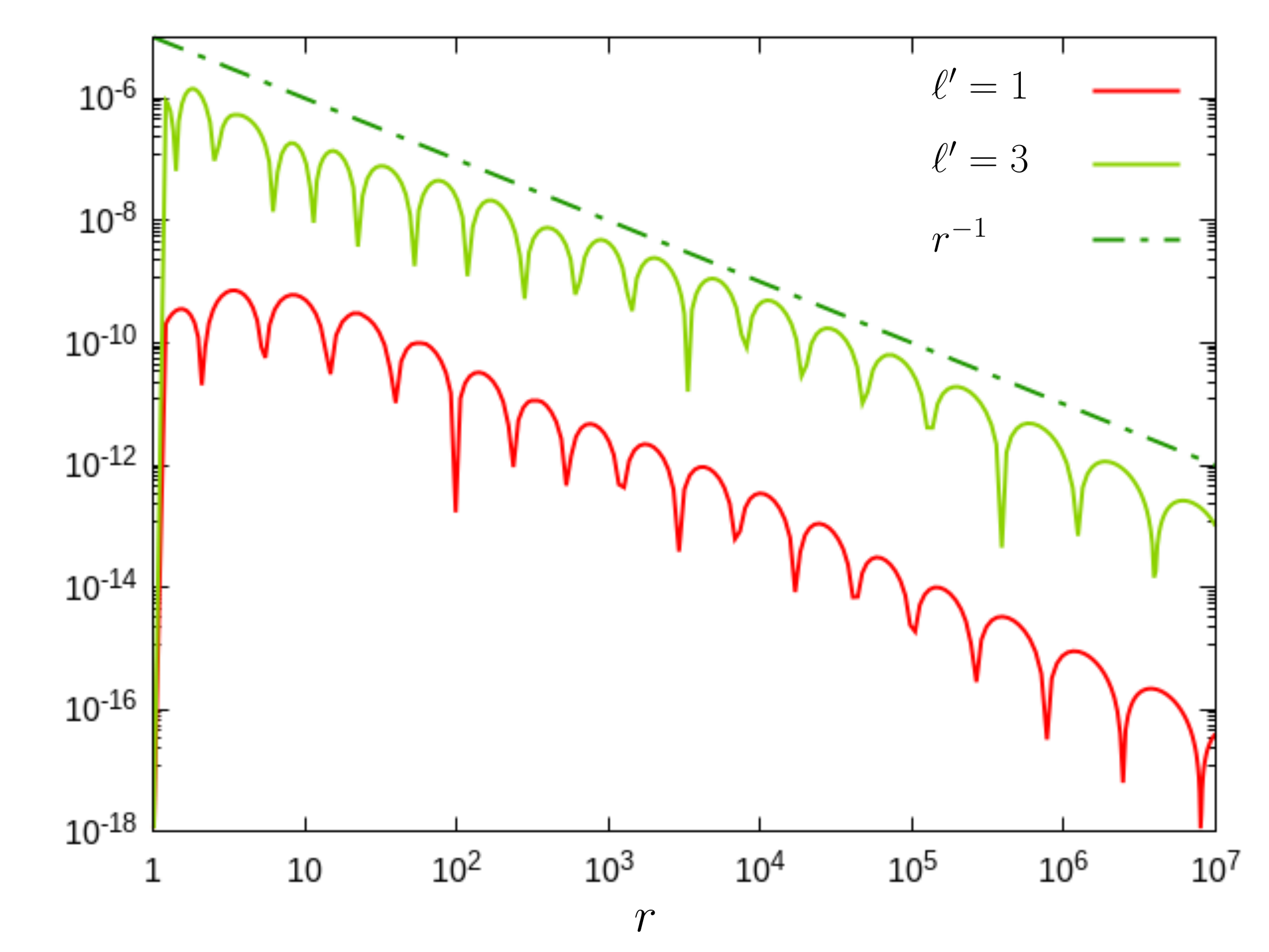}
				\vskip-0.4cm
				\caption{\scriptsize The decay rate of the mode  $\dkk_5{}^0$ of $\dkk$ is $r^{-1}$.}
				\label{fig:decahk5}
			\end{subfigure}
		}
	\end{centering}
\vskip0.1cm
	\caption{\footnotesize The non-linear perturbative form of the algebraic-hyperbolic system was integrated numerically by applying the initial data $\dKK|_{\mathscr{S}_{r_0}}=-10^{-1}\cdot {}_0{Y_{\ell'}}{}^0$ with $\ell'=1,2,3$. All the modes $\dkk_\ell{}^0$, with $\ell=1,2,3,4,5,\, m=0$, of $\dkk$ decay at least as fast as $r^{-1}$.}
	\label{fig:decahk}
\end{figure}

The $r$-dependence of the absolute value of various modes of the constrained variables are plotted on figures \ref{fig:decahK} and \ref{fig:decahk}. In the present case these modes are evolved by applying the non-linear perturbative form of the algebraic-hyperbolic form of the constraints. The initial data specified at $\mathscr{S}_{r_0}$, with $r_0=1$, was chosen, as in the previous subsection, to take the form $\dKK|_{\mathscr{S}_{r_0}}= - \alpha \cdot {}_0{Y_{\ell'}}{}^0$, with $\alpha=0.1$ and with $\ell'=1,2,3$, and $\dkk|_{\mathscr{S}_{r_0}}=0$. 
The decay rates relevant for the $\ell$-modes, with $\ell=0,1,2,3,4,5$, of $\dKK$ and,with $\ell=1,2,3,4,5$, of $\dkk$ are indicated on the panels of  Figs.\,\ref{fig:decahK} and \ref{fig:decahk}, respectively, such that on each panel there are thee graphs corresponding to the choices $\ell'=1,2,3$ made in specifying the initial data at $\mathscr{S}_{r_0}$.

\medskip

As in the parabolic-hyperbolic case, Figs.\,\ref{fig:decahK} and \ref{fig:decahk} make it transparent that the individual modes get at almost immediately to their asymptotic decay rates. This clearly verifies that the separation of the excitation modes from the Schwarzschild background provide us a much more transparent picture. Here there is again a very important additional message conveyed by the individual panels of Figs.\,\ref{fig:decahK} and \ref{fig:decahk}. Visibly, all the $\ell=1,2,3,4,5$ modes of $\dKK{\,}_\ell{}^0$ of $\dKK$ decay at the rate $r^{-2}$ and all the $\ell=1,2,3,4,5$ modes of $\dkk{\,}_\ell{}^0$ of $\dkk$ decay at least as fast as $r^{-1}$, by which, in virtue of Table\,\ref{table:falloff}, they all suit immediately to the strong asymptotic flatness requirement. Nevertheless, the decay rate of the monopole part $\dKK{\,}_0{}^0$ of $\dKK$ is slightly less than $r^{-3/2}$ which, in the case of the yielded strictly near Schwarzschild initial data configurations, is too slow to allow asymptotic flatness even in the weak sense.

\subsection{Revisiting the notion of near-Schwarzschild configurations}
\label{subsec: revisit}

One of the most striking consequence of the results presented in the previous subsections is that the monopole part $\dKK{\,}_0{}^0$ of $\dKK$ appears to decay far too slow to fit to asymptotically flat initial data configurations. One might conclude that this is simply the consequence of the superiority of the elliptic method in regard that the evolutionary method is not capable to impose asymptotic fall off conditions on the to be solutions. However convincing such a claim may sound we should not forget that there is a one-to-one correspondence between the space of solutions produced by the elliptic and by either of the evolutionary methods. Accordingly, it seems to be more adequate to question the appropriateness of the fixing of the freely specifiable fields in \cite{Beyer:2017njj,Beyer:2019kty}, and also in the former subsections of the present paper---in investigating strictly near-Schwarzschild configurations. By inspecting the determination of strictly near Schwarzschild initial data configurations given in subsection \ref{subsec: near-Schwarzscild} it gets clear immediately that enormous freedom remained to be explored.

\medskip

In doing so recall first that the evolutionary method provides a type of flexibility which is not available in applying the elliptic method in solving the constraint equations. Namely, in advance of solving the elliptic equations one has to fix all the freely specifiable fields globally, i.e.\,promptly on the entire of the three-dimensional initial data surface, $\Sigma$, whereas the use of either of the evolutionary methods always allows us to choose the freely specifiable fields right in the interim of the ``time integration'' process, i.e.\,they can be updated, leaves by leaves, as the desired functional form of the constrained fields require.

\medskip

Exactly this freedom was used in subsections \ref{subsub: Florian-ansatz-radial} and \ref{subsub: ansatz-alg-hyp-radial}, and it will also be used in the following two subsections to demonstrate that asymptotically flat initial data configurations can be produced by either of the evolutionary forms of the constraints.
Note, however, that here--instead of trying to be extremely ambitious---we simply aim to demonstrate that the applied particular relaxations do, indeed, provide us the desired asymptotically flat configurations.

\subsubsection{Parabolic-hyperbolic system}\label{subsub: Florian-ansatz}

As an immediate application of the above outlined ideas this subsection is to demonstrate that---without modifying the original form of the parabolic-hyperbolic system, given in \cite{Racz2016},---simply by applying the ansatz, proposed in \cite{Beyer:2020kty},
\begin{equation}\label{eq: Florian-ansatz}
	\boldsymbol{\kappa}=\mathcal{R} \cdot \mathbf{K}\,,
\end{equation}
\begin{equation}\label{eq: Florian-ansatz-specR}
	\mathcal{R}= -\frac12 + \frac{M}{4\,(M+r)}\,,
\end{equation}
and simply updating leaves by leaves the freely specifiable variable $\boldsymbol{\kappa}$, in equations \eqref{eq:phN}, \eqref{eq:phK} and \eqref{eq:phk}, we do really get a strongly asymptotically flat initial data configuration.

\medskip

Before proceeding and presenting the corresponding numerical results note first that the non-linear perturbative form of the parabolic-hyperbolic system is integrated which has been verified to be superior, in precision, to the full evolutionary system. This allows us to monitor the $r$-dependence of the
individual modes as opposed to the plots in \cite{Beyer:2020kty} show the $r$-dependence of the norms of the solutions determined by using the full form of the  parabolic-hyperbolic system. Note also that the interval of ``time integration'' is considerably longer than the one used in \cite{Beyer:2020kty}, and that the initial data we use is significantly stronger than  the one applied in  \cite{Beyer:2020kty}.

\medskip

\begin{figure}[ht!]
	\vskip-0.1cm
	\begin{centering}
		{\tiny
			\begin{subfigure}{0.48\textwidth}
				\includegraphics[width=\textwidth]{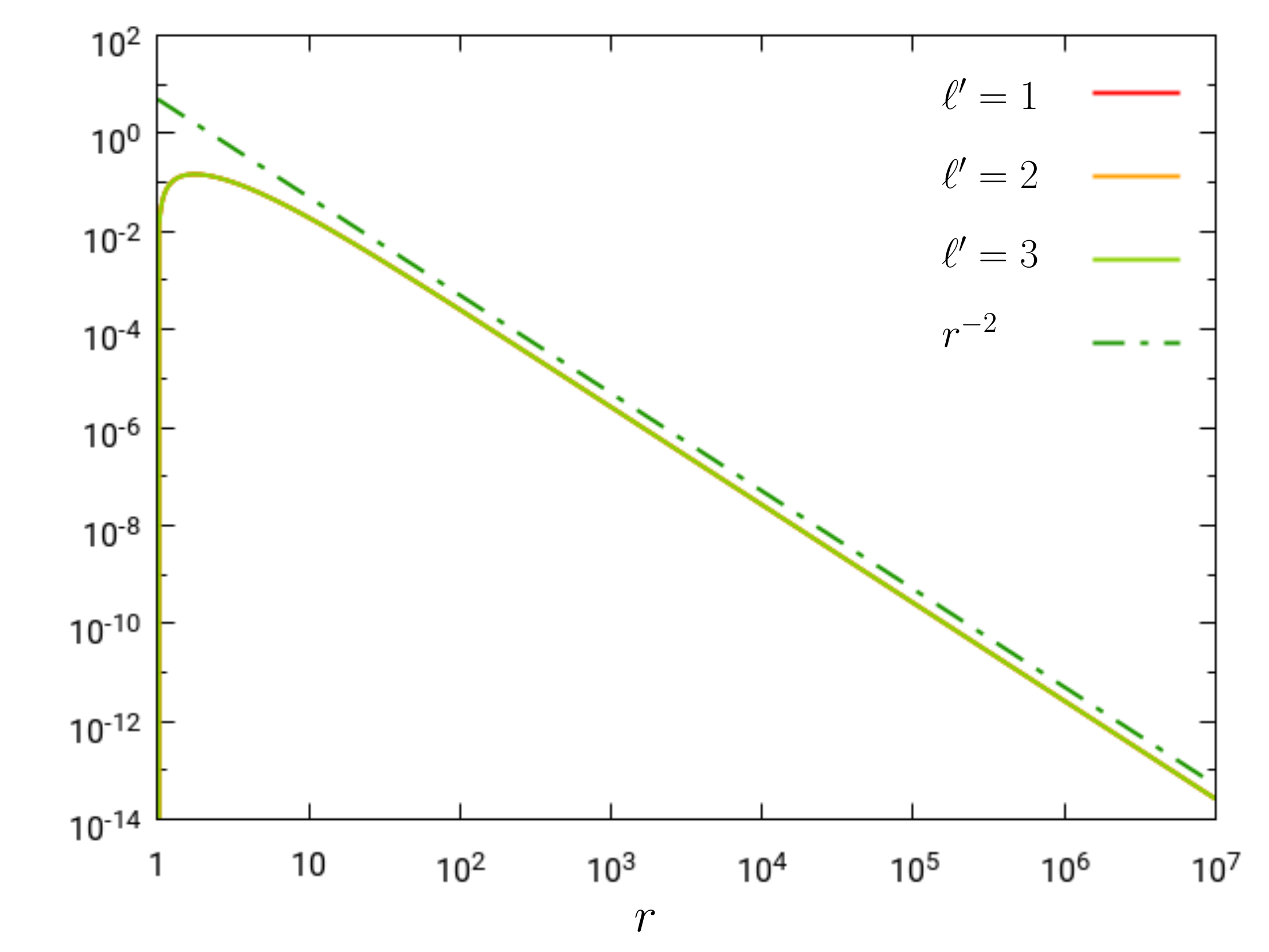}
				\vskip-0.4cm
				\caption{\scriptsize The decay rate of the mode $\dKK{\,}_0{}^0$ of $\dKK$
					is $r^{-2}$. }
				\label{fig:Florian-decphK0}
			\end{subfigure}
			\begin{subfigure}{0.48\textwidth}\vskip0.2cm
				\includegraphics[width=\textwidth]{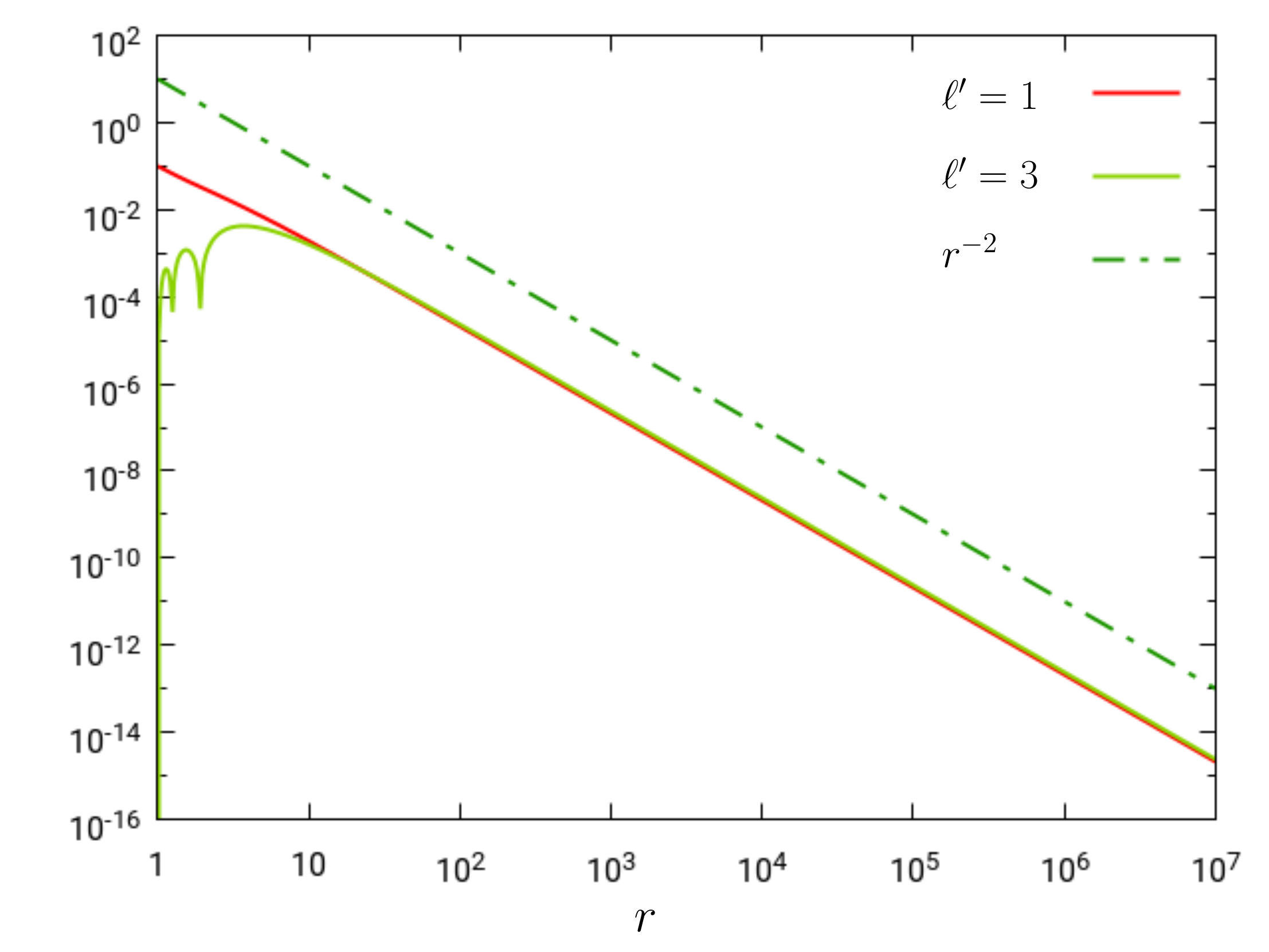}
				\vskip-0.4cm
				\caption{\scriptsize The decay rate of the mode $\dKK{\,}_1{}^0$ of $\dKK$
					is $r^{-2}$.}
				\label{fig:Florian-decphK1}
			\end{subfigure}
			\begin{subfigure}{0.48\textwidth}\vskip0.2cm
				\includegraphics[width=\textwidth]{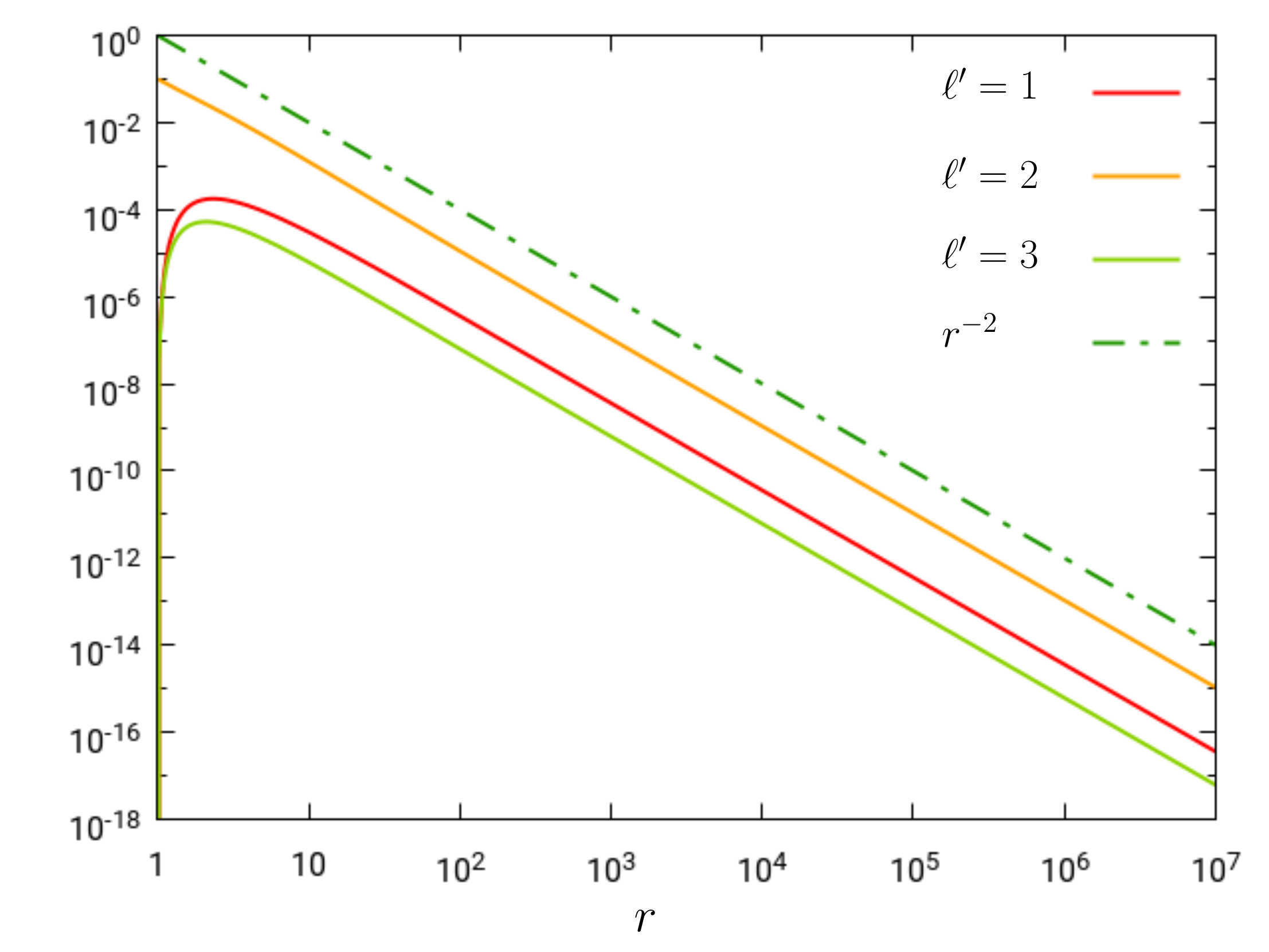}
				\vskip-0.4cm
				\caption{\scriptsize The decay rate of the mode $\dKK{\,}_2{}^0$ of $\dKK$
					is $r^{-2}$.}
				\label{fig:Florian-decphK2}
			\end{subfigure}
			\hskip.5cm
			\begin{subfigure}{0.48\textwidth}\vskip0.2cm
				\includegraphics[width=\textwidth]{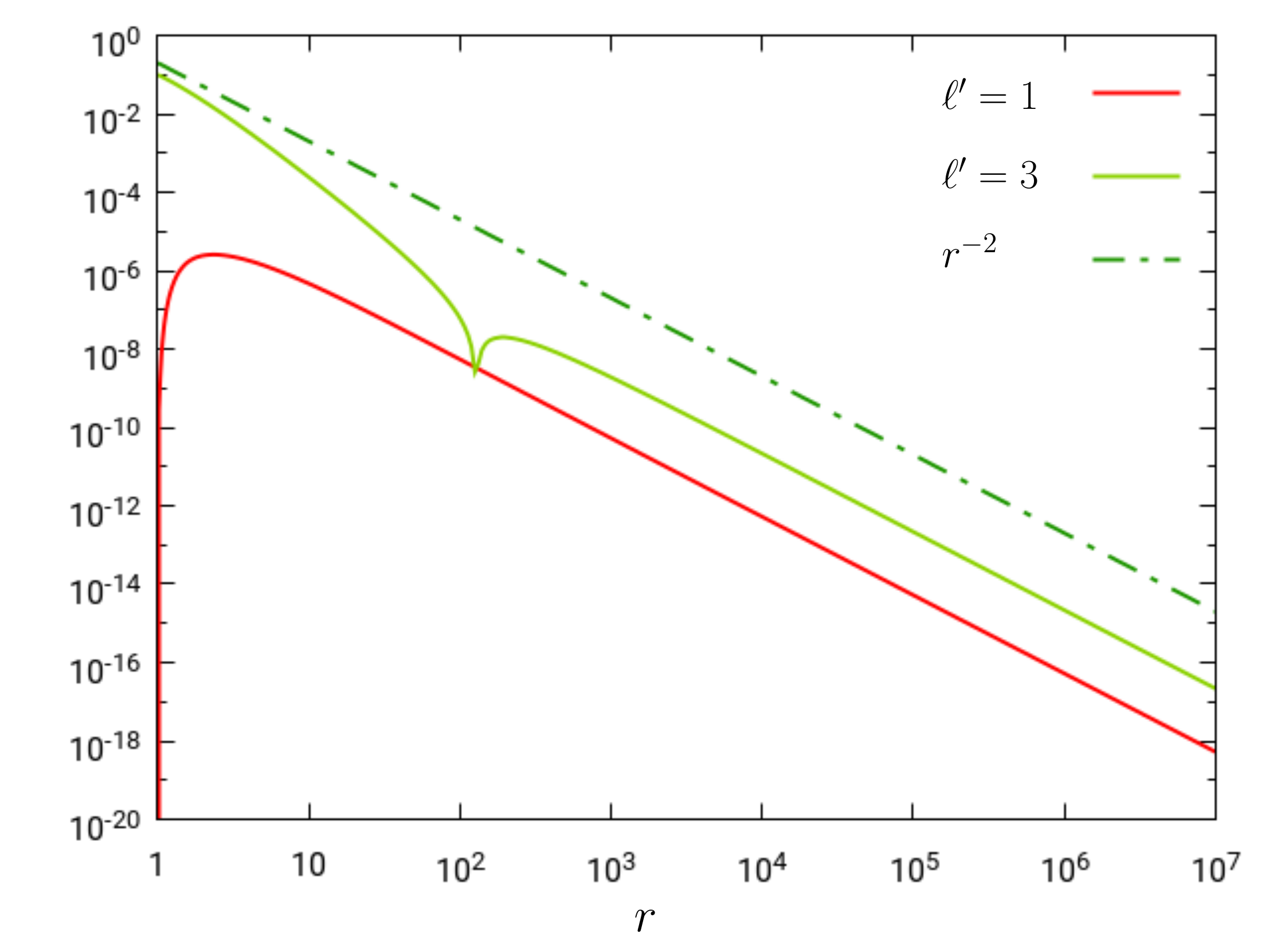}
				\vskip-0.4cm
				\caption{\scriptsize The decay rate of the mode $\dKK{\,}_3{}^0$ of $\dKK$ is $r^{-2}$.}
				\label{fig:Florian-decphK3}
			\end{subfigure}
			\begin{subfigure}{0.48\textwidth}\vskip0.2cm
				\includegraphics[width=\textwidth]{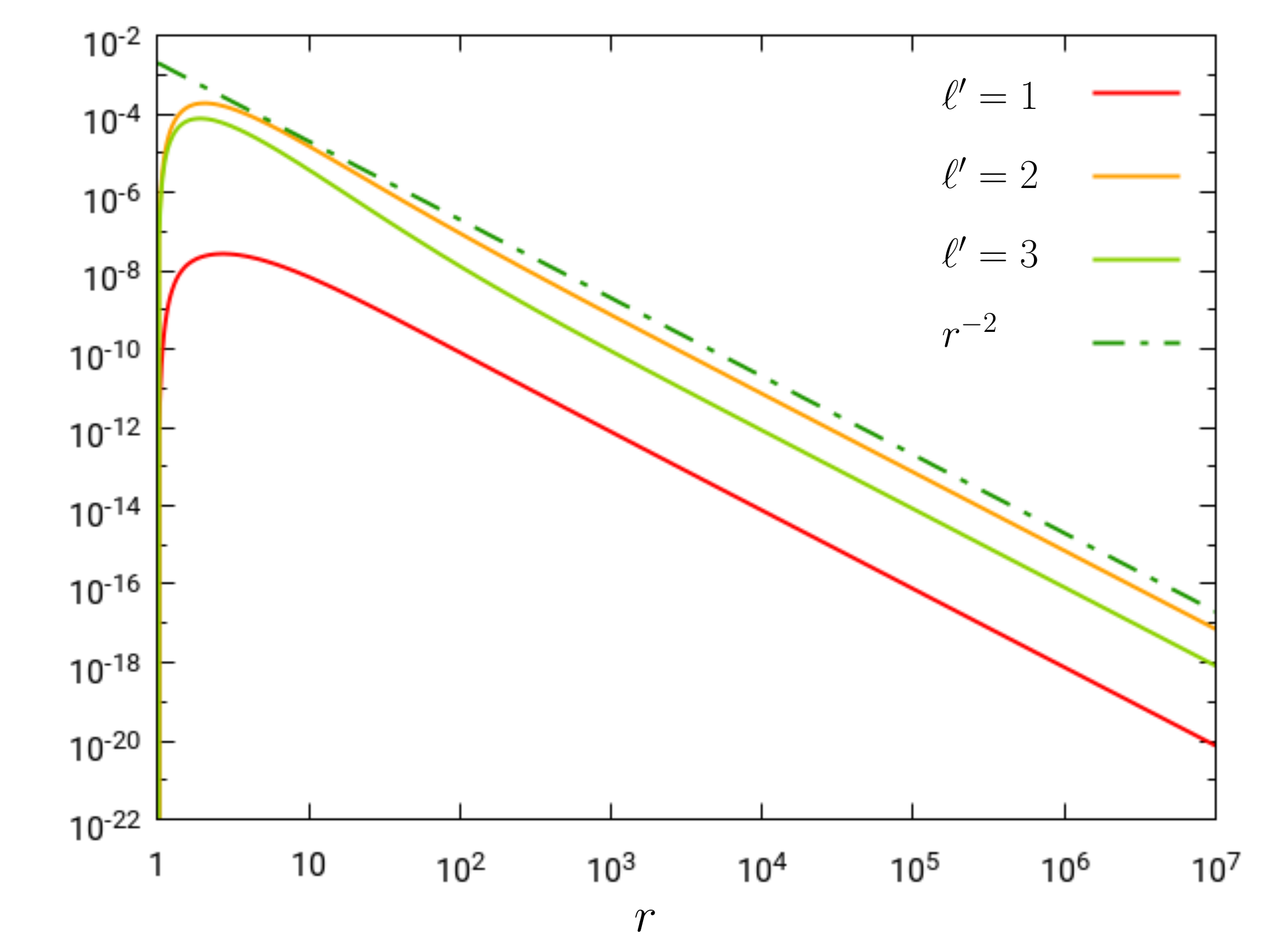}
				\vskip-0.4cm
				\caption{\scriptsize The decay rate of the mode $\dKK{\,}_4{}^0$ of $\dKK$ is $r^{-2}$.}
				\label{fig:Florian-decphK4}
			\end{subfigure}
			\hskip.5cm
			\begin{subfigure}{0.48\textwidth}\vskip0.2cm
				\includegraphics[width=\textwidth]{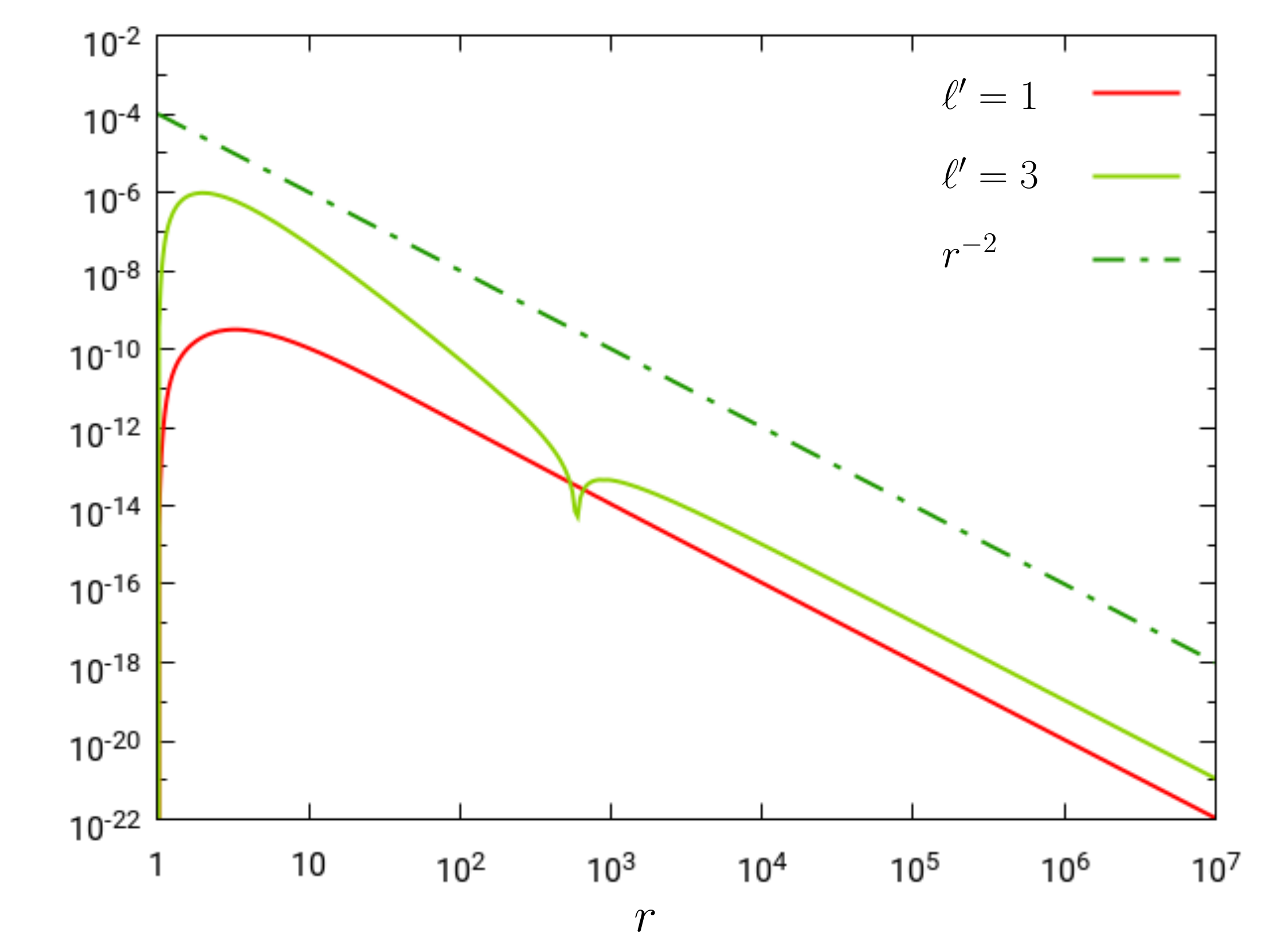}
				\vskip-0.4cm
				\caption{\scriptsize The decay rate of the mode $\dKK{\,}_5{}^0$ of $\dKK$ is $r^{-2}$.}
				\label{fig:Florian-decphK5}
			\end{subfigure}
		}
	\end{centering}
	\vskip-0.1cm
	\caption{\footnotesize  The non-linear perturbative form of the parabolic-hyperbolic system was integrated numerically such that $\boldsymbol{\kappa}$ is always updated in accordance with the ansatz \eqref{eq: Florian-ansatz}, and also by applying the initial perturbation $\dKK|_{\mathscr{S}_{r_0}}=-10^{-1}\cdot {}_0{Y_{\ell'}}{}^0$ with $\ell'=1,2,3$. Each of the modes $\dKK{\,}_\ell{}^0$, with $\ell=0,1,2,3,4,5$, including the monopole part $\dKK{\,}_0{}^0$, decays at least as fast as $r^{-2}$ allowing thereby the initial data configuration to be of strongly asymptotically flat.}
	\label{fig:Florian-decphK}
\end{figure}

\begin{figure}[ht!]
	\vskip-0.1cm
	\begin{centering}
		{\tiny
			\begin{subfigure}{0.48\textwidth}
				\includegraphics[width=\textwidth]{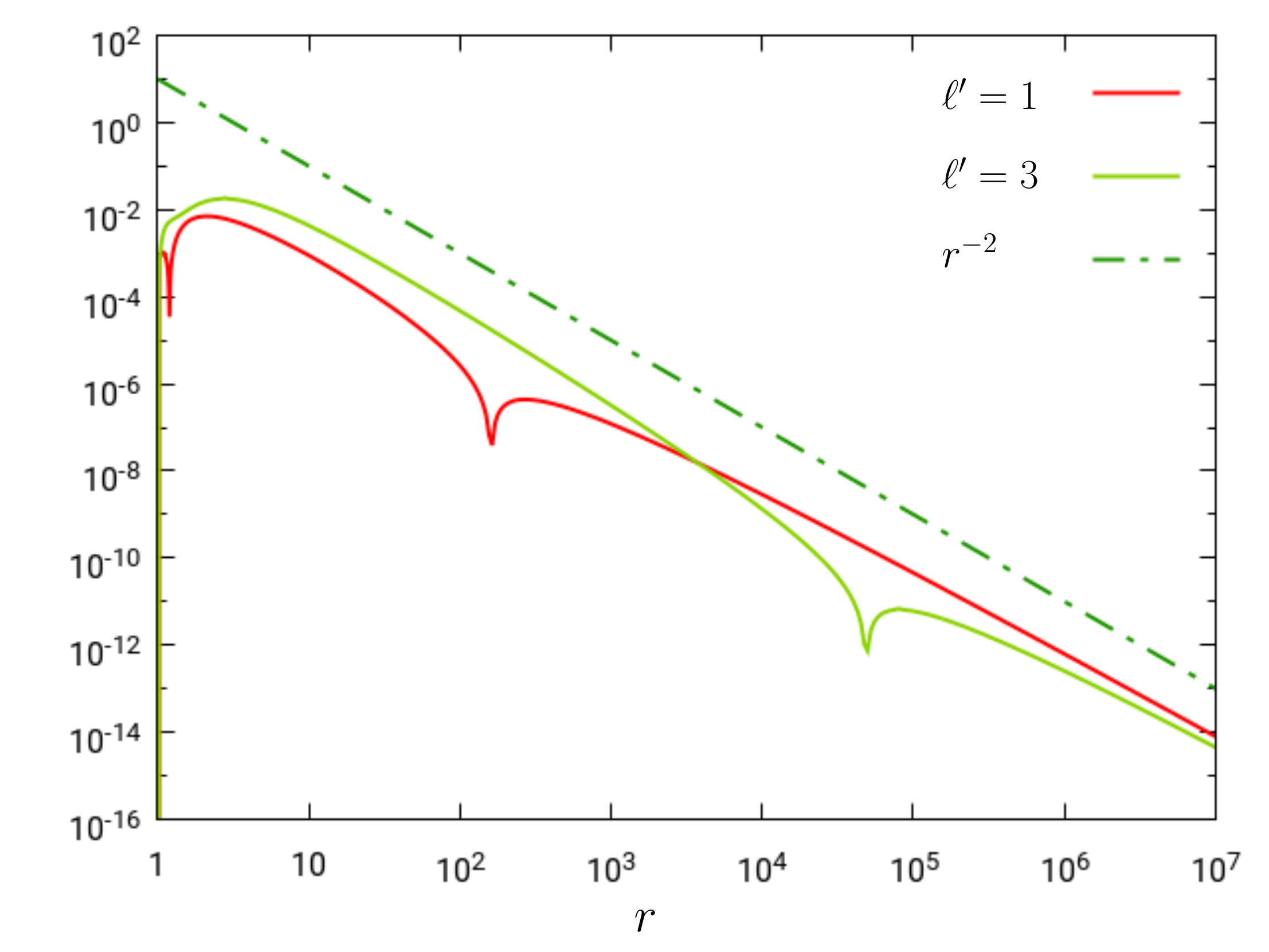}
				\vskip-0.4cm
				\caption{\scriptsize The decay rate of the mode $\dkk{\,}_1{}^0$ of $\dkk$ is $r^{-2}$. }
				\label{fig:Florian-decphk1}
			\end{subfigure}
			\begin{subfigure}{0.48\textwidth}
				\includegraphics[width=\textwidth]{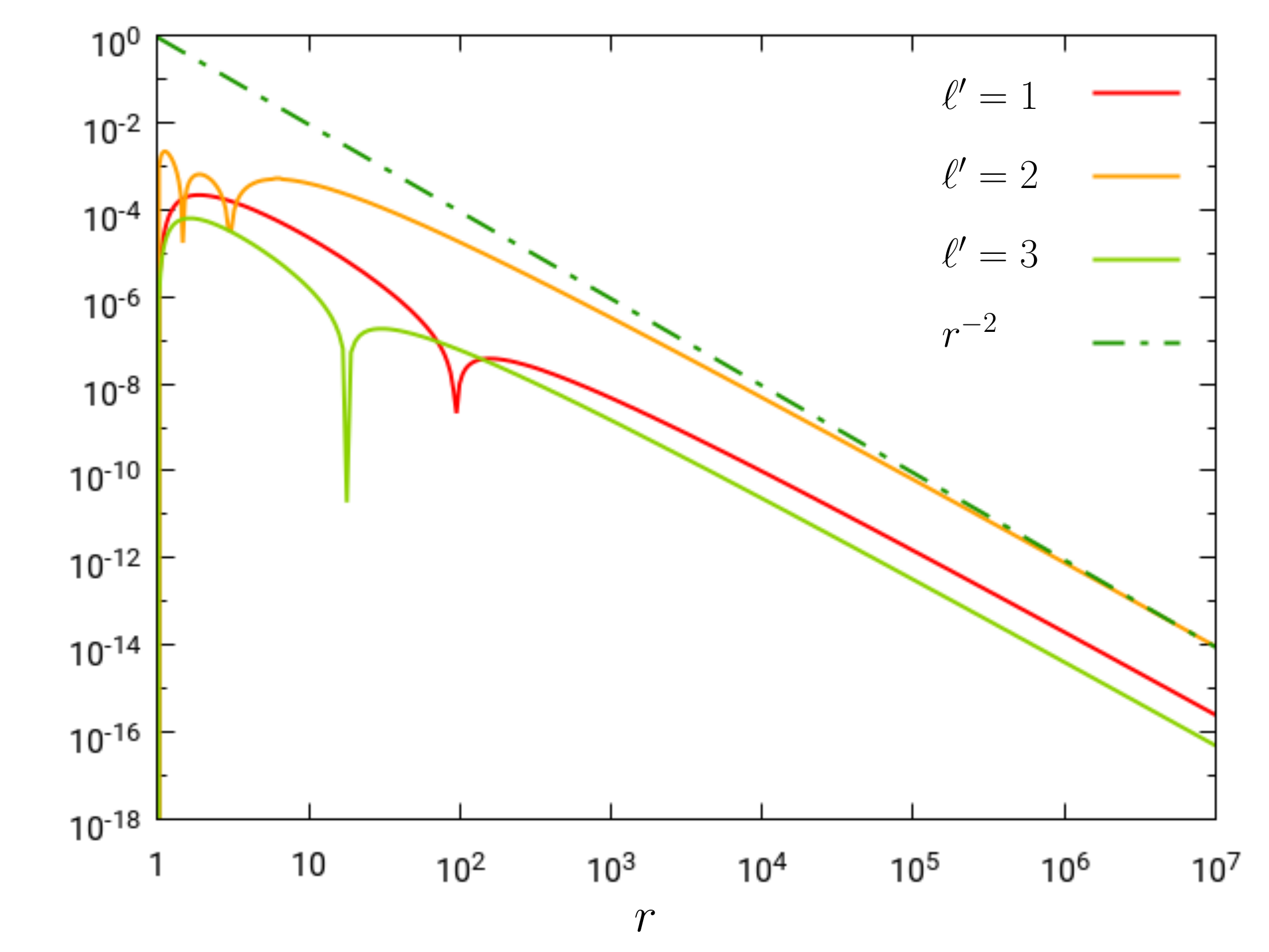}
				\vskip-0.4cm
				\caption{\scriptsize The decay rate of the mode $\dkk{\,}_2{}^0$ of $\dkk$ is $r^{-2}$.   }
				\label{fig:Florian-decphk2}
			\end{subfigure}
			\hskip.5cm
			\begin{subfigure}{0.48\textwidth}\vskip0.2cm
				\includegraphics[width=\textwidth]{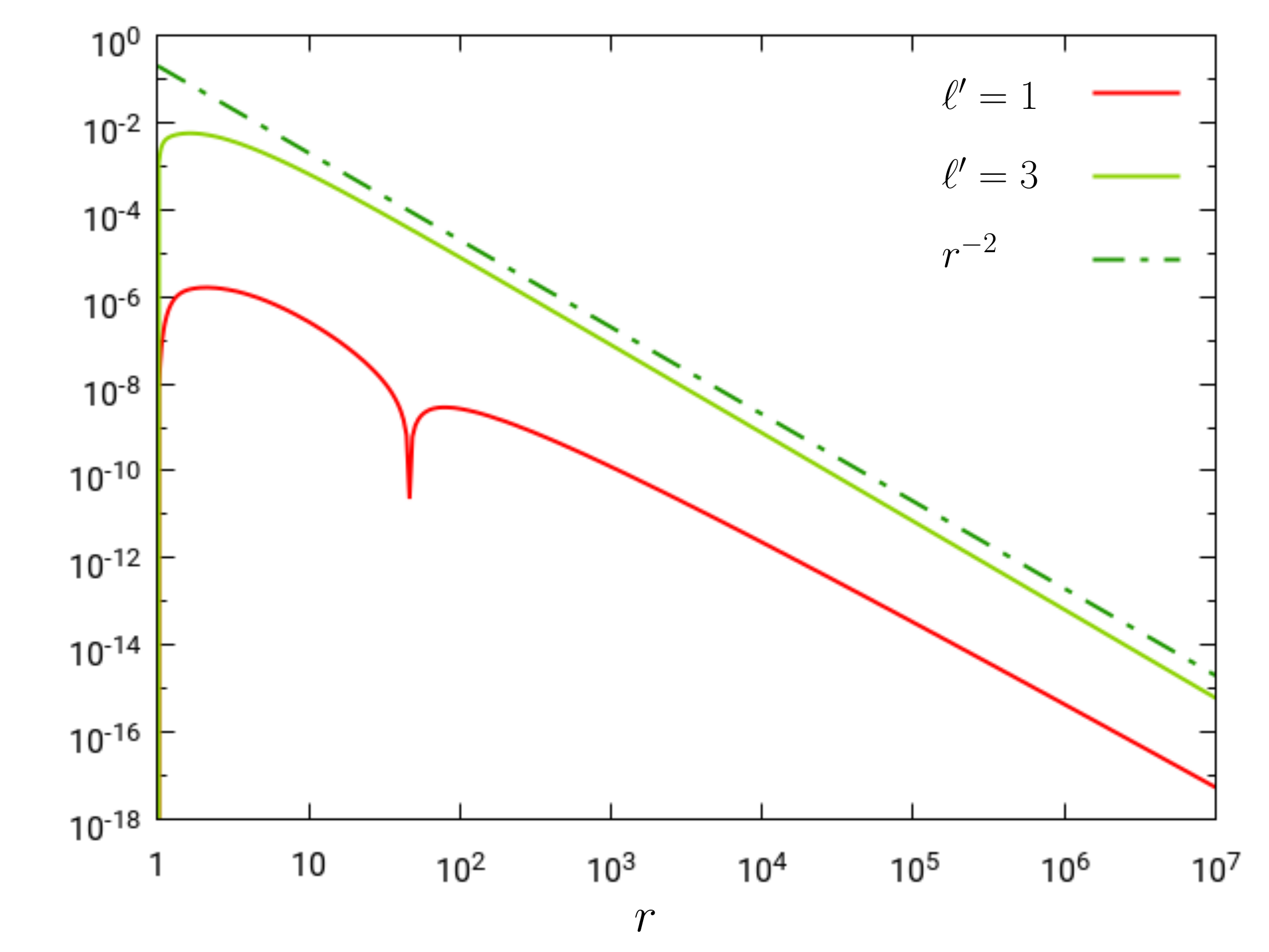}
				\vskip-0.4cm
				\caption{\scriptsize The decay rate of the mode $\dkk{\,}_3{}^0$ of $\dkk$ is $r^{-2}$.}
				\label{fig:Florian-decphk3}
			\end{subfigure}
			\begin{subfigure}{0.48\textwidth}\vskip0.2cm
				\includegraphics[width=\textwidth]{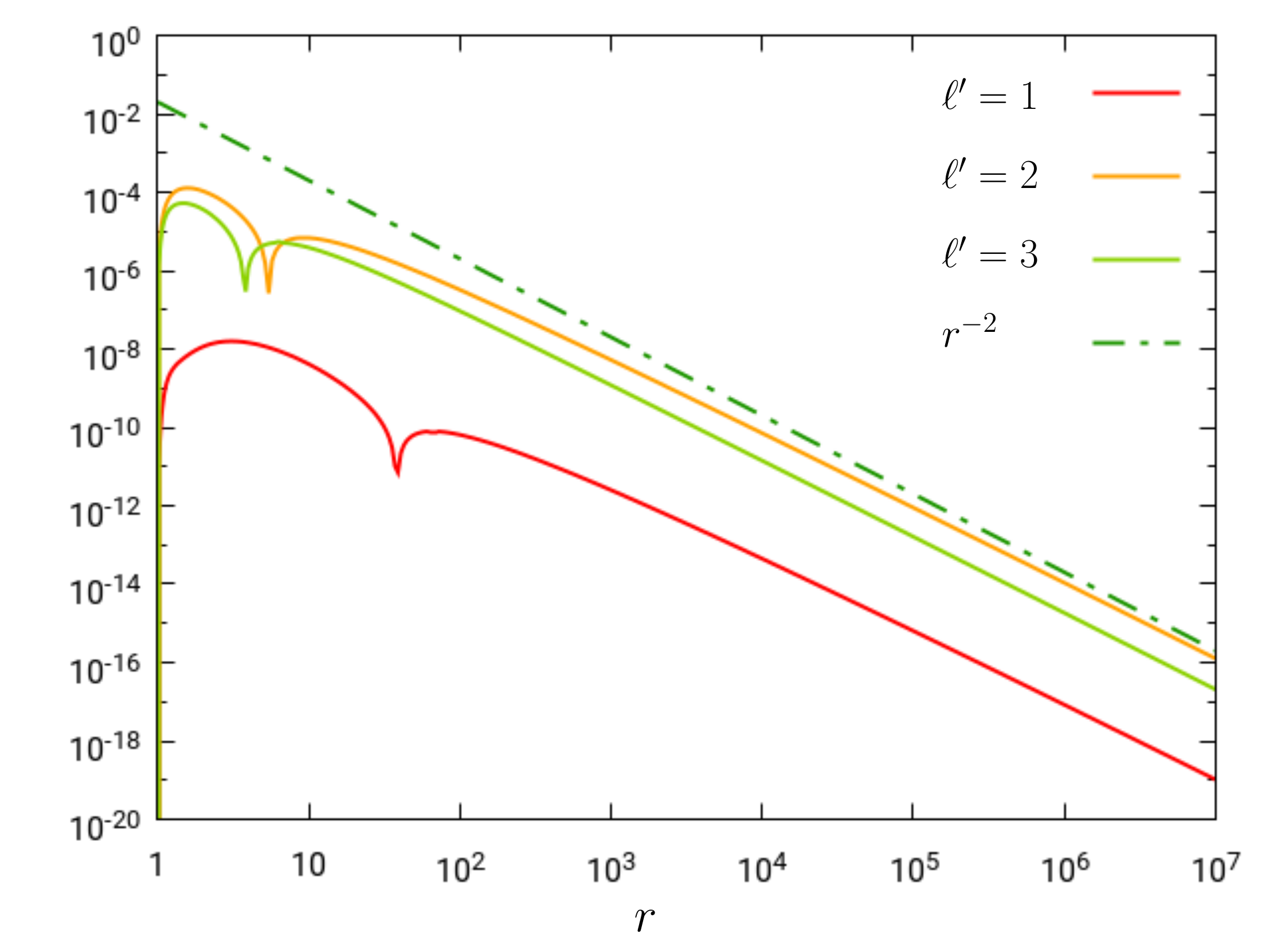}
				\vskip-0.4cm
				\caption{\scriptsize  The decay rate of the mode $\dkk{\,}_4{}^0$ of $\dkk$ is $r^{-2}$.}
				\label{fig:Florian-decphk4}
			\end{subfigure}
			\hskip.5cm
			\begin{subfigure}{0.48\textwidth}\vskip0.2cm
				\includegraphics[width=\textwidth]{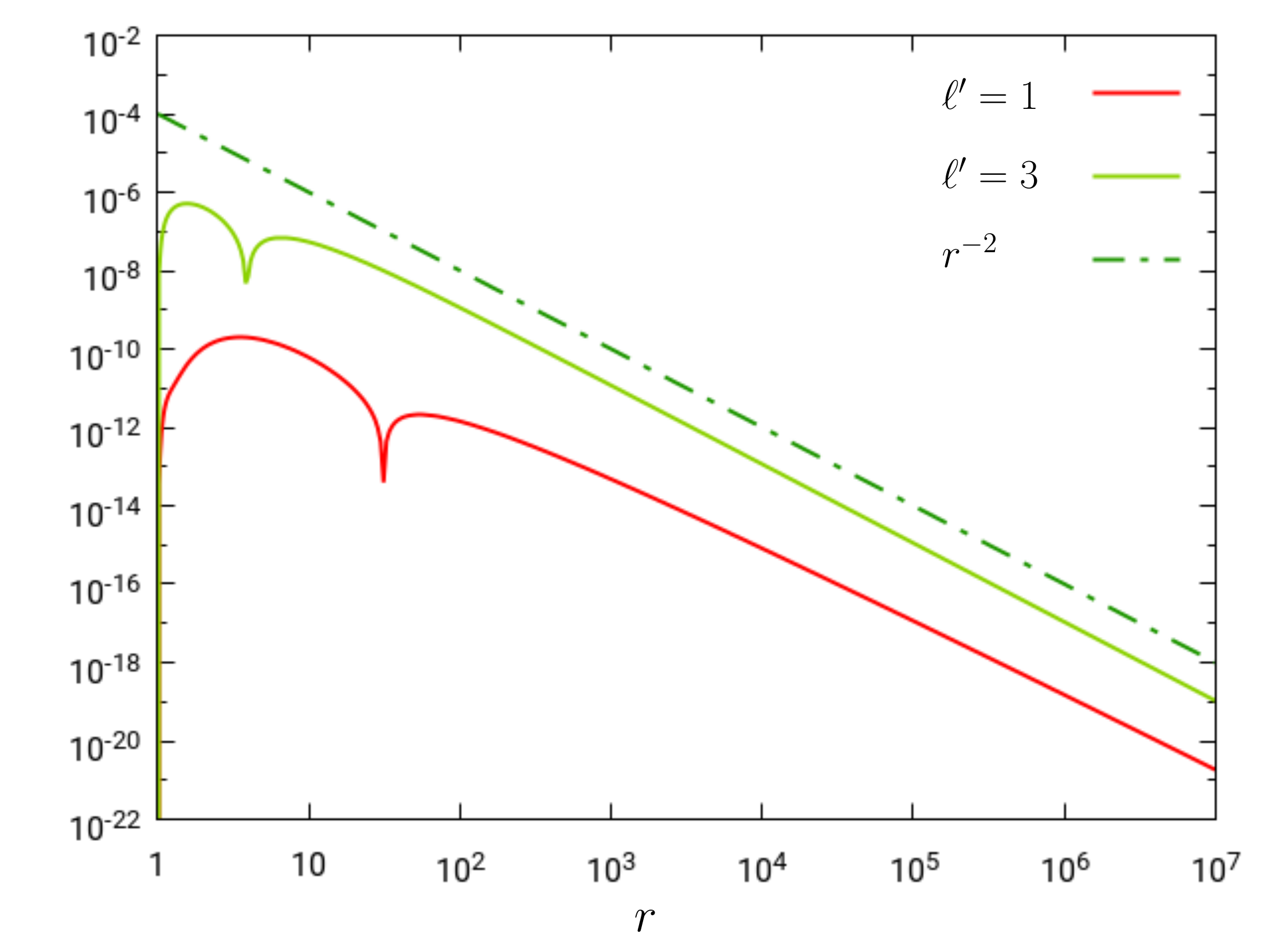}
				\vskip-0.4cm
				\caption{\scriptsize The decay rate of the mode $\dkk{\,}_5{}^0$ of $\dkk$ is $r^{-2}$.}
				\label{fig:Florian-decphk5}
			\end{subfigure}
		}
	\end{centering}
	\vskip0.1cm
	\caption{\footnotesize  The non-linear perturbative form of the parabolic-hyperbolic system was integrated numerically such that $\boldsymbol{\kappa}$ is always updated in accordance with the ansatz \eqref{eq: Florian-ansatz}, and also by applying the initial perturbation $\dKK|_{\mathscr{S}_{r_0}}=-10^{-1}\cdot {}_0{Y_{\ell'}}{}^0$ with $\ell'=1,2,3$. If each of the modes $\dkk{\,}_\ell{}^0$ with $\ell=1,2,3,4,5$ would fall off only with the rate  $r^{-1}$ that could already guarantee that the initial data configuration is strongly asymptotically flat. Remarkably, supposedly due to the use of the anzatz \eqref{eq: Florian-ansatz}, each of the modes $\dkk{\,}_\ell{}^0$ decay much faster around $r^{-2}$.}
	\label{fig:Florian-decphk}
\end{figure}

\begin{figure}[ht!]
	\vskip-0.1cm
	\begin{centering}
		{\tiny
			\begin{subfigure}{0.48\textwidth}
				\includegraphics[width=\textwidth]{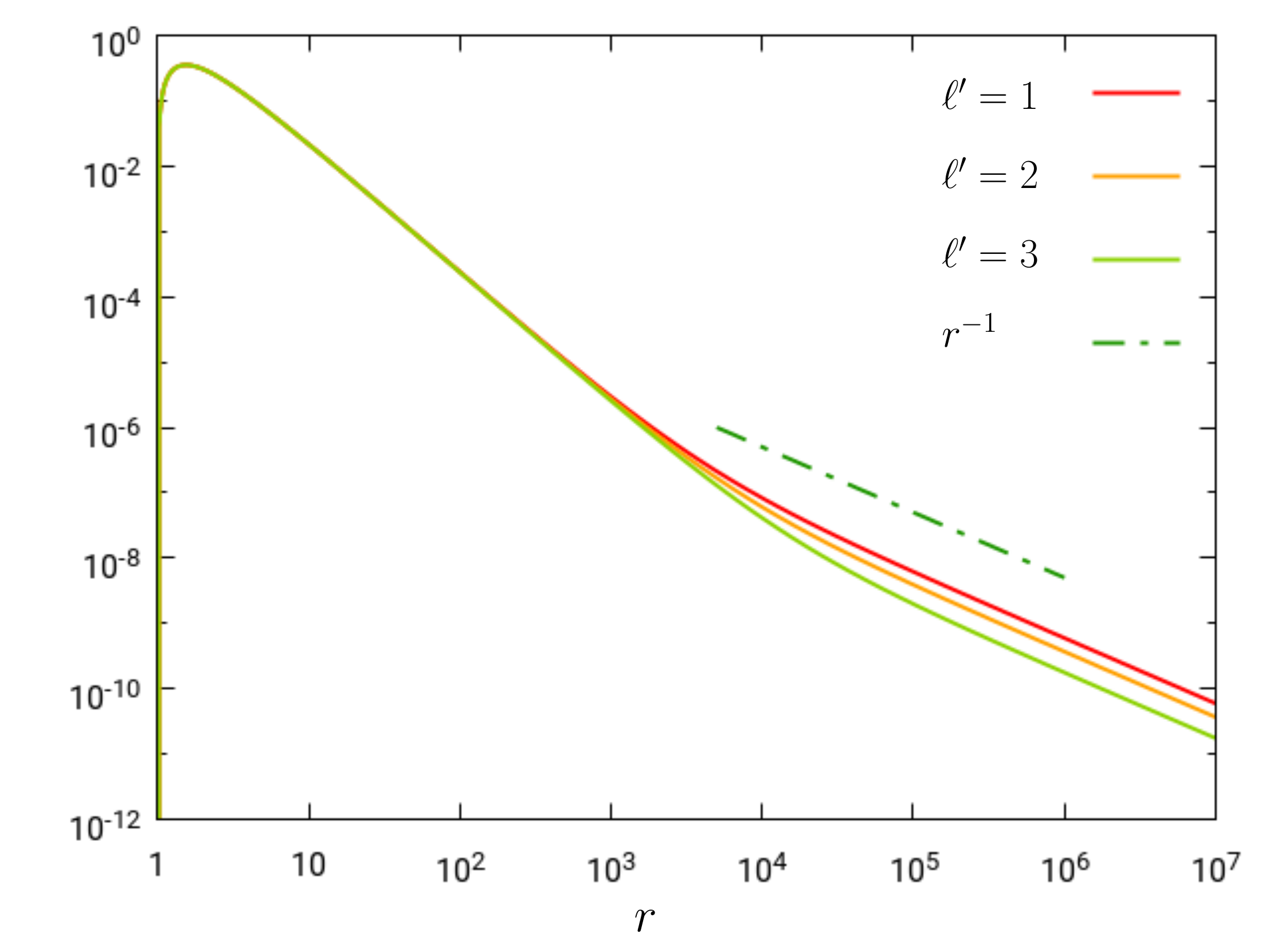}
				\vskip-0.4cm
				\caption{\scriptsize The decay rate of the monopole part $\dNNh_0{}^0$ of $\dNNh$.}
				\label{fig:Florian-decphN0}
			\end{subfigure}
			\begin{subfigure}{0.48\textwidth}
				\includegraphics[width=\textwidth]{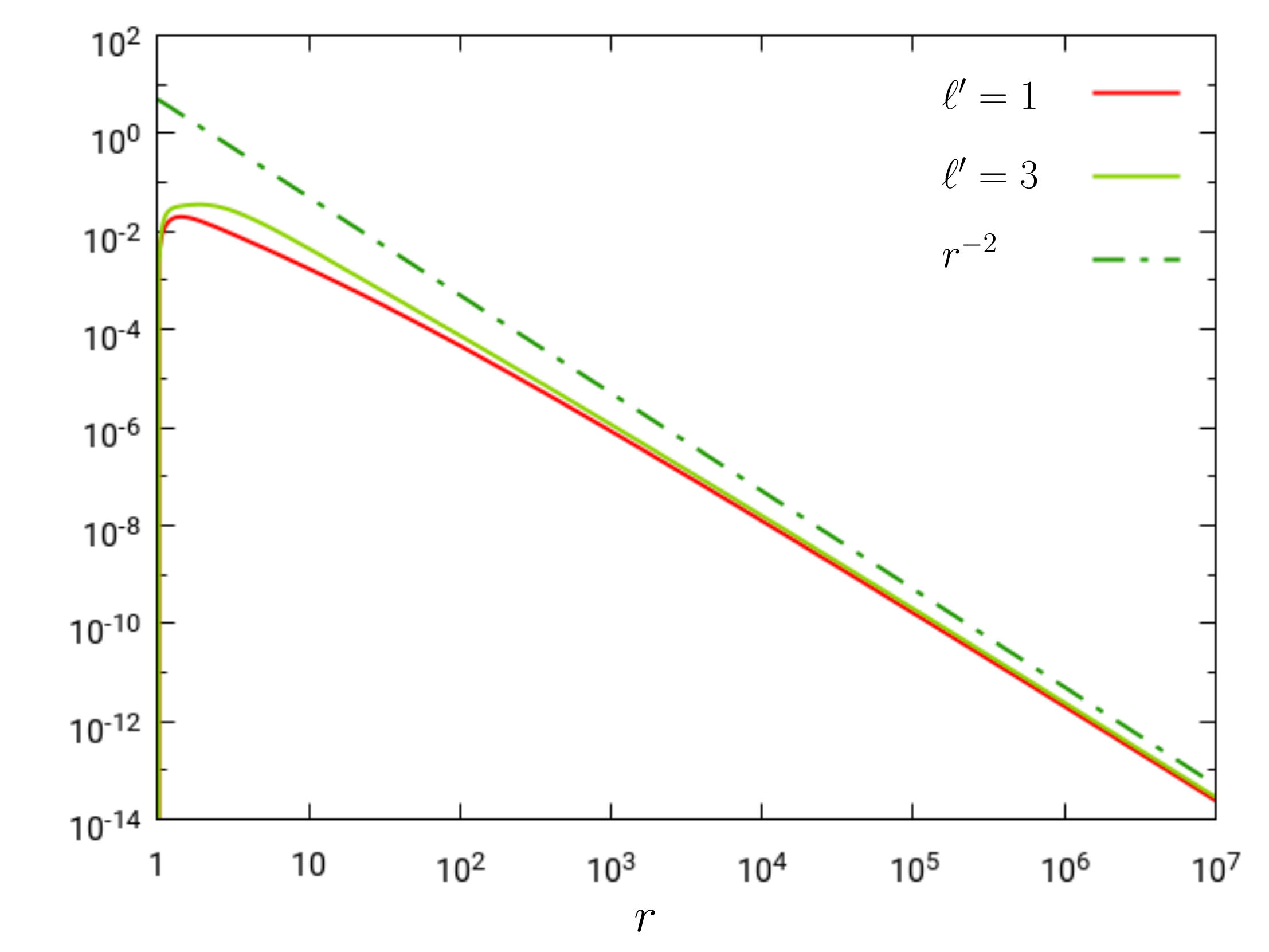}
				\vskip-0.4cm
				\caption{\scriptsize  The decay rate of the mode $\dNNh_1{}^0$ of $\dNNh$ is $r^{-2}$.}
				\label{fig:Florian-decphN1}
			\end{subfigure}
			\begin{subfigure}{0.48\textwidth}\vskip0.1cm
				\includegraphics[width=\textwidth]{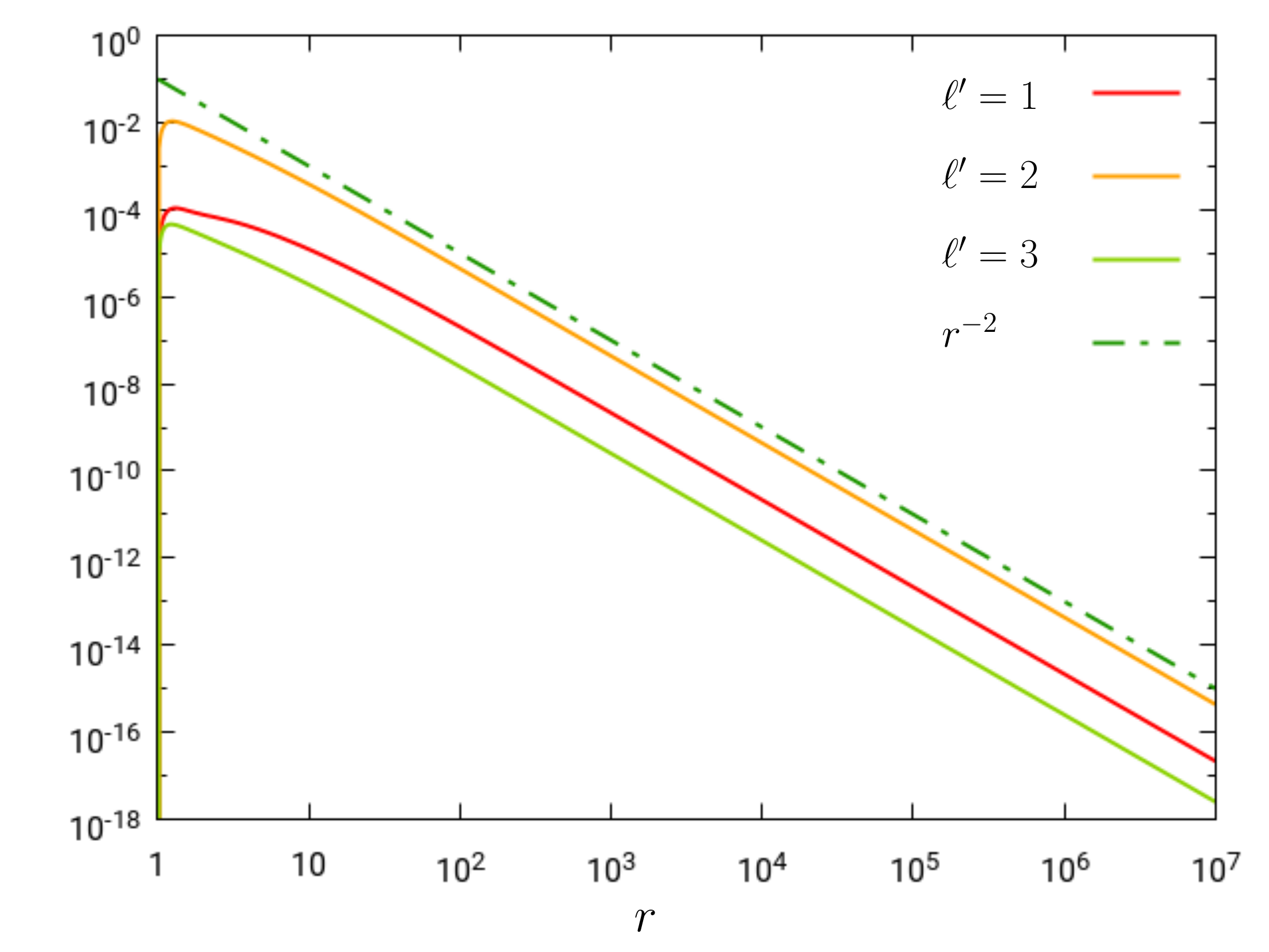}
				\vskip-0.4cm
				\caption{\scriptsize The decay rate of the mode $\dNNh_2{}^0$ of $\dNNh$ is $r^{-2}$.}
				\label{fig:Florian-decphN2}
			\end{subfigure}
			\hskip.5cm
			\begin{subfigure}{0.48\textwidth}\vskip0.1cm
				\includegraphics[width=\textwidth]{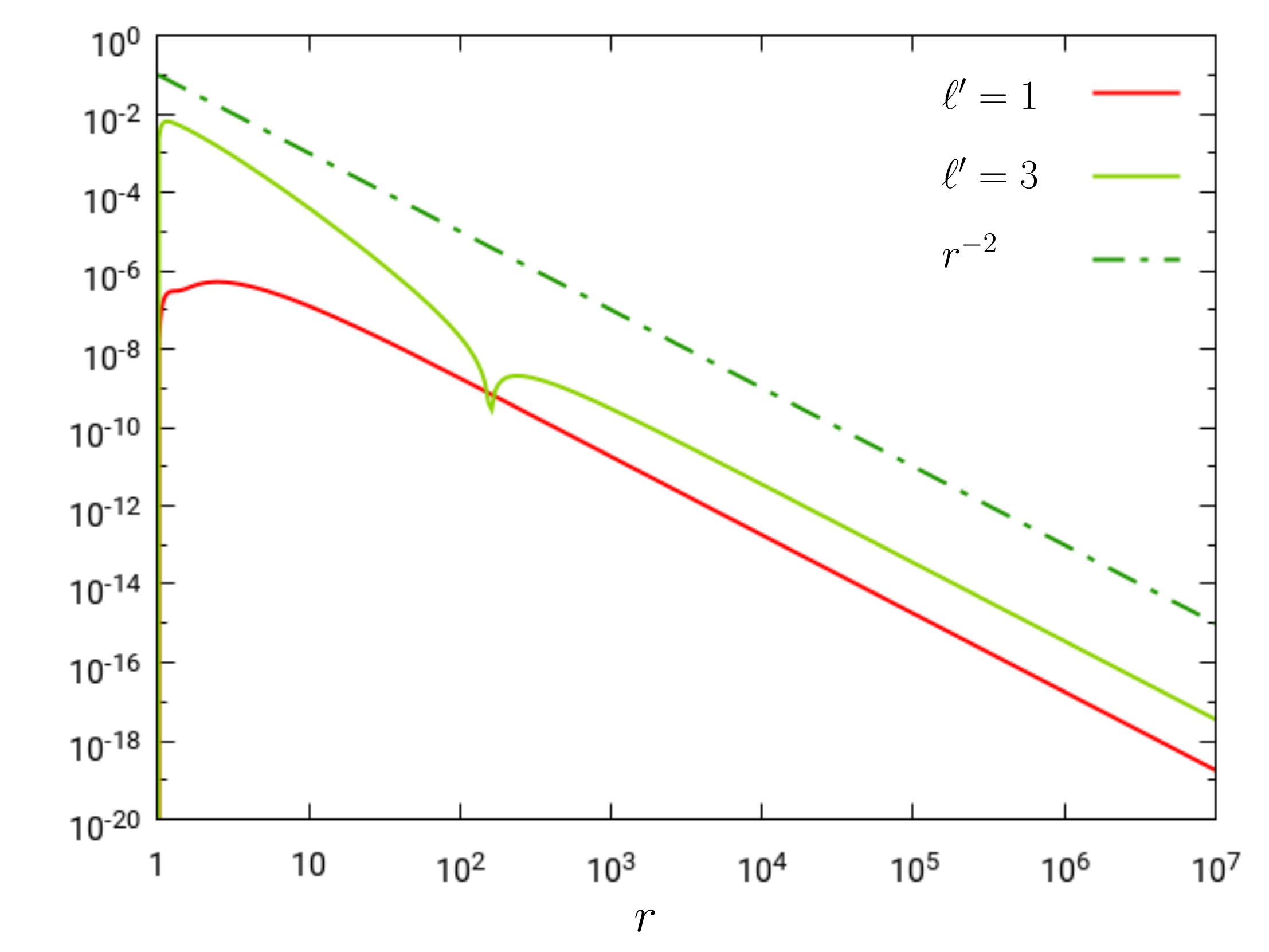}
				\vskip-0.4cm
				\caption{\scriptsize The decay rate of the mode $\dNNh_3{}^0$ of $\dNNh$ is $r^{-2}$.}
				\label{fig:Florian-decphN3}
			\end{subfigure}
			\begin{subfigure}{0.48\textwidth}\vskip0.1cm
				\includegraphics[width=\textwidth]{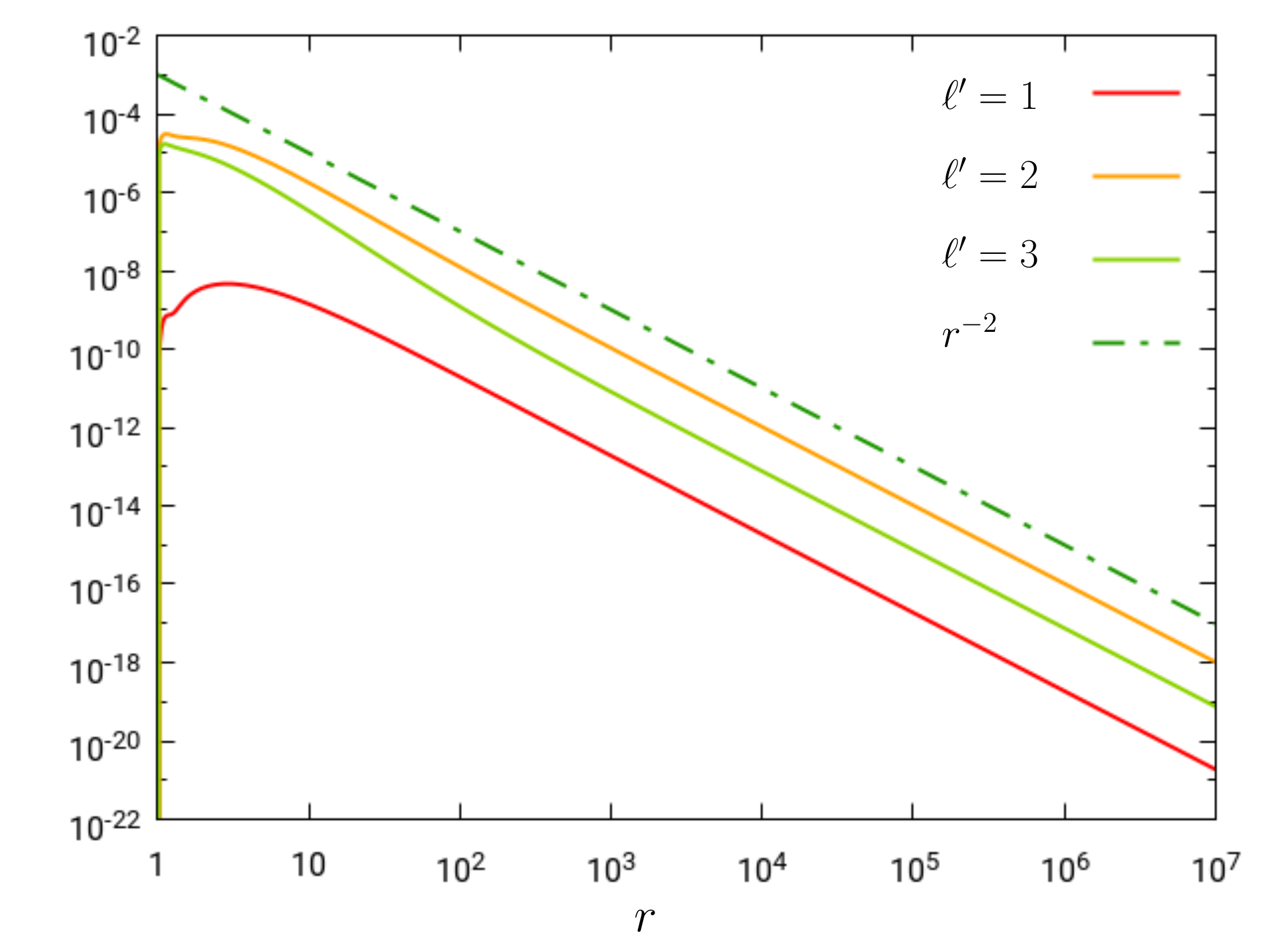}
				\vskip-0.4cm
				\caption{\scriptsize The decay rate of the mode $\dNNh_4{}^0$ of $\dNNh$ is $r^{-2}$.}
				\label{fig:Florian-decphN4}
			\end{subfigure}
			\hskip.5cm
			\begin{subfigure}{0.48\textwidth}\vskip0.1cm
				\includegraphics[width=\textwidth]{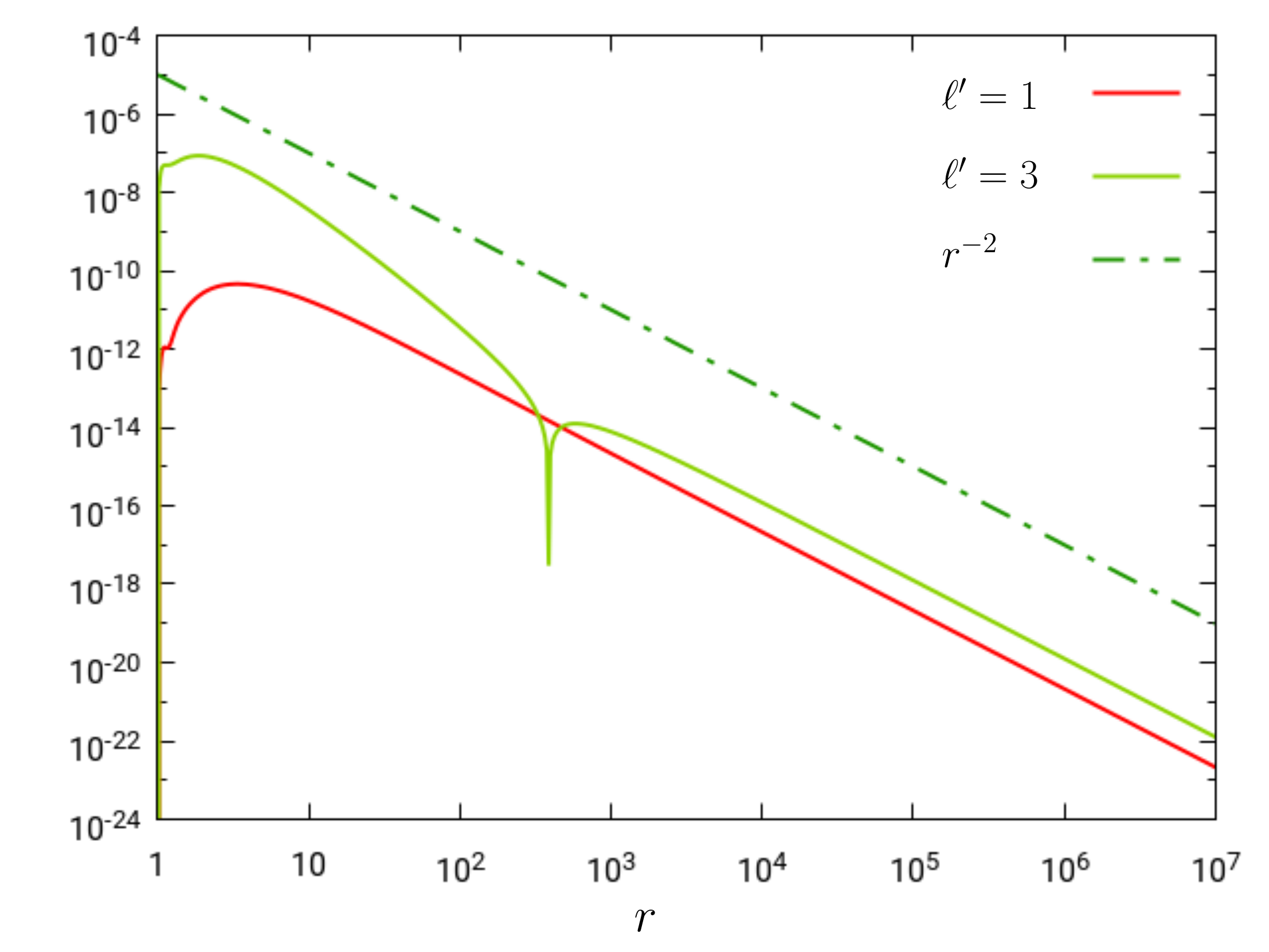}
				\vskip-0.4cm
				\caption{\scriptsize The decay rate of the mode $\dNNh_5{}^0$ of $\dNNh$ is $r^{-2}$. }
				\label{fig:Florian-decphN5}
			\end{subfigure}
		}
	\end{centering}
	\vskip-0.1cm
	\caption{\footnotesize The non-linear perturbative form of the parabolic-hyperbolic system was integrated numerically such that $\boldsymbol{\kappa}$ is always updated in accordance with the ansatz \eqref{eq: Florian-ansatz}, and also by applying the initial perturbation $\dKK|_{\mathscr{S}_{r_0}}=-10^{-1}\cdot {}_0{Y_{\ell'}}{}^0$ with $\ell'=1,2,3$. Each of the modes $\dNNh_\ell{}^0$, with $\ell=1,2,3,4,5$, of $\dNNh$ start to decay as $r^{-2}$ which is, likewise it was in case of the modes $\dkk_\ell{}^0$, is much faster than it is needed for $\dNNh$ to fit strongly asymptotically flat initial data configuration. Notably the monopole mode $\dNNh_0{}^0$ also starts with  $r^{-2}$, nevertheless, around $r=10^3$ it switches to the rate $r^{-1}$ as it should happen if the ADM mass of the composite system is slightly differs from the ADM mass of the background.}
	\label{fig:Florian-decphN}
\end{figure}

The functional form of the initial data applied here does also differ from the one used in \cite{Beyer:2020kty}. For the sake of easy comparability we apply exactly the same initial data as was used in producing Figs.\,\ref{fig:decphK}, \ref{fig:decphk}, \ref{fig:decphN}. Accordingly, the initial data specified at $\mathscr{S}_{r_0}$, with $r_0=1$, was chosen to be of the form $\dKK|_{\mathscr{S}_{r_0}}= - 10^{-1} \cdot {}_0{Y_{\ell'}}{}^0$, with $\ell'=1,2,3$, and with $\dkk|_{\mathscr{S}_{r_0}}=0$ and $\dNNh|_{\mathscr{S}_{r_0}}=0$.
Three different solutions corresponding to the choices $\ell'=1,2,3$ are plotted in case of each of the monitored modes. The $r$-dependence of various $\ell$-modes of $\dKK$, $\dkk$ and $\dNNh$, along with the pertinent decay rates, are indicated on Figs.\,\ref{fig:Florian-decphK}, \ref{fig:Florian-decphk} and \ref{fig:Florian-decphN}.

\medskip

As it is depicted by Fig.\,\ref{fig:Florian-decphK}, each of the modes $\dKK{\,}_\ell{}^0$, with $\ell=0,1,2,3,4,5$,---notably, including the monopole part $\dKK{\,}_0{}^0$ too---decays at least as fast as $r^{-2}$. This, in virtue of Table\,\ref{table:falloff}, implies that $\KK$ decays in accordance with the requirements allowing the initial data configuration to be of strongly asymptotically flat. The main message conveyed by Fig.\,\ref{fig:Florian-decphk} is analogous. Each of the modes $\dkk{\,}_\ell{}^0$, with $\ell=1,2,3,4,5$, decay as $r^{-2}$, that is much faster than the decay rate $r^{-1}$  which, in virtue of Table\,\ref{table:falloff}, could already guarantee that the initial data configuration is strongly asymptotically flat. Note that the panels in  Fig.\,\ref{fig:Florian-decphN} simply further strengthen the above conclusions. Each of the modes $\dNNh_\ell{}^0$, with $\ell=1,2,3,4,5$, of $\dNNh$ decay as $r^{-2}$ which, in virtue of Table\,\ref{table:falloff}, is again much faster than $\NNh$ has to produce to fit to a strongly asymptotically flat initial data configuration. Notably the monopole mode $\dNNh_0{}^0$ also starts to fall off as $r^{-2}$ which, however, around $r=10^3$ switches to the rate $r^{-1}$. This implies that the use of the ansatz \eqref{eq: Florian-ansatz} affects slightly the value of the ADM mass of the background.

\medskip

All in all Figs.\,\ref{fig:Florian-decphK}, \ref{fig:Florian-decphk} and \ref{fig:Florian-decphN} verifies that all the conditions listed in subsection\,\ref{subsec: asymp-flatness}, in particular, in Table\,\ref{table:falloff}, hold, i.e.\,the time integration of the  parabolic-hyperbolic form of the constraint system, along with the continuous updating of $\boldsymbol{\kappa}$ in accordance with the ansatz \eqref{eq: Florian-ansatz} and \eqref{eq: Florian-ansatz-specR}, does indeed yield strongly asymptotically flat initial data.

\subsubsection{Algebraic-hyperbolic system}\label{subsub: alg-hyp-ansatz}

This subsection is to indicate that there is a large freedom to control the asymptotic behavior of the monopole part $\dKK{\,}_0{}^0$ of $\dKK$---in a way analogous to the one described in the previous subsection---even if the algebraic-hyperbolic form of the constraints is to be solved. Specifically, by inspecting the algebraic-hyperbolic system, \eqref{eq:ah1}--\eqref{eq:ah4}, it gets immediately transparent that a minimal alteration of the selection rules of strictly near-Schwarzschild configurations can be achieved by choosing the freely specifiable variable ${\interior{\bf K}}{}_{AB}$, or alternatively $\interior{\mathbf{K}}_{qq}={\interior{\bf K}}{}_{AB}q^Aq^B$---which is vanishing for strictly near-Schwarzschild configurations---in a suitable way.

\medskip

In doing so notice first that the algebraic form of the Hamiltonian constraint, comprised by \eqref{eq:ah3} and \eqref{eq:ah4}, reads as
\begin{equation}\label{eq: alg-Ham-constr}
	\kkappa=\tfrac{1}{2}\KK^{-1}\Big[\dd^{-1}\big(2\aaa\,\kk\kkb-\bb\,\kkb^2-\bbb\,\kk^2\big)-\tfrac12\KK^2-\big({}^{\scriptscriptstyle(\!3\!)}\!R-\Kc_{AB}\Kc{}^{AB}\big)\Big]\,.
\end{equation}
Recall also that, in virtue of subsection \ref{subsub: ansatz-alg-hyp-radial}, in order to get the desired fall off property for the monopole part $\KK_0{}^0$ of $\KK$ one should guarantee the relation
\begin{equation}\label{eq: ansatz-ah-gen00}
\boldsymbol{\kappa}_0{}^0= \mathcal{R}\cdot {\bf K}_0{}^0+\mathcal{S}%\,,
\end{equation}
to hold, for some smooth negative functions $\mathcal{R}=\mathcal{R}(r)$ and $\mathcal{S}=\mathcal{S}(r)$ such that, for sufficiently large values of $r$, $\mathcal{R}$ and $\mathcal{S}$ tends to a constant value $-1/2$ and zero, respectively. In our numerical simulations $\mathcal{R}$ and $\mathcal{S}$ were fixed as
\begin{equation}\label{eq: delta-rho}
\mathcal{R}= -\frac12\,\frac{r+M}{r+2\,M}\,, \qquad \mathcal{S}=-\frac{\delta\cdot M^{\rho}}{(r+2\,M)^{\rho}}\,,
\end{equation}
where $\delta>0$ and $\rho$ designed to take value from the interval $3/2<\rho\leq 2$.
Note that in \eqref{eq: alg-Ham-constr} the contraction $\Kc_{AB}\Kc{}^{AB}$, because of the vanishing of $\bb$ in the present case, can be seen to take the form
\begin{equation}
\Kc_{AB}\Kc{}^{AB}=\tfrac12\,\aaa^{-2} \Kcqq{\overline{\Kcqq}}\,,
\end{equation}
where $\Kcqq=\Kc_{AB}q^Aq^B$. Note also that the source terms $q^i\dot{\widehat{n}}^l\interior{\mathbf{K}}{}_{li}$ and $q^i\widehat{D}^l\interior{\mathbf{K}}{}_{li}$ in \eqref{eq:ah2}---which, in the generic case, are given by (5.8) and (5.9) of \cite{Racz:2017krc}, respectively,---simplify considerably. For instance, in virtue of (A.12) of \cite{Racz:2017krc}, $q^B\dot{\widehat{n}}^A \Kc_{AB}$ vanishes because $\widehat{N}$ depends only on the radial coordinate $r$, whereas (A.13) of \cite{Racz:2017krc}, reduces to
\begin{equation}\label{eq: alg-qDKc}
q^A{\widehat{D}^B}\Kc_{AB}=\tfrac{1}{2} \,\aaa^{-1} \,\ethb\Kcqq\,.
\end{equation}

For the sake of simplicity the freely specifiable field $\Kcqq$ was assumed to have only a single mode taking the form
\begin{equation}
\Kcqq=(\Kcqq)_2{}^0\cdot {}_{2}Y_2{}^0.
\end{equation}
Note that since the above outlined procedure admits a regular spherical solution, with $\mathcal{R}= -\tfrac12\,\frac{r+M}{r+2\,M}$, $\mathcal{S}=0$ and $\Kcqq=0$, this can be used as a reference solution in the non-linear perturbative approach.

\medskip

Turning now to the specific numerical investigations note first that the use of non-vanishing $\Kc_{AB}$ creates an unexpectedly strong influence on each of the $\ell$-modes of constraint fields $\KK$ and $\kk$.  Simplifications could be gained by exciting only the even modes of $\KK$ and $\kk$. In addition we also have to admit that because of some unexpected numerical instabilities we could get sufficiently long time evolution in the interim of the numerical integration of \eqref{eq:ah1} and \eqref{eq:ah2} only when very tiny excitations in the monopole and the quadrupole part of $\KK$ was applied. 

\medskip

\begin{figure}[ht!]
	\vskip-0.1cm
	\begin{centering}
		{\tiny
			\begin{subfigure}{0.48\textwidth}
				\includegraphics[width=\textwidth]{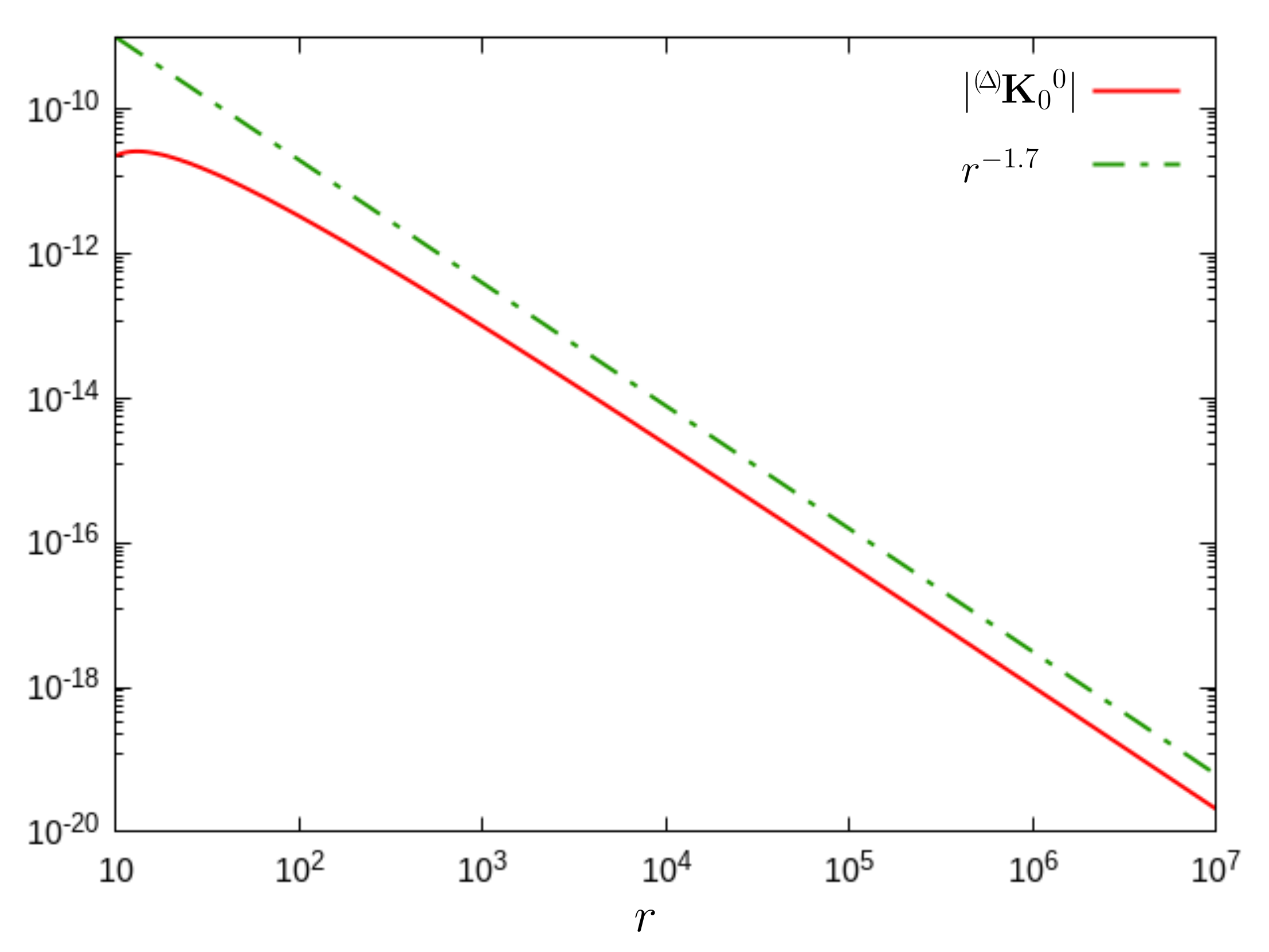}
				\vskip-0.4cm
				\caption{\scriptsize The decay rate of the mode $\dKK{\,}_0{}^0$ of $\dKK$
					is $r^{-1.7}$. }
				\label{fig:a-h-decphK0}
			\end{subfigure}
			\begin{subfigure}{0.48\textwidth}\vskip0.2cm
				\includegraphics[width=\textwidth]{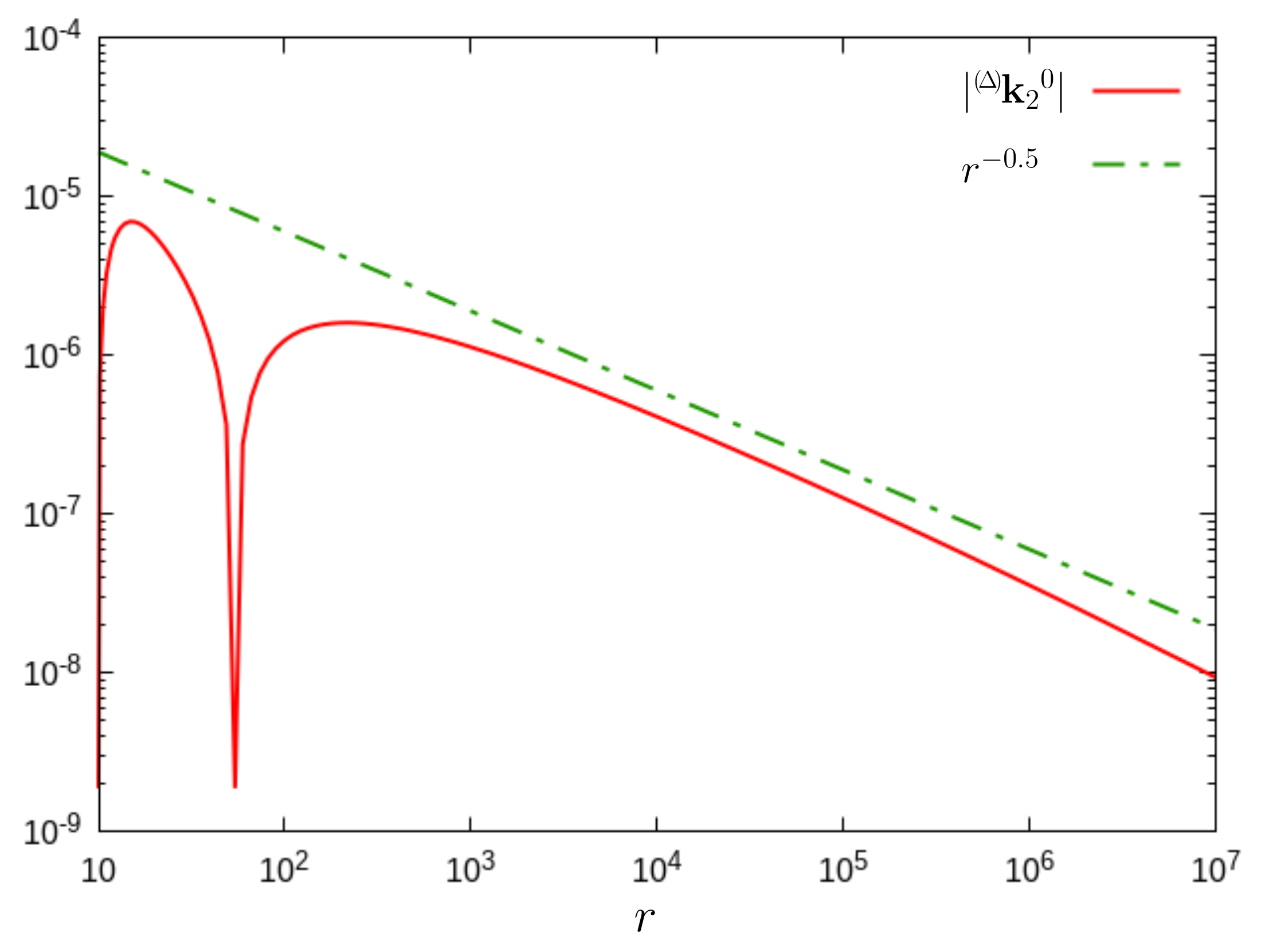}
				\vskip-0.4cm
				\caption{\scriptsize The decay rate of $\dkk{\,}_2{}^0$ 
					is better than $r^{-1/2}$.}
				\label{fig:a-h-decphK1}
			\end{subfigure}
			\begin{subfigure}{0.48\textwidth}\vskip0.2cm
				\includegraphics[width=\textwidth]{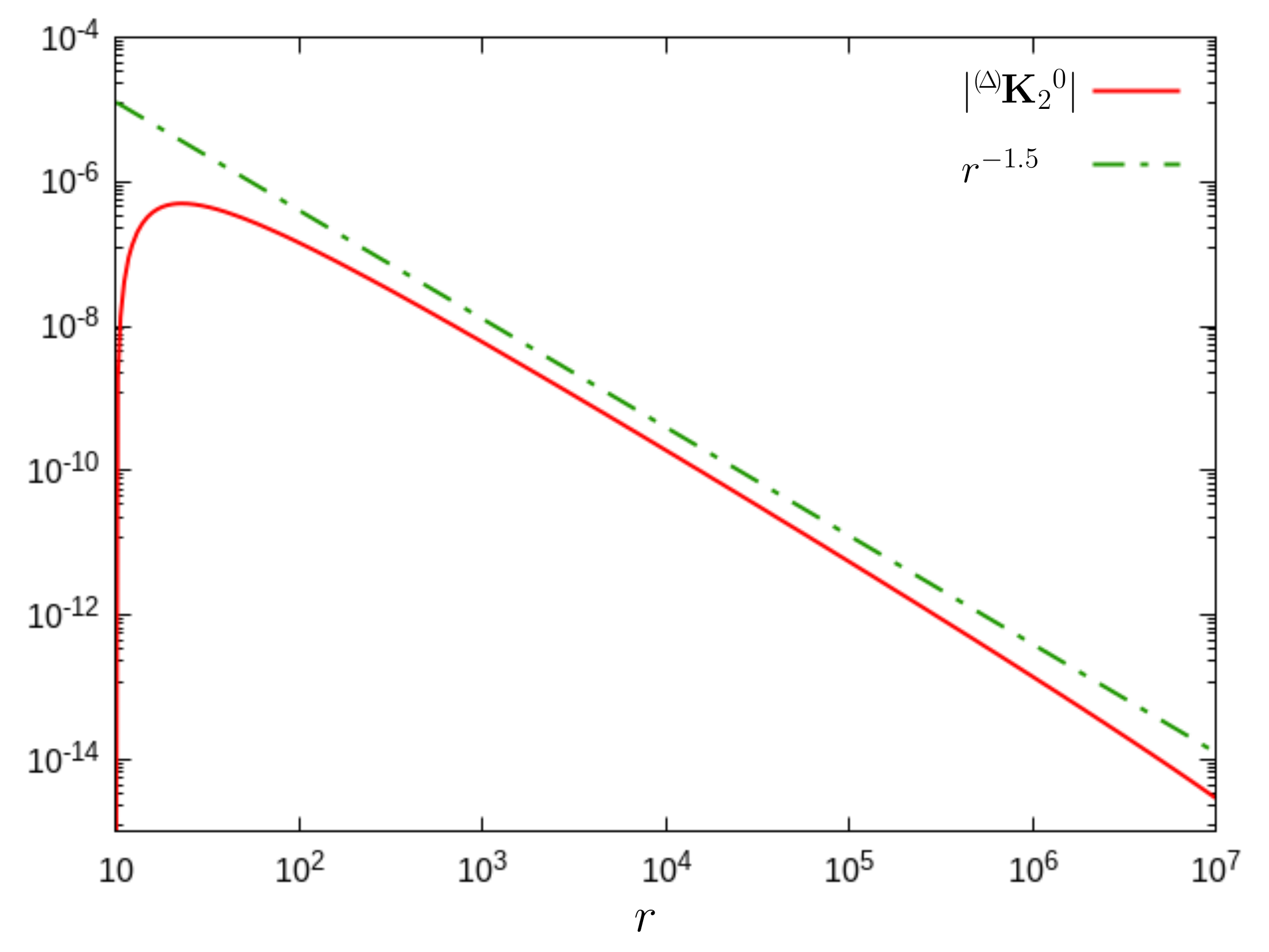}
				\vskip-0.4cm
				\caption{\scriptsize The decay rate of $\dKK{\,}_2{}^0$ 
					is better than $r^{-3/2}$.}
				\label{fig:a-h-decphK2}
			\end{subfigure}
			\hskip.5cm
			\begin{subfigure}{0.48\textwidth}\vskip0.2cm
				\includegraphics[width=\textwidth]{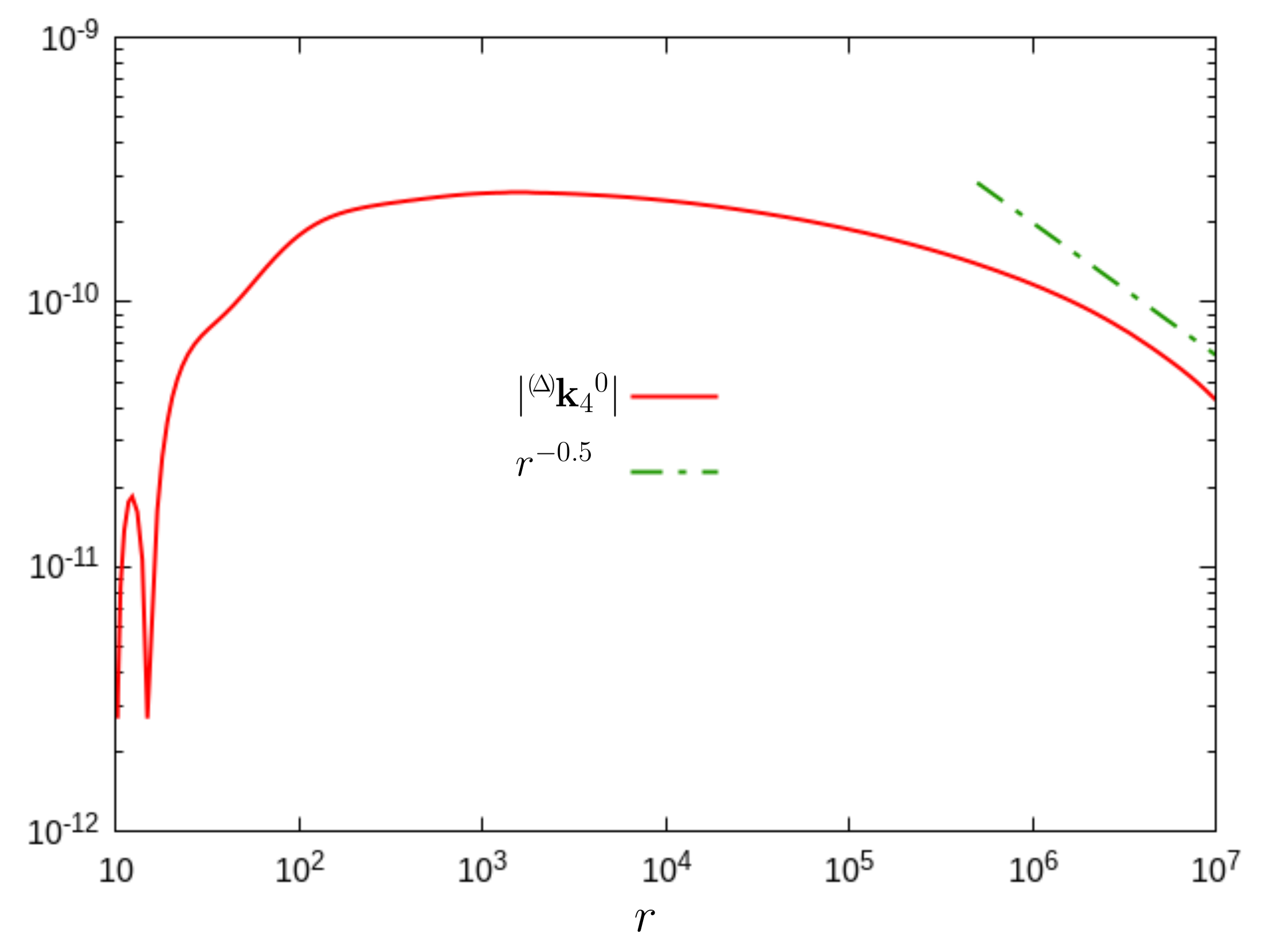}
				\vskip-0.4cm
				\caption{\scriptsize The decay rate of $\dkk{\,}_4{}^0$ 
					is better than $r^{-1/2}$.}
				\label{fig:a-h-decphK3}
			\end{subfigure}
			\begin{subfigure}{0.48\textwidth}\vskip0.2cm
				\includegraphics[width=\textwidth]{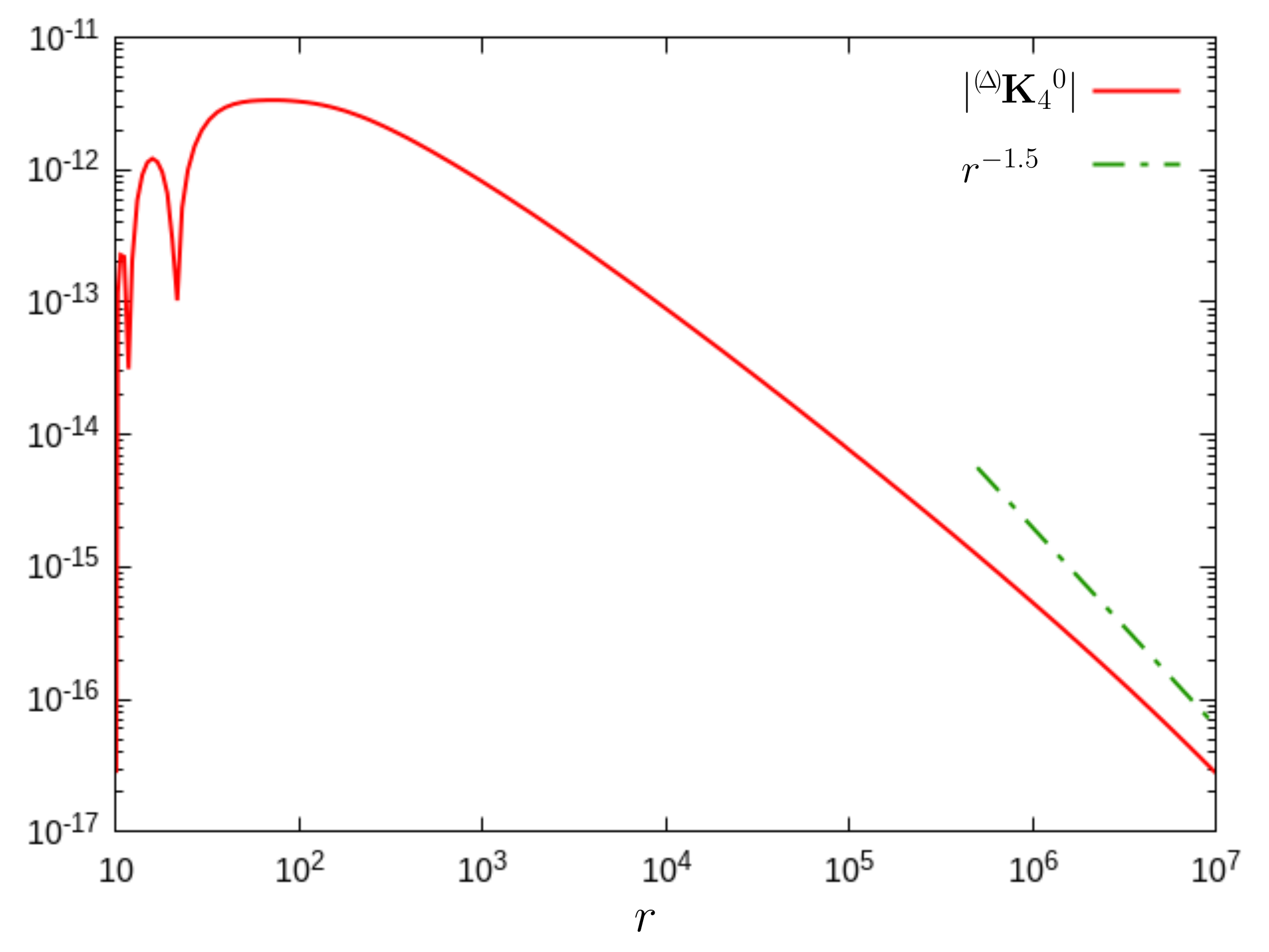}
				\vskip-0.4cm
				\caption{\scriptsize The decay rate of $\dKK{\,}_4{}^0$ 
					is depicted.}
				\label{fig:a-h-decphK4}
			\end{subfigure}
		}
	\end{centering}
	\vskip-0.1cm
	\caption{\footnotesize  The non-linear perturbative form of the algebraic-hyperbolic system was integrated numerically by applying the initial perturbation $\dKK|_{\mathscr{S}_{r_0}} = 6.5\cdot 10^{-12}\cdot Y_0{}^0+10^{-15}\cdot Y_2{}^0$. As it is depicted on the left panels the monopole part $\dKK{\,}_0{}^0$ of $\dKK$ decays with the rate $r^{-1.7}$, whereas the $\ell=2,4$ modes of $\dKK$ decay slightly better that $r^{-3/2}$. On the right panels the $\ell=2,4$ modes of $\dkk$ are plotted. These are much slower decaying, nevertheless, they both appear to decay slightly better that $r^{-1/2}$.
		Accordingly, the asymptotic behavior of $\dKK$ and $\dkk$ allows the corresponding initial data configuration to be weakly asymptotically flat.}
	\label{fig:dec-ah-KK-kk}
\end{figure}

\begin{figure}[ht!]
	\vskip-0.1cm
	\begin{centering}
		{\tiny
			\begin{subfigure}{0.48\textwidth}
				\includegraphics[width=\textwidth]{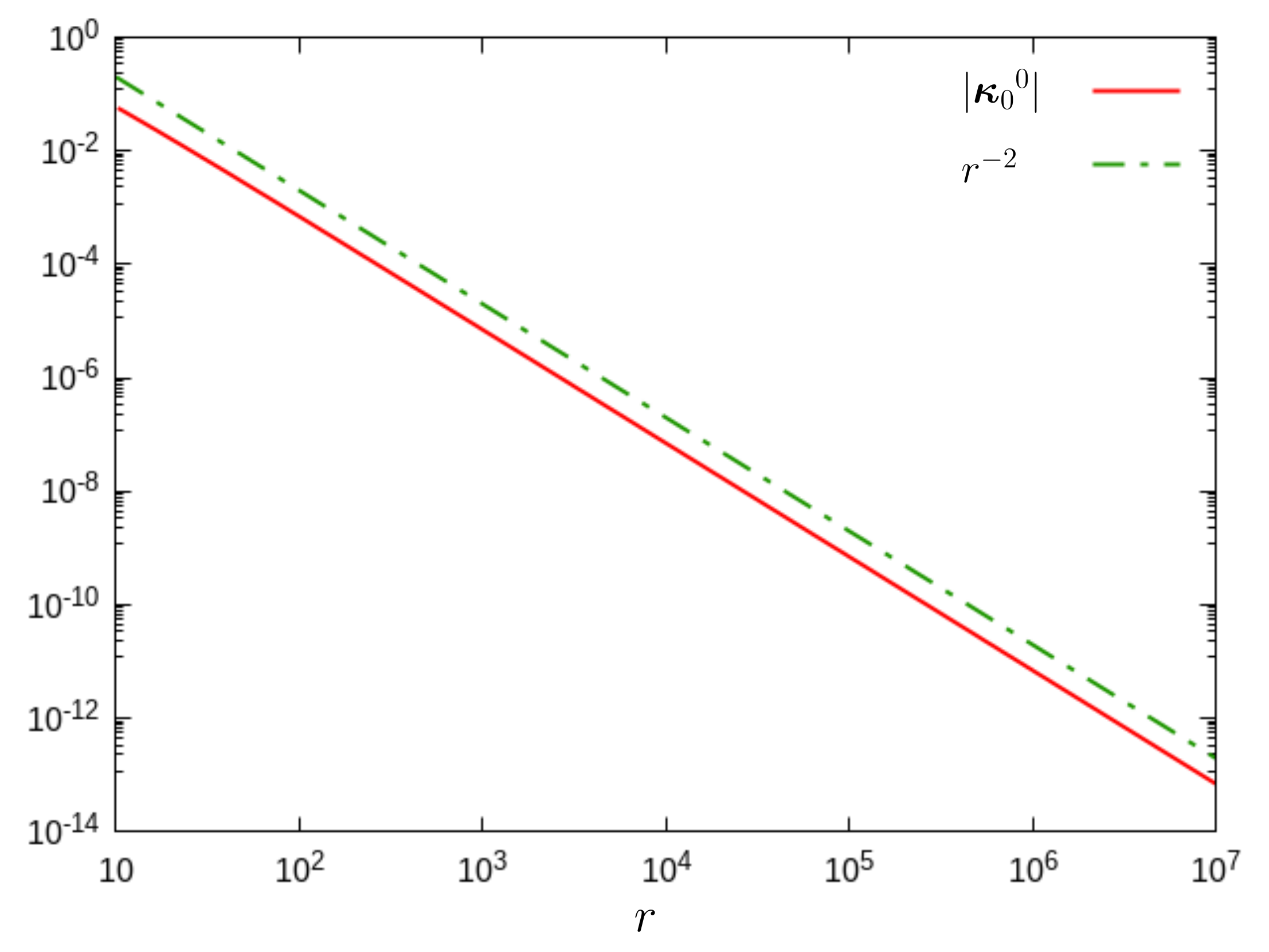}
				\vskip-0.4cm
				\caption{\scriptsize The decay rate of $\boldsymbol{\kappa}{\,}_0{}^0$ is close to $r^{-2}$. }
				\label{fig:decahkap0}
			\end{subfigure}
			\begin{subfigure}{0.48\textwidth}\vskip0.2cm
				\includegraphics[width=\textwidth]{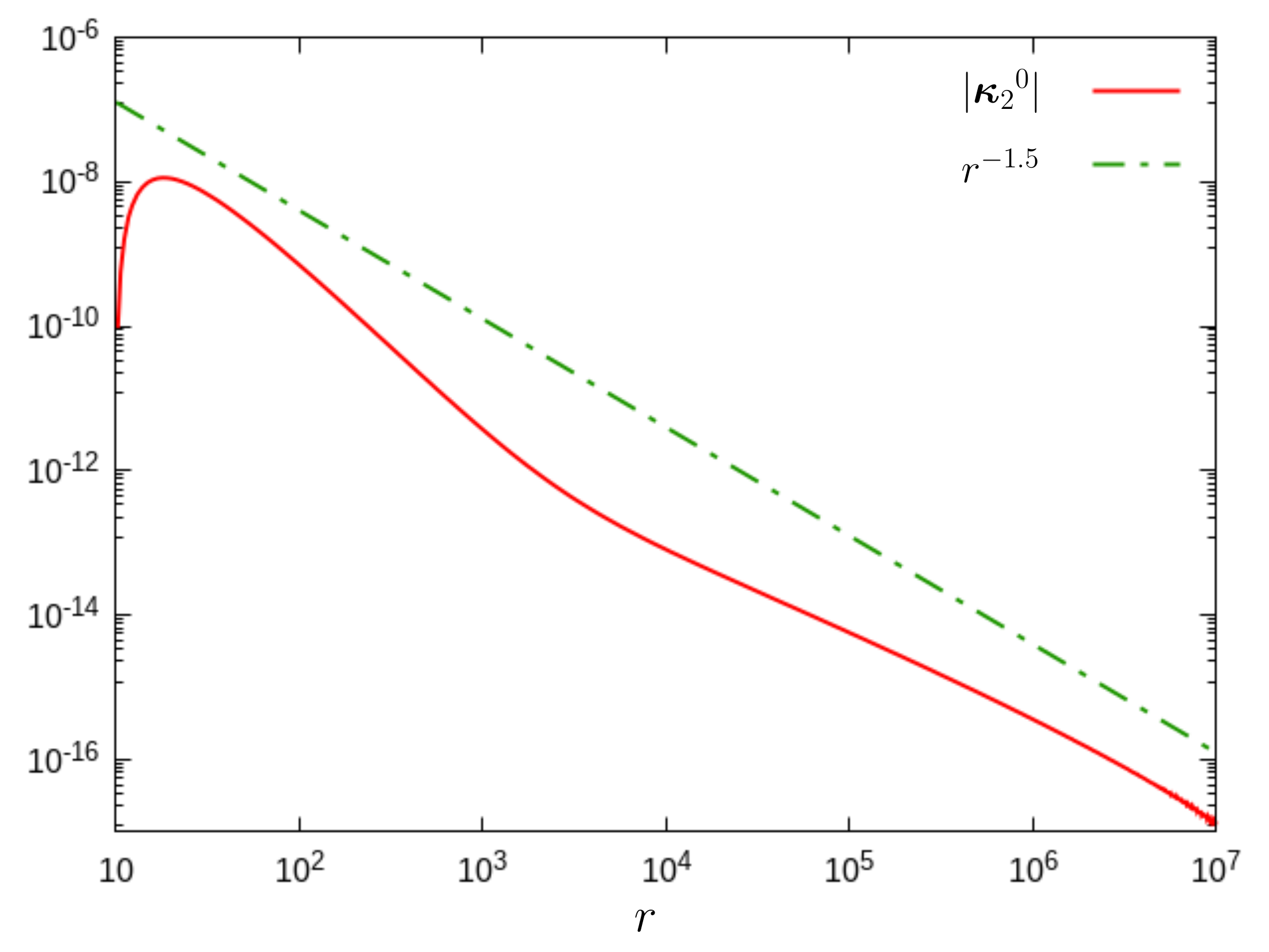}
				\vskip-0.4cm
				\caption{\scriptsize The decay rate of $\boldsymbol{\kappa}{\,}_2{}^0$ 
					is everywhere better than $r^{-3/2}$.}
				\label{fig:decahkap1}
			\end{subfigure}
			\begin{subfigure}{0.48\textwidth}\vskip0.2cm
				\includegraphics[width=\textwidth]{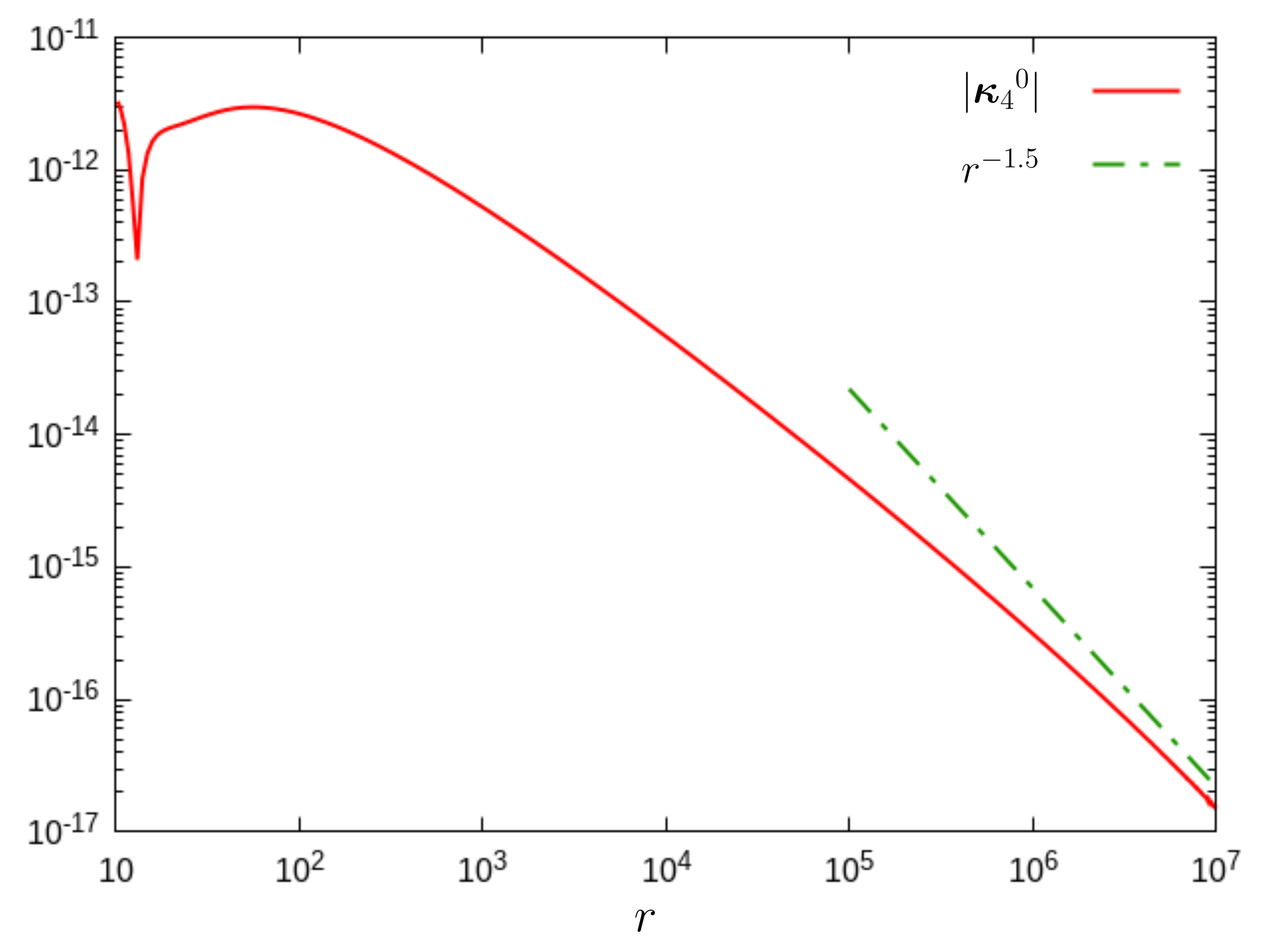}
				\vskip-0.4cm
				\caption{\scriptsize The decay rate of the mode $\boldsymbol{\kappa}{\,}_4{}^0$ 
					is better than $r^{-3/2}$.}
				\label{fig:decahkap2}
			\end{subfigure}
			\hskip.5cm
			\begin{subfigure}{0.48\textwidth}\vskip0.2cm
				\includegraphics[width=\textwidth]{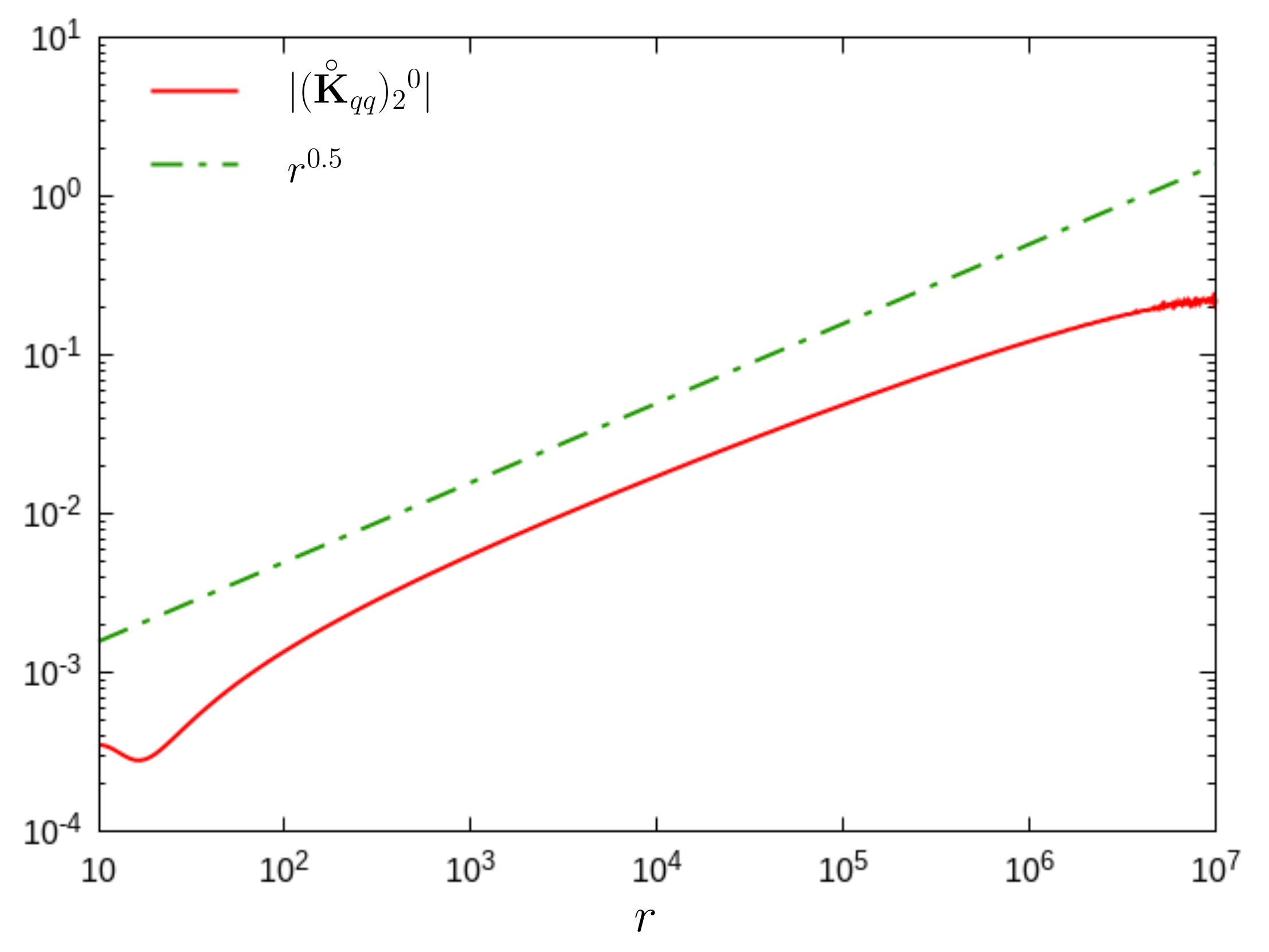}
				\vskip-0.4cm
				\caption{\scriptsize The growth rate of $\Kcqq$ is less than $r^{1/2}$.}
				\label{fig:dec-ah-Kcqq}
			\end{subfigure}
		}
	\end{centering}
	\vskip-0.1cm
	\caption{\footnotesize  The non-linear perturbative form of the algebraic-hyperbolic system was integrated numerically by applying the initial perturbation $\dKK|_{\mathscr{S}_{r_0}} = 6.5\cdot 10^{-12}\cdot Y_0{}^0+10^{-15}\cdot Y_2{}^0$. It is shown that all the $\ell=0,2,4$ modes of $\boldsymbol{\kappa}$ decay definitely better than $r^{-3/2}$. It is also clearly depicted that the only non-vanishing $s=2, \ell=2, m=0$ mode of ${\overset{\circ}{\mathbf{K}}}_{qq}={\overset{\circ}{\mathbf{K}}}_{AB}q^Aq^B$ does grow slower than $r^{1/2}$. Thereby, the asymptotic behavior of $\boldsymbol{\kappa}$ and ${\overset{\circ}{\mathbf{K}}}_{qq}$ allows the corresponding initial data configuration to be weakly asymptotically flat.}
	\label{fig:dec-ah-kap-Kcqq}
\end{figure}

After all of these technicalities it is time to inspect the numerical results depicted on Figs.\,\ref{fig:dec-ah-KK-kk} and \ref{fig:dec-ah-kap-Kcqq}. In producing the solution, characterized by these figures, the parameters $\delta$ and $\rho$, in \eqref{eq: delta-rho}, were fixed as $\delta=10^{-9}$ and $\rho=1.7$. In addition, the tiny initial data $\dKK|_{\mathscr{S}_{r_0}} = 6.5\cdot 10^{-12}\cdot Y_0{}^0+10^{-15}\cdot Y_2{}^0$ was used. On the left panels in Fig.\,\ref{fig:dec-ah-KK-kk} clearly indicated that the monopole part $\dKK{\,}_0{}^0$ of $\dKK$, as expected in virtue of \eqref{eq: ansatz-ah-gen00}, decays as $r^{-1.7}$, whereas all the higher mode excitations with $\ell=2,4$  of $\dKK$ decay close but slightly better than $r^{-3/2}$. Analogously, on the right panel Fig.\ref{fig:dec-ah-KK-kk} the excited $\dkk_{\ell}{}^0$ modes of $\dkk$ with $\ell=2,4$ are seen to decay slightly better than $r^{-1/2}$. Note also that, as it is depicted by first three panels in Fig.\,\ref{fig:dec-ah-kap-Kcqq}, the $\ell=0,2,4$ modes of $\boldsymbol{\kappa}$ decay a bit better than the corresponding $\ell$-modes of $\dKK$ do. Finally, on the last panel \ref{fig:dec-ah-Kcqq} in Fig.\,\ref{fig:dec-ah-kap-Kcqq} it is also verified that $\Kcqq$ grows slower than $r^{1/2}$ which allows it to fit to a weakly asymptotically flat initial data set. 
	Accordingly, Table\,\ref{table:falloff}, along with the fall off properties of the constrained fields, guarantees that the pertinent initial data set is indeed weakly asymptotically flat.

\medskip

All in all, the panels in Figs.\,\ref{fig:dec-ah-KK-kk} and \ref{fig:dec-ah-kap-Kcqq} verify that by invoking the freely specifiable field $\Kc_{AB}$, supplemented by the analytic procedure described in the first part of this subsection to relate $\Kc_{AB}$ to the constraint fields leaves by leaves, the yielded initial data set satisfies all the conditions listed in subsection\,\ref{subsec: asymp-flatness}, in particular, in Table\,\ref{table:falloff}, required to hold for weakly asymptotically flat initial data configurations. Accordingly, this example indicates that, by choosing the freely specifiable field $\Kc_{AB}$ suitably, the time integration of the algebraic-hyperbolic system can also be used to produce asymptotically flat initial data configurations.

\section{Final remarks}\label{sec: final-remarks}
\setcounter{equation}{0}

The two novel alternative evolutionary formulations---the parabolic-hyperbolic and the algebraic-hyperbolic forms---of the constraint equations were used to investigate the asymptotic behavior of ``near Schwarz\-schild'' vacuum initial data sets. In doing so a delicate combination of analytic and numerical investigations had to be applied. To generate near Schwarzschild configurations by making use of the evolutionary methods---after fixing eight of the completely freely specifiable variables---, as an input data only the constrained variables can be chosen freely on a two-sphere at some finite location. In principle, this should not allow to have any control on the asymptotic behavior of the pertinent solutions. Therefore,  it is of fundamental interest to know if  asymptotically flat initial data can be produced by the evolutionary forms of the constraints and if so how this can be done.

\medskip

For this reason, in this paper, a systematic investigation of the asymptotic behavior of near Schwarz\-schild vacuum initial data sets is carried out by making use of numerical solutions to the evolutionary form of the constraints. Special attention was given to the asymptotic behavior of the yielded initial data configurations. To get a really adequate picture in addition to applying the full set of evolution equations the use of their non-linear perturbative forms---their explicit form was worked out in details and they are presented in appendix of the present paper---turned out to be of fundamental importance.

\medskip

One of our most important findings is that, for strictly near Schwarzschild initial data specifications, apart from the  monopole part of the trace of the extrinsic curvature's tensorial projection ${\rm\bf K}=\widehat \gamma{}^{kl}K_{kl}$ all modes of the constrained variables---$\KK$ and $\kk$ in the algebraic-hyperbolic case and $\NNh$, $\KK$ and $\kk$ in the  parabolic-hyperbolic case---decay sufficiently fast to support the strong asymptotic flatness of the solutions to the evolutionary form of the constraints. Nevertheless, the monopole part of ${\rm\bf K}$ was found to decay far too slow to allow asymptotically flat initial data configurations. Therefore, getting control on the decay rate of ${\rm\bf K}$ is really of fundamental importance in producing asymptotic flatness of initial data configurations.

\medskip

In section \ref{subsec: revisit} it is demonstrated---by means of specific examples relevant for the  parabolic-hyperbolic and for the algebraic-hyperbolic equations, separately---that the evolutionary forms of the constraints can also be used to produce asymptotically flat initial data configurations. As demonstrated in \cite{Beyer:2020kty}, the time integration of the parabolic-hyperbolic system is capable to produce strongly asymptotically flat near Schwarzschild initial data configurations. This, as verified in subsection \ref{subsub: Florian-ansatz}, requires only to set the otherwise freely specifiable scalar field $\boldsymbol{\kappa}$ to be proportional to the constraint variable ${\bf K}$.
It is important to emphasize that there is no direct analogue of this process in the interim of the time integration of the algebraic-hyperbolic system involving $\boldsymbol{\kappa}$ as one of the dynamical variables. Nevertheless, as outlined in subsection \ref{subsub: alg-hyp-ansatz}, by invoking the trace free part $\Kc_{ij}$ of the tensorial projection $\KK_{ij}$ of the three-dimensional extrinsic curvature tensor $K_{ij}$, the time integration of the algebraic-hyperbolic system can also be used to produce asymptotically flat initial data configurations, though the one presented in the this paper was found to be merely weakly asymptotically flat and it was limited to permissibly tiny perturbations.

\medskip

In closing the present paper it is worth recalling, as our most important message to be conveyed, that both the parabolic-hyperbolic and the algebraic-hyperbolic forms of the constraints can be used to generate asymptotically flat initial data sets. However preliminary these results are they definitely open the arena for future numerical investigations, and, more importantly, they also provide a firm invitation for complementary analytic studies.

%%%%%%%%%%%%%%%%%%%% ACKNOWLEDGMENTS %%%%%%%%%%%%%%%%%%%%%%%%%%%%%%%%%%%

\section*{Acknowledgments}

The first author was supported in parts by the NKFIH grants K-115434 and by the COST Action Gravitational waves, black holes and fundamental physics CA16104. IR was supported by the POLONEZ programme of the National Science Centre of Poland (under the project No. 2016/23/P/ST1/04195) which has received funding from the European Union`s Horizon 2020 research and innovation programme under the Marie Sk{\l}odowska-Curie grant agreement No.~665778.

\includegraphics[scale=0.21, height = 32pt]{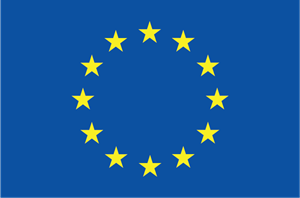}

\section*{Appendix: The non-linear perturbative equations}
\appendix

This appendix is to give the explicit non-linear perturbative forms of the parabolic-hyperbolic and  algebraic-hyperbolic systems relevant for the  splittings \eqref{eq: splitting-ph} and \eqref{eq: splitting-ah}.

\subsection*{The non-linear perturbative form of the parabolic-hyperbolic system}
\label{sec:dev_ph}
\renewcommand{\theequation}{P-H.\arabic{equation}}
\setcounter{equation}{0}

    %This was published in \cite{Nakonieczna:2017eev}

    \begin{align}
        \instar{K} \Big[ \partial_r\dNNh & - \tfrac12\, \NNt\big(\ethb\,\dNNh\big)- \tfrac12\, \NNtb\big(\eth\,\dNNh\big) \Big]  \nonumber\\ &
        - \tfrac12\,\dd^{-1} \bigg\{ \left( \dNNh^2 + 2\:\bgNNh\,\dNNh \right) \Big[ \aaa \left( \eth\ethb\,\dNNh - \BB\big(\ethb\,\dNNh\big) \right)  \nonumber\\ &
        - \bb \left( \ethb^2\dNNh - \tfrac12\, \AAAb\big(\ethb\,\dNNh\big) - \tfrac12\, \CCb\big(\eth\,\dNNh\big) \right) + cc. \Big]
        \nonumber\\ &
        + \left( \dNNh^2 + 2\:\bgNNh\,\dNNh \right) \Big[ \aaa \left( \eth\ethb\:\bgNNh - \BB\big(\ethb\:\bgNNh\big) \right)
        \nonumber\\ &
        - \bb \left( \ethb^2\:\bgNNh - \tfrac12\, \AAAb\big(\ethb\:\bgNNh) - \tfrac12\, \CCb\big(\eth\:\bgNNh\big) \right) + cc. \Big]
        %\nonumber\\ &
        + \bgNNh^2 \Big[ \aaa \left( \eth\ethb\,\dNNh - \BB\big(\ethb\,\dNNh\big) \right)
        \nonumber\\ &
        - \bb \left( \ethb^2\dNNh - \tfrac12\, \AAAb\big(\ethb\,\dNNh\big) - \tfrac12\, \CCb\big(\eth\,\dNNh\big) \right) + cc. \Big] \bigg\} - {}^{\scriptscriptstyle(\!0\!)}\!\mathcal{A}\,\dNNh
        \nonumber\\ &
         - \bgNNh^3\,{}^{\scriptscriptstyle(\!\Delta\!)\!}\mathcal{B} - \left( \dNNh^3 + 3\:\bgNNh\,\dNNh^2 + 3\:\bgNNh^2\,\dNNh \right) \left( {}^{\scriptscriptstyle(\!\Delta\!)\!}\mathcal{B} + {}^{\scriptscriptstyle(\!0\!)\!}\mathcal{B} \right)=0,
    \end{align}
    where
    \begin{align}
        \mathcal{A} =\ {}^{\scriptscriptstyle(\!0\!)}\!\mathcal{A}, \hskip1.1cm & \\
        {}^{\scriptscriptstyle(\!\Delta\!)\!}\mathcal{B} = \,\mathcal{B}-{}^{\scriptscriptstyle(\!0\!)\!}\mathcal{B} = &  -\kkappa\,\dKK - \tfrac{1}{4} \left( \dKK^2 + 2\:\bgKK\,\dKK \right)  \nonumber\\
        &+ \tfrac12\,\dd^{-1} \Big[ 2\aaa \left( \dkk\,\dkkb + \bgkk\,\dkkb + \dkk\,\bgkkb \right) \nonumber\\
        &- \bb \left( \dkkb{}^2 + 2\:\bgkkb\,\dkkb \right) - \bbb \left( \dkk^2 + 2\:\bgkk\,\dkk \right) \Big].
    \end{align}

    \begin{align}
        \partial_r\dkk &  - \tfrac12 \,\NNt\big(\ethb\,\dkk\big) -\tfrac12 \,\NNtb\big(\eth\,\dkk\big) \nonumber\\
        &- \tfrac12\, \left[ \dNNh\,\big(\eth\,\dKK\big) + \bgNNh\,\big(\eth\,\dKK\big) + \dNNh\,\big(\eth\,\bgKK\big) \right] + \dff\,=0,
    \end{align}
    where
    \begin{align}
        \dff = \ff & -{}^{\scriptscriptstyle(\!0\!)\!}\ff = -\tfrac12\, \left[ \dkk \big(\eth\,\NNtb\big) + \dkkb \big(\eth\,\NNt\big) \right]  \nonumber\\
        & + \tfrac12\, \left[ \dKK\big(\eth\,\dNNh\big) + \dKK\big(\eth\,\bgNNh\big) + \bgKK\big(\eth\,\dNNh\big) \right]
        - \kkappa\,(\eth\,\dNNh) - \big(\eth\kkappa\big)\dNNh + \instar{K}\,\dkk  \nonumber\\
        & + \tfrac12\,\dd^{-1} \left[  \big[\aaa\big(\eth\,\dNNh\big) - \bb\big(\ethb\,\dNNh\big)\big]\big(\interior{\mathbf{K}}_{ij}{q^i\overline q{}^j}\big)
        +\big[\aaa\big(\ethb\,\dNNh\big)
        - \bbb\big(\eth\,\dNNh\big)\big]\big(\interior{\mathbf{K}}_{ij}{q^i q^j}\big) \right]  \nonumber\\
        & + \big(q^i\widehat{D}^l \Kc_{li}\big)\dNNh\,.
    \end{align}

    \medskip

    Finally,
    \begin{align}
        \partial_r\dKK & - \tfrac12\, \NNt\big(\ethb\,\dKK\big) -\tfrac12\, \NNtb\big(\eth\,\dKK\big) \nonumber\\
        &- \tfrac12\,\dd^{-1} \bigg\{ \dNNh \left[ \aaa \Big( \eth\,\dkkb + \ethb\,\dkk \Big)
        - \bb \big(\ethb\,\dkkb\big) - \bbb\big(\eth\,\dkk\big) \right] \nonumber\\
        &+ \bgNNh \left[ \aaa \Big( \eth\,\dkkb + \ethb\,\dkk \Big)
        - \bb\big(\ethb\,\dkkb\big) - \bbb\big(\eth\,\dkk\big) \right] \nonumber\\
        &+ \dNNh \left[ \aaa\left(\eth\,\bgkkb + \ethb\,\bgkk\right) - \bb\big(\ethb\,\bgkkb\big) - \bbb\big(\eth\,\bgkk\big) \right] \bigg\}
        + \dFF\,=0,
    \end{align}
    where
    \begin{align}
        \dFF =\FF -{}^{\scriptscriptstyle(\!0\!)\!}\FF =  &
        \tfrac{1}{4}\,\dd^{-1} \Big\{ \dNNh \Big[ 2\,\aaa\,\BB\,\dkkb - \bb \left( \CCb\,\dkk + \AAAb\,\dkkb \right) \nonumber\\
        &+ 2\,\aaa\,\BB\,\bgkkb - \bb \left( \CCb\,\bgkk + \AAAb\,\bgkkb \right) + cc. \Big] \nonumber\\
        &+ \bgNNh \Big[ 2\,\aaa\,\BB\,\dkkb - \bb \left( \CCb\,\dkk + \AAAb\,\dkkb \right) + cc. \Big]\Big\} \nonumber\\
        & -\dd^{-1}\Big[ \left(\aaa\,\dkkb - \bbb\,\dkk\right) \left( \eth\,\dNNh + \eth\,\bgNNh \right) \nonumber\\
        &+ \left( \aaa\,\bgkkb -\bbb\,\bgkk \right) \eth\dNNh + cc. \Big] + \tfrac12 \,\instar{K}\,\dKK\,.
    \end{align}

\subsection*{The non-linear perturbative form of the algebraic-hyperbolic system}
    \label{sec:dev_ah}
\renewcommand{\theequation}{A-H.\arabic{equation}}
\setcounter{equation}{0}

    \begin{align}
        \partial_r\dKK & -\tfrac12\,\NNt\,\ethb\,\dKK-\tfrac12\,\NNt\,\eth\,\dKK \nonumber \\
        & -\tfrac{1}{2}\,\NNh\,\dd^{-1}\Big[\aaa(\eth\,\dkkb+\ethb\,\dkk)-\bb\,\ethb\,\dkkb-\bbb\,\eth\,\dkk\Big]+\dFF=0,
    \end{align}
    \begin{align}
        \partial_r\dkk & - \tfrac12\,\NNt\,\ethb\,\dkk-\tfrac12\,\NNtb\,\eth\,\dkk \nonumber \\
        & +{\NNh}\,{\bgKK}^{-1}\Big\{\dkappa\,\eth\,\bgKK-\dd^{-1}\big[(\aaa\dkk-\bb\,\dkkb)\eth\,\bgkkb+(\aaa\dkkb-\bbb\,\dkk)\eth\,\bgkk\big]\Big.\nonumber \\
        &+\Big.\kkappa\,\eth\,\dKK-{\dd}^{-1}\big[(\aaa\kk-\bb\,\kkb)\eth\,\dkkb+(\aaa\kkb-\bbb\,\kk)\eth\,\dkk\big]\Big\}
        \nonumber \\  &-{\NNh\,\dKK}({\KK\,\bgKK})^{-1}\big\{\kkappa\,\eth\,\KK-{\dd}^{-1}\big[(\aaa\kk-\bb\,\kkb)\eth\,\kkb+(\aaa\kkb-\bbb\,\kk)\eth\,\kk\big]\big\}+\dff=0
    \end{align}
    where the lower order source terms are
    \begin{align}
        \dFF= \ & \tfrac{1}{4}\,\NNh\,\dd^{-1}\Big[2\,\aaa\,\BB\,\dkkb-\bb(\CCb\,\dkk+\AAAb\,\dkkb)+cc.\Big]\nonumber \\
       & -\dd^{-1}\Big[(\aaa\dkkb-\bbb\,\dkk)\eth\,\NNh+cc.\Big]
        -\big(\dkappa-\tfrac12\,\dKK\big)\instar{K}
    \end{align}
    and
    \begin{align}
        \dff= \ & -\tfrac12\,\Big[\dkk\,\eth\NNtb+\dkkb\,\eth\NNt\Big]\nonumber \\
        &  +\tfrac{1}{2}\NNh\,(\dd\,\bgKK)^{-1}\Big[(\aaa\dkk-\bb\,\dkkb)(\BBb\,\bgkk+\BB\,\bgkkb)+(\aaa\kk-\bb\,\dkk)(\BBb\,\dkk+\BB\,\dkkb)\nonumber \\
        & +(\aaa\dkkb-\bbb\,\dkk)(\CC\,\bgkkb+\AAA\bgkk)+(\aaa\kkb-\bbb\,\kk)(\CC\,\dkkb+\AAA\dkk)\Big]\nonumber \\
        & -\tfrac12\,{\NNh\,\dKK}\,{(\dd\cdot\KK\,\bgKK)^{-1}}\Big[(\aaa\kk-\bb\,\kkb)(\BBb\,\kk+\BB\,\kkb)+(\aaa\kkb-\bbb\,\kk)(\CC\,\kkb+\AAA\kk)\Big]\nonumber \\
        & -\left(\dkappa-\tfrac12\,\dKK\right)\eth\NNh-\tfrac12\,{\NNh\,\dKK}\,{(\KK\,\bgKK)^{-1}}\,\eth\kkappa_0+\instar{K}\,\dkk %-{}^{\scriptscriptstyle(\!\Delta\!)}\hspace{-1.2pt}(q^i\dot{\widehat{n}}{}^l\Kc_{li})+{}^{\scriptscriptstyle(\!\Delta\!)}\hspace{-1.2pt}(q^i\widehat{D}^l\Kc_{li})
    \end{align}

%%%%%%%%%%%%%%%%%%%% REFERENCES %%%%%%%%%%%%%%%%%%%%%%%%%%%%%%%%%%%%%%%%

\end{document}